\documentclass[12pt]{article}

\usepackage{fancyhdr}

\usepackage[margin=0.75in]{geometry}
\usepackage{graphicx}
\usepackage{float}
\usepackage{subfig}
\usepackage{booktabs}
\usepackage{xcolor}
\usepackage{multirow}
\usepackage{wrapfig}
\usepackage{adjustbox}
\usepackage{threeparttable}
\usepackage{listings}
\usepackage[linesnumbered, ruled]{algorithm2e}
\usepackage{placeins}

\definecolor{ForestGreen}{RGB}{63,142,38}
\definecolor{UnibasMint}{RGB}{30,165,165}

\newcommand{\aliA}[1]{{\color{black}#1}}
\newcommand{\aliB}[1]{{\color{black}#1}}
\newcommand{\aliC}[1]{{\color{black}#1}}
\newcommand{\aliD}[1]{{\color{black}#1}}
\newcommand{\discuss}[1]{{\color{black}#1}}

\newcommand{\flo}[1]{{\color{black}#1}}
\newcommand{\ali}[1]{{\color{black}#1}}

\newcommand{\simdag}{\mbox{SG-SD}}
\newcommand{\msg}{\mbox{SG-MSG}}
\newcommand{\smpi}{\mbox{SG-SMPI}}
\newcommand{\simgrid}{\mbox{SG}}
\newcommand{\sil}{\textit{SimAS}}
\newcommand{\dlbtool}{\textit{DLB\_tool}}
\newcommand{\dlb}{\textit{DLS4LB}}
\newcommand{\lsim}{\textit{LoopSim}}

\newcommand{\cut}[1]{}

\begin{document}
\sloppy

\title{SimAS: A \underline{Sim}ulation-assisted Approach for the Scheduling \underline{A}lgorithm \underline{S}election under Perturbations}

\author{Ali Mohammed and Florina M. Ciorba\\
	Department of Mathematics and Computer Science\\
	University of Basel, Switzerland\\
}	




\maketitle

\clearpage
\tableofcontents
\clearpage
\abstract
{
\label{sec:abstract}


Many scientific applications consist of large and \mbox{computationally-intensive} loops, such as N-body, Monte Carlo, and computational fluid dynamics }
These loops contain \mbox{computationally-intensive} operations, resulting in heavy loop bodies. 
Dynamic loop \mbox{self-scheduling}~(DLS) techniques are used to \aliD{parallelize and to balance the load during the} execution of such applications. 
Load imbalance \aliD{arises from} variations in \aliD{the} loop iteration (or tasks) execution times,  \aliD{caused by} problem, algorithmic, or systemic characteristics. 
The variations in systemic characteristics are referred to as perturbations, and can be caused by other applications or processes that share the same resources, or a temporary system fault or malfunction and include, decreased delivered computational speed, reduced available network bandwidth, or larger network latencies.
DLS achieves a balanced load execution of scientific applications on high-performance computing~(HPC) systems.
Therefore, the selection of the most efficient DLS technique is critical to achieve the best application performance.
The following question motivates this work:~\emph{``Given an application, an HPC system, and their characteristics and interplay, which DLS technique will achieve improved performance under unpredictable perturbations?''} 
Existing studies focus on variations in the delivered computational speed only as the source of perturbations in the system.
However, perturbations in available network bandwidth or latency are inevitable on production HPC systems. 
Also, scheduling solutions based on optimization techniques, e.g., evolutionary algorithms, can not adapt to perturbations during execution.
The alternative of using machine learning for DLS selection requires training and learning either prior to execution or during previous \mbox{time-steps} in \mbox{time-stepping} applications.
\emph{\aliD{A} Simulator-assisted scheduling} (\sil{}) is introduced as a new \mbox{control-theoretic-inspired approach} to dynamically select DLS techniques that improve the performance of applications \aliD{executing} on heterogeneous HPC systems under perturbations.
The present work examines the performance of seven applications on a heterogeneous system under \emph{all \aliC{the} above system perturbations}.\cut{in the delivered computational speed, available network bandwidth and latency.}
\sil{} \aliD{is evaluated as a} \emph{proof of concept} using native and simulative experiments.\cut{the SimGrid simulation toolkit.}
The performance results confirm the \aliD{original} hypothesis that \emph{no single DLS technique can deliver \aliD{the absolute} best performance in all scenarios}, whereas the \mbox{\sil{}-based} DLS selection \aliD{resulted in} improved application performance in most experiments.
\paragraph*{Keywords.}
	Performance, loop scheduling, load balancing, heterogeneous computing systems, perturbations, simulation
%

\clearpage
\section{Introduction}
\label{sec:intro}

Scientific applications are often characterized by large and computationally-intensive parallel loops. 
The performance of these applications on high-performance computing (HPC) systems may degrade due to load imbalance caused by problem, algorithmic, or systemic characteristics. 
Application (problem or algorithmic) characteristics include \ali{the irregularity of the} number of computations per loop iterations due to conditional statements, \ali{where} systemic characteristics include variations in delivered computational speed of processing elements~(PEs), available network bandwidth or latency.
Such variations are referred to as perturbations, and can also be caused by other applications or processes that share the same resources, or a temporary system fault or malfunction.
%
Dynamic loop \mbox{self-scheduling}~(DLS) is a widely-used approach for improving the execution of parallel applications using self-scheduling, that is, \emph{dynamic} assignment of the loop iterations to free and requesting processing elements.
A wide range of DLS techniques exists and can be divided into \emph{nonadaptive} and \emph{adaptive} techniques. 
The nonadaptive DLS techniques account for the variability in loop iterations execution times due to application characteristics \aliD{via modeling their assumptions}. 
The nonadaptive DLS techniques include \emph{self-scheduling}~(SS), \emph{fixed size chunking}~(FSC)~\cite{FSC}, \aliC{\emph{modified fixed size chunking}~(mFSC)~\cite{banicescu:2013:a},} \emph{guided \mbox{self-scheduling}}~(GSS)~\cite{GSS}, \aliC{\emph{trapezoid \mbox{self-scheduling}~(TSS)~\cite{tzen1993trapezoid}}}, \emph{factoring}~(FAC)~\cite{FAC}, and \emph{weighted factoring}~(WF)~\cite{WF} \aliD{among others}.  
The adaptive DLS techniques account for irregular system characteristics \aliD{that are only known during execution} by adapting the amount of work assigned (chunk size) per PE request according to the application performance measured during execution.
Adaptive DLS techniques include \emph{adaptive weighted factoring}~(AWF)~\cite{AWF}, its variants \emph{batch}~(AWF-B), \emph{chunk}~(AWF-C), \emph{batch-like}~(AWF-D), \aliD{and} \emph{chunk-like}~(AWF-E)~\cite{AWFBC}, \aliD{as well as} \emph{adaptive factoring}~(AF)~\cite{AF}, \aliD{among others}. 

An \emph{a priori} selection of the most appropriate DLS technique for a given application and system is \aliD{not trivial}, given the various sources of load imbalance and the different load balancing properties of the DLS techniques. 
This observation raises the following question and motivates the present work: \emph{``Given an application, an HPC system, and their characteristics and interplay, which DLS technique will achieve improved performance under unpredictable perturbations?''}
Earlier work studied the flexibility of DLS~(\aliD{taken as} robustness to \aliD{variable} delivered computational speed) and the selection of \aliD{the most} robust DLS using machine learning~\cite{sukhija:2014:a} \aliA{with} the SimGrid~(\simgrid{})~\cite{SimGrid} simulation toolkit.
The selection of DLS techniques for synthetic time-stepping scientific workloads using reinforcement learning was also studied using \simgrid{}~\cite{Boulmier:2017a}.\cut{ where smart agents learn from application performance in previous time-steps to select a DLS that would improve the performance.} 
The aforementioned work focuses on one source of perturbations, \aliD{namely the variation in the delivered computing speed,} in time-stepping applications to learn from previous \aliD{time} steps. 
That approach may not be applicable to \aliD{non-iterative} applications.
Scheduling solutions using static optimizations, e.g., using evolutionary \ali{and genetic} algorithms, can not dynamically adapt to the perturbations encountered during execution.
Modern HPC systems are often heterogeneous production systems typically shared by many users.
Therefore, perturbations in the available network bandwidth and latency \aliD{are unavoidable} in such systems. 

\begin{figure}[]
		\centering
		\centering
		\subfloat[Variation of PSIA performance with various perturbations per scheduling technique.]{
			\includegraphics[clip, trim=0cm 0cm 0cm 0cm,scale=0.52]{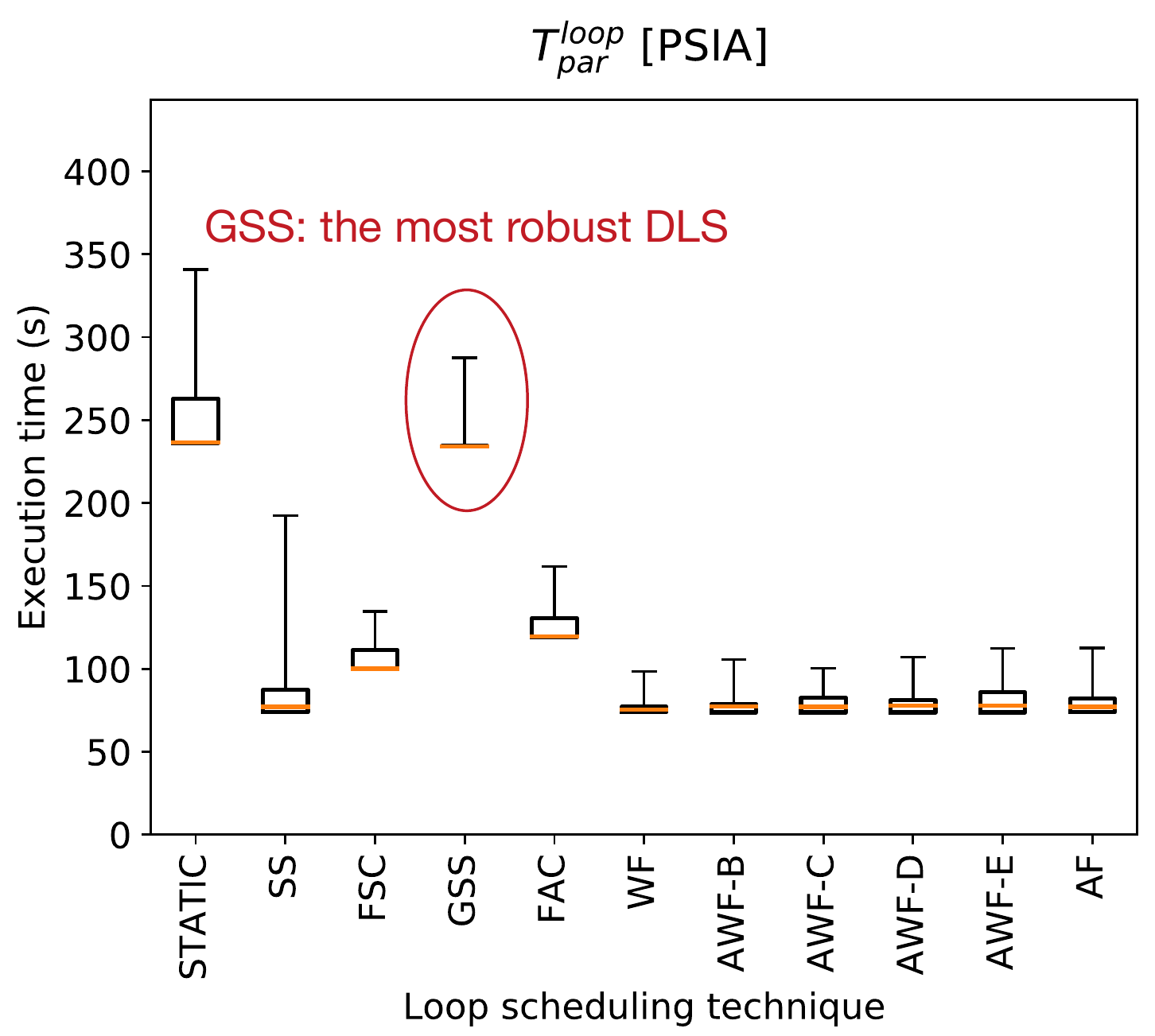}%
			\label{subfig:psia_boxes}%
		} \hspace{0cm}
		\subfloat[Application performance under various perturbations using different scheduling techniques.]{%
			\includegraphics[clip, trim=0cm 0cm 0cm 0cm,scale=0.52]{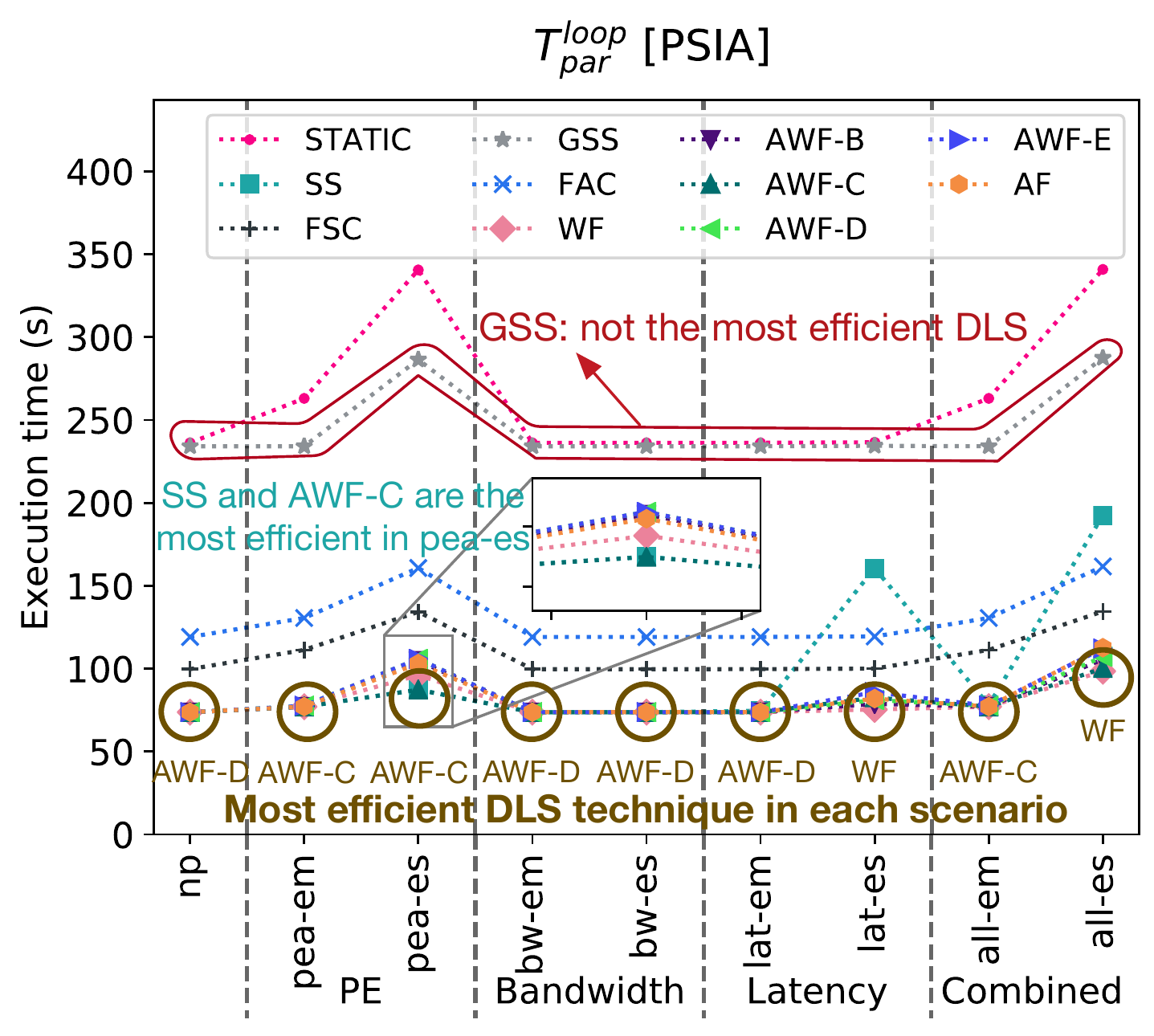}
			\label{subfig:psia_lines}%
		} 
		\\	
		\caption{\textbf{Simulative} performance of PSIA~(c.f. Section~\ref{subsec:app}) under perturbations in computing availability, network bandwidth, and latency (c.f. Section~\ref{subsec:perts}). 
				The most robust DLS technique (GSS in ~\figurename{~\ref{subfig:psia_boxes}}) delivers consistent performance under various system perturbations.  However, GSS does not achieve the best performance under all perturbations as shown in~\figurename{~\ref{subfig:psia_lines}}.
				As shown in the figure, \emph{no single technique delivers the best performance in all execution scenarios}~\cite{Mohammed:2018p}.} 
		\label{fig:motiv}
\end{figure}

The study of the performance of scientific applications with DLS under perturbations revealed that the most robust DLS technique, identified as the DLS technique that results in the least variation of the application execution time under various perturbations, does not always achieve the best performance in all execution scenarios~\cite{Mohammed:2018p}.
\figurename{~\ref{fig:motiv}} shows the simulative performance of PSIA~(c.f. Section~\ref{subsec:app}) on $696$~cores of miniHPC~(c.f. Section~\ref{subsec:comp}) under perturbations~(c.f. Section~\ref{subsec:perts}). 
According to these results, GSS is the most robust DLS technique due to the minimal variation of its performance under perturbations~(\figurename{\ref{subfig:psia_boxes}}), however, it results in poor application performance under perturbations.
Even the next most robust DLS technique, WF, is outperformed by SS and AWF-C in certain perturbation scenarios, as can be seen in~\figurename{\ref{subfig:psia_lines}}.
These results suggest that even if the most robust DLS technique is known a priori, which could be challenging, the application performance degrades in different execution scenarios due to perturbations. 
Therefore, a methodology for the dynamic selection of DLS techniques is needed to achieve the highest possible performance in all execution scenarios.

In the present work, in an effort to \emph{select} the most appropriate DLS \emph{dynamically} for a given application and system under perturbations, the \mbox{\underline{Sim}ulation-assisted} scheduling \underline{A}lgorithm \underline{S}election, \sil{}, approach is proposed.
The performance of two scientific applications~(PSIA~\cite{Eleliemy:2017b} and Mandelbrot~\cite{mandelbrot1980fractal}) executed in \mbox{single-sweep} and \mbox{time-stepping} modes, and five synthetic workloads is studied on a heterogeneous HPC system using \mbox{nonadaptive} and adaptive DLS techniques, in the presence of perturbations in computing speed, network bandwidth, and network latency. 
The amount of operations in each loop iteration of the five synthetic workloads is assumed to follow five different probability distributions, namely: constant, uniform, normal, exponential, and gamma probability distributions.
The synthetic workloads are used to cover a broader spectrum of application load imbalance profiles beyond what is encountered in practice.

The present work makes the following contributions:
(1)~Proposes a novel \emph{simulator-assisted scheduling} (\sil{}\footnote{https://github.com/unibas-dmi-hpc/SimAS}) approach for dynamically selecting a DLS technique during execution based on the application characteristics and the present (monitored or predicted)\cut{ (or predicted for the near future)} state of the computing system; 
(2)~\aliC{Extends a dynamic load balancing tool (\dlbtool{}) \aliD{from the literature}~\cite{carino2007tool} for parallelizing scientific applications into \dlb{}\footnote{https://github.com/unibas-dmi-hpc/DLS4LB} with four more DLS techniques, namely SS, FSC, WF, and TSS. 
In addition, the \dlb{} is extended to support \sil{} \aliD{as the fourteenth option} to select DLS techniques dynamically during execution; }
(3)~\aliC{Evaluates the performance of two real applications (PSIA and Mandelbrot) and five synthetic workloads \aliD{using DLS techniques} under perturbations via native and simulative experiments;} 
(4)~Confirms the \aliD{original} hypothesis that no single DLS ensures the best performance in all execution scenarios considered.

This work is structured as follows. 
Section~\ref{sec:background} contains a brief review of the selected DLS techniques, the \simgrid{} simulation toolkit, as well as the work related to the performance of scheduling scientific applications with DLS in the presence of perturbations. 
The proposed \sil{} approach\cut{ for selecting a DLS technique in the presence of perturbations} \ali{is} discussed in Section~\ref{sec:sil}. 
\aliC{The factorial design of experiments, together with details \aliD{about} the DLS and \sil{} implementation \aliD{into the \dlb{}}, the HPC system \aliD{characteristics}, and \aliD{the} perturbations \aliD{injected} in native and simulative experiments are presented in Section~\ref{sec:exp}. 
The analysis of applications load imbalance and the evaluation of the performance of \aliD{the} applications under perturbations are discussed in Section~\ref{sec:evalaution}. }
The work concludes and outlines potential future work in Section~\ref{sec:conc}.

\section{Background and Related Work}
\label{sec:background}
\cut{In this section, a background on loop scheduling techniques and the \simgrid{} simulation toolkit is provided. Certain relevant work on robust loop scheduling is presented and discussed.}
 
\textbf{Loop scheduling.}
The aim of loop scheduling is to achieve a balanced load execution among parallel PEs with minimum scheduling overhead.
Loop scheduling can be divided into \emph{static} and \emph{dynamic}. 
In static loop scheduling, the loop iterations are divided and assigned to PEs before execution; both division and assignment remain fixed during execution. 
This work considers static (block) scheduling, denoted STATIC, each PE being assigned a chunk size equal to the number of iterations $N$ divided by the number of PEs $P$.
STATIC incurs \emph{minimum} scheduling overhead, compared to dynamic loop scheduling, and may lead to load imbalance for non-uniformly distributed tasks and/or on perturbed systems.

In \emph{dynamic loop scheduling}~(DLS), free and requesting PEs are assigned, via self-scheduling, loop iterations during execution. 
The DLS techniques can be categorized into \emph{nonadaptive} and \emph{adaptive} techniques. 
The nonadaptive DLS techniques considered in this work are: SS~\cite{SS}, FSC~\cite{FSC}, mFSC~\cite{banicescu:2013:a}, GSS~\cite{GSS}, TSS~\cite{tzen1993trapezoid}, FAC~\cite{FAC}, and WF~\cite{WF}.
While STATIC represents one scheduling extreme, SS represents the other scheduling extreme.
In SS, \ali{the} size of each chunk is one loop iteration.
This yields a high load balance with potentially very large scheduling overhead.
\ali{FSC} assigns loop iterations in chunks of fixed sizes, where the chunk size depends on the standard deviation of loop iteration execution times~$\sigma$ as an indication of its variation and the incurred scheduling overhead~$h$.
FSC requires this information ($h$ and $\sigma$) to be known before the execution to calculate the chunk size.
\aliC{mFSC alleviates the requirement of pre-calculating $h$ and $\sigma$, and calculates a fixed chunk size that results in a number of chunks equal to that produced by FAC (\aliD{described below}).}
GSS assigns loop iterations in chunks of decreasing sizes, where the size of a chunk is equal to the number of remaining unscheduled loop iterations $R$ \aliB{divided by} the number of PEs $P$.
\aliC{Similar to GSS, TSS assigns chunks of loop iterations in decreasing sizes. 
Unlike GSS, the chunk sizes decrease linearly, to ease the chunk calculation operation and \aliD{to} minimize the scheduling overhead.}
FAC employs a probabilistic modeling of loop characteristics \aliD{that takes into account the mean of iterations execution time $\mu$ and their standard deviation $\sigma$)} to calculate batch sizes that maximize the probability of achieving a load balanced execution.
\flo{A PE's chunk size is equal to the batch size divided by $P$.}
When $\mu$ and $\sigma$ are unavailable, \aliD{a practical implementation of FAC, \cut{denoted FAC2,}{}} assigns half of the remaining loop iterations $R$ \aliD{to} a batch. 
WF divides a batch \aliD{of iterations} into unequally-sized chunks, proportional to the relative PE speeds (\aliD{called} weights).
The PE weights need to be determined prior to the execution and \aliD{are assumed not to} change \aliD{during execution.}
This work considers the \emph{practical implementations} of FAC and WF. 
All nonadaptive DLS techniques account for variations in \aliD{the} iteration execution times due to application characteristics. 

The \emph{adaptive} DLS techniques \aliD{monitor} the performance of the application during execution and adapt the chunk calculation accordingly. 
Adaptive DLS techniques include AWF~\cite{AWF}, its variants~\cite{AWFBC}: AWF-B, AWF-C, AWF-D, AWF-E, and AF~\cite{AF}, \aliD{among others}.
AWF is designed for time-stepping applications.
It improves WF by \aliD{adapting} the relative weights of PEs during execution by \aliD{monitoring} their performance in each time-step. 
AWF-B relieves the time-stepping requirement in AWF, and measures the performance after each \emph{batch} to update the PE weights.
AWF-C is similar to AWF-B where weight updates are performed upon the completion of each \emph{chunk}, instead of a batch.
AWF-D is similar to AWF-B and considers the total chunk time (equal to the \aliD{sum of the} iteration execution times \aliD{in the chunk} plus the associated overhead of \aliD{the} PE to acquire the chunk) and all the book keeping operations to calculate and update the PE weights. 
AWF-B and AWF-C only consider the chunk iterations execution times.
AWF-E is similar to AWF-C by updating the PE weights on \emph{every chunk}. 
Yet AWF-E is also similar to AWF-D by also considering the total chunk time.
Unlike FAC, \ali{AF} dynamically estimates the values of $\mu$ and $\sigma$ during execution and updates them based on the measured performance of the PEs \aliD{on the executed loop iterations}.
%
 
\emph{Loop scheduling in simulation.}
SimGrid~\cite{SimGrid}~(\simgrid{}) is a versatile event-based simulation toolkit designed for the study of the behavior of large-scale distributed systems. 
It provides ready to use application programming interfaces~(APIs) to represent applications and computing systems through different interfaces: MSG~(\msg{}), SimDag~(\simdag{}), and SMPI~(\smpi{}).
%
%
\simgrid{} uses a simple \aliD{and} fast CPU computation model and verified \aliD{and more complex} network models~\cite{velho2009accuracy}, which render it well suited for the study of computationally-intensive \aliD{parallel and} distributed scientific applications.

Various studies have used \simgrid{} to \aliD{evaluate} the performance of applications with DLS techniques in different scenarios~\cite{Boulmier:2017a,sukhija:2014:a}.
\ali{To attain high trustworthiness in the performance results obtained with \simgrid{},} 
the implementation of the nonadaptive DLS techniques in \simdag{} has been verified~\cite{HPCS} by reproducing the results presented in the work that introduced factoring~\cite{FAC}.
The accuracy of \aliB{the performance results obtained by simulative} experiments against native experiments has recently \aliD{also} been quantified~\cite{Mohammed:2018a}.
The \aliD{present} work employs the \simdag{} interface to study the performance of scientific applications on a heterogeneous \aliD{computing} platform under perturbations. 

%

\noindent\textbf{Related work.}
\aliC{Scheduling of applications on large HPC systems involves many sources of uncertainties, e.g., task execution times and perturbations in the computing system. 
Therefore, many studies have focused on robust schedules that maintain certain performance requirements despite fluctuations in the behavior of the system~\cite{ali2004measuring}. 
Robust scheduling of tasks with uncertain execution and communication times was studied~\cite{canon2010evaluation}~\cite{yang2003rumr} using a multi-objective evolutionary algorithm and \aliD{using} dynamic scheduling, respectively. 
Moreover, the flexibility of dynamic loop scheduling techniques was examined~\cite{sukhija:2013:b} in an effort to select the most flexible technique using machine learning. 
However, a robust scheduling technique \aliD{may} not \aliD{always} guarantee the best performance in all possible execution scenarios \aliD{and for all application parameters (e.g. problem size and data distribution)}. 
Thus, dynamically selecting the best performing DLS technique is of paramount importance, given the broad spectrum of available DLS techniques, \aliD{each with unique} properties. 
Selecting the best performing DLS technique for \mbox{time-stepping} applications, using \aliD{reinforcement} learning was introduced~\cite{Boulmier:2017a} by adapting to the variations in the delivered computational speed during previous time-steps. 
In addition, machine learning and decision trees were used to select the best performing DLS technique dynamically from a portfolio of DLS techniques~\cite{sukhija:2014:a} and for multi-threaded applications parallelized with OpenMP~\cite{zhang2005runtime} or \aliD{with} Charm++~\cite{Menon:2017}.}
A \mbox{knowledge-based} rule was design~\cite{} to select a scheduling technique that improves the performance of the application.
Application and system characteristics need to be feed to the rule to select an appropriate scheduling technique for a portfolio of scheduling techniques. 
However, perturbations during execution was not considered, and the selected scheduling technique is not changed during execution.

Scheduling solutions based on optimization techniques, \aliD{such as}, genetic and evolutionary algorithms, can not adapt to perturbations during execution.
\ali{None of the aforementioned efforts} considered perturbations in network bandwidth or latency. 
\ali{This work complements the previous efforts} by studying the performance of scientific applications using nonadaptive and adaptive DLS techniques under different perturbation \aliD{scenarios} (variations in delivered computational speed, network bandwidth, and network latency) \ali{on} a heterogeneous computing system.
A new approach, namely \emph{simulator-assisted scheduling} (\sil{}) is introduced, to dynamically select DLS techniques that improve the performance of applications on heterogeneous system under multiple sources of perturbations \aliD{known mostly during execution}.
\section{Simulator-Assisted Scheduling Approach (\sil{})}
\label{sec:sil}

\begin{figure}[]
		\centering
		\centering
		\subfloat[A generic control system.]{
			\includegraphics[clip, trim=0cm 0cm 0cm 0cm,scale=0.75]{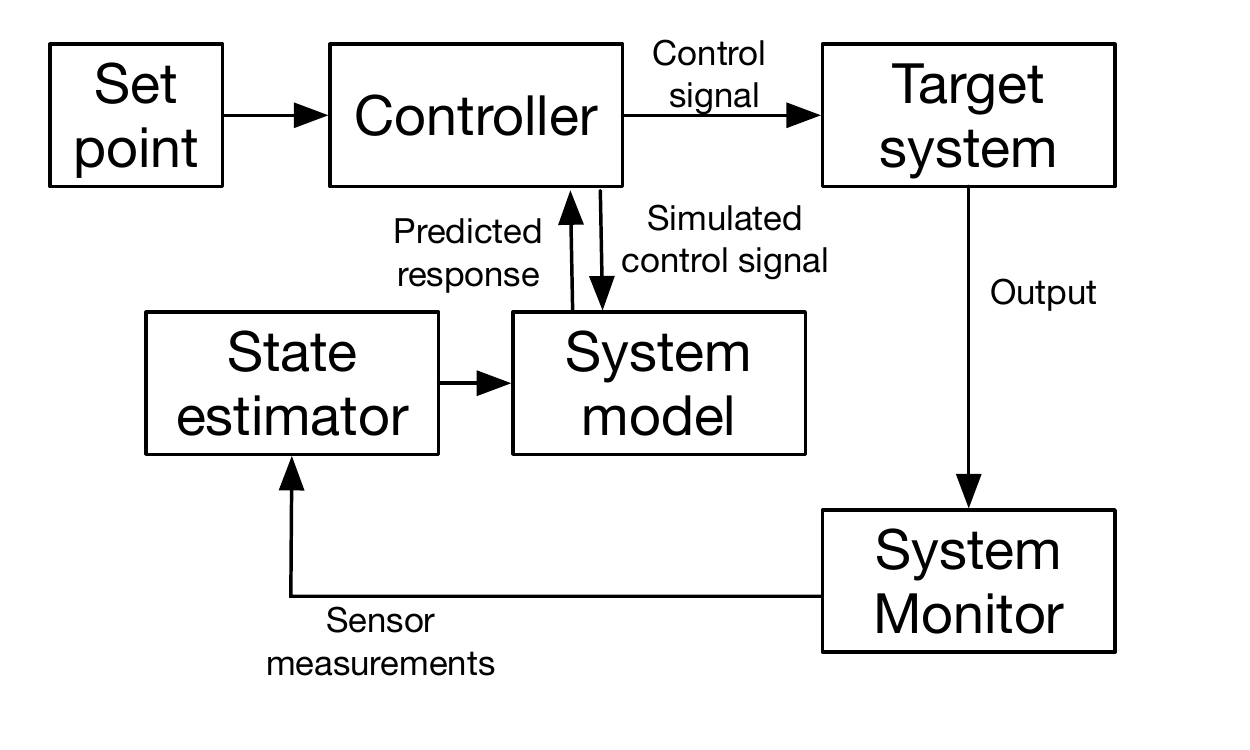}%
			\label{subfig:control}%
		}\\ 
		\subfloat[Proposed \sil{} approach for loop scheduling.]{%
			\includegraphics[clip, trim=0cm 0cm 0cm 0cm,scale=0.8]{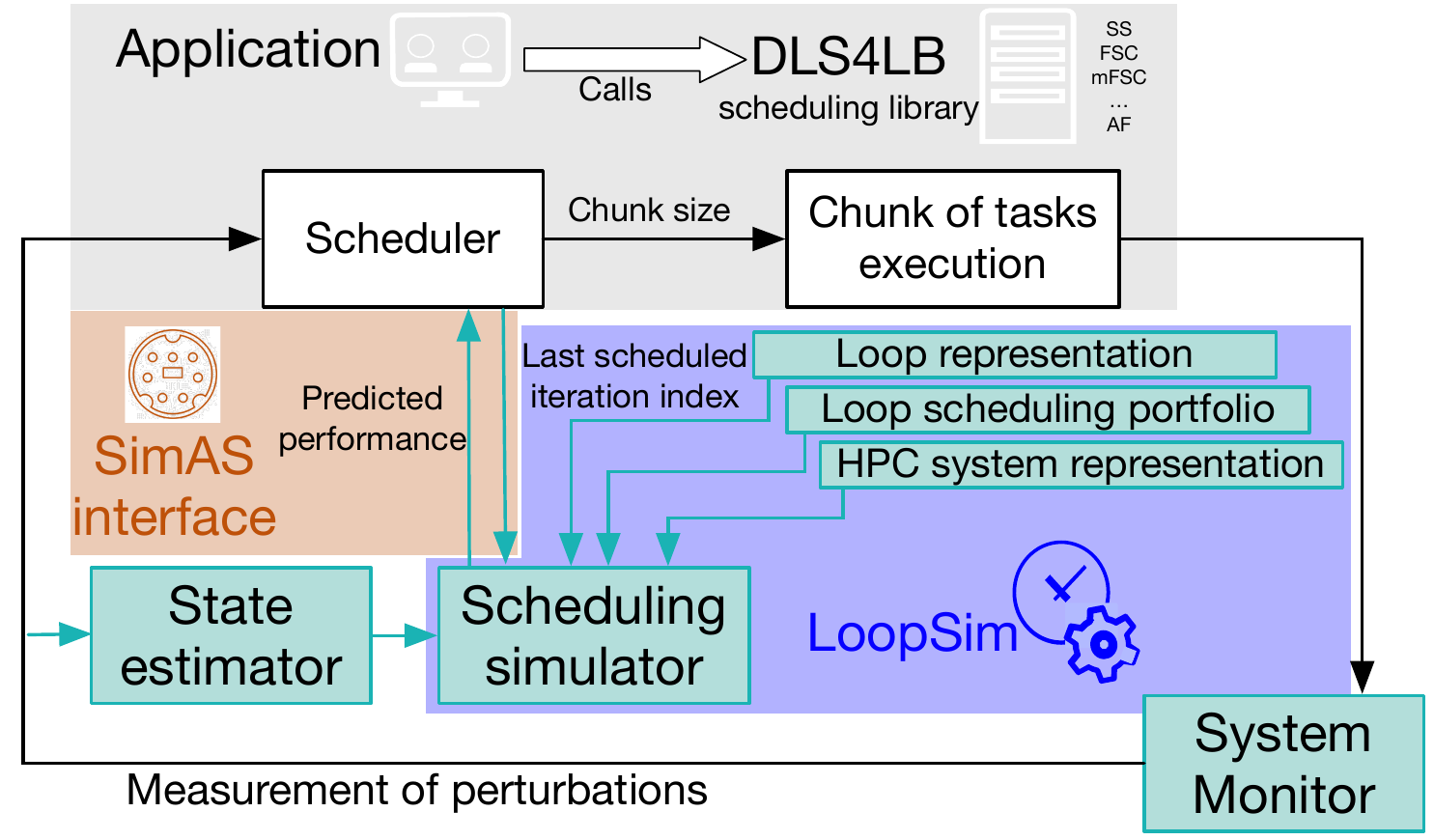}
			\label{subfig:sil}%
		} 
		\\	
		\caption{The proposed \emph{\underline{Sim}ulation-assisted scheduling \underline{A}lgorithm \underline{S}election} (\sil{}) approach for the selection of DLS techniques. \sil{} (b) is analogous to a typical control system~(a). The components highlighted in {\color{UnibasMint}mint color} represent the \sil{} additions to a typical loop scheduling system. {The \dlb{} {library}~(c.f. Section~\ref{subsec:dls}) is used for the parallel task scheduling and execution, \lsim{} ~(c.f. Section~\ref{subsec:sim}) is used to predict the application performance with different DLS techniques under perturbations. 
				\sil{}~(c.f. Section~\ref{subsec:sil}) is integrated with \dlb{} to communicate with \lsim{} and to select DLS techniques dynamically during execution.}} 
		\label{fig:approach}
	
\end{figure}

The \sil{} is inspired by control theory, where a controller (scheduler) is used to achieve and maintain a desired state (load balance) of the system (parallel loop execution), as illustrated in \figurename{~\ref{fig:approach}}(a)~and~(b).
The \sil{} concept is motivated by the well-known control strategy model predictive control~(MPC)~\cite{rawlings2000tutorial}. 
The MPC~controller predicts the performance of the system with different control signals to optimize system performance.
As shown in \figurename{~\ref{subfig:sil}}, a call to \sil{} is inserted inside \ali{a} typical scheduling loop. 
\sil{} leverages state-of-the-art simulation toolkits to estimate the performance of an application in a given execution scenario.
The system monitor and estimator components read the system state during the execution and update the computing system representation accordingly.
The above steps may \aliA{be repeated} several times during the execution of the loop, and \aliD{the \sil{} call} frequency can be aligned with the perturbations frequency or intensity.

%
%
The advantage of \sil{} is that it leverages the use of already developed state-of-the-art simulators to predict the performance dynamically during execution. 
The \aliD{prediction} accuracy of a simulator is strongly influenced by the representation of both the applications and the systems in simulation as well as by the subsystem models \aliD{it comprises}~\cite{Mohammed:2018a}. 
Given that the main concern of this work is load imbalanced computationally-intensive applications with replicated data, the influence of the memory subsystem (e.g. complex memory hierarchy) on their performance is minimal. 
Therefore, application performance can \aliD{accurately be predicted} via simulation.
For instance, the percent error between native and simulative executions for a given application (PSIA~\cite{Eleliemy:2017b}) using the \simdag{} interface was found to be between $0.95\%$ and $2.99\%$~\cite{Mohammed:2018a}.
The percent error is calculated as 
\begin{equation}
\%E = (1 - \frac{T_{sim}}{T_{native}}) \times 100
\end{equation}
	, where $T_{native}$ and $T_{sim}$ are the native and simulative performance, respectively.
Moreover, it was found that the performance simulations with SimGrid captures the native applications performance features and identifies the most efficient DLS technique for PSIA and Mandelbrot applications~\cite{Mohammed_FGCS}.
It is expected that the accuracy and speed of the simulators employed by \sil{} will improve as \aliA{they are continuously being developed and refined}. 
The cost of frequent calls to \sil{} can be amortized by launching parallel \sil{} instances to concurrently derive predictions for various DLS.
Alternatively, this cost can be entirely mitigated by asynchronously calling \sil{}, concurrently to the application execution.
Upon completion, \sil{} returns \aliD{as recommendations} \emph{best suited DLS technique} to the calling application, which can then \aliA{directly} use the recommended DLS to improve \aliD{its} performance. 

The system monitor and estimator components can be implemented with a number of system monitoring tools~\cite{Ciorba:2016}, such as \texttt{collectl}. 
\aliA{Such} tools can periodically be instantiated to measure PE and network loads and to update the system representation in the simulator \aliD{for the next call to \sil{}.}
The measured chunk execution times can also be used to estimate the current PE computational speeds.
\cut{An overview of system monitoring and tools can be found in the literature~\cite{Ciorba:2016}.}
The PE loads can be estimated and predicted using autoregressive integrated moving average~\cite{mehrotra2015power}.\cut{where a control theory inspired method is introduced for power aware scheduling of tasks on heterogeneous systems.}

%
\section{Experimental Design and Setup}
\label{sec:exp}
In this work we employ a factorial design of experiments, due to the numerous parameters and values to explore.
The design of \aliD{the} factorial experiments is presented in the following (cf. Table~\ref{tbl:ex}), along with details \aliD{of} the DLS techniques implementation and \sil{}, the computing system under test and \aliD{its injected} perturbations in native and simulative experiments.


	\begin{table}[h]
		\centering
		\caption{Design of factorial experiments}
		\label{tbl:ex}
		\resizebox{\textwidth}{!}{%
		\begin{threeparttable}
			{	\renewcommand{\arraystretch}{1.15} 
			\begin{tabular}{lll}
				\toprule
				\textbf{Factors}                                           & \textbf{Values}                                                                                                      & \textbf{Properties}                                                                                                                                        \\ \midrule
	\multirow{2}{*}{  \begin{tabular}[c]{@{}l@{}} \\   \\  \\ \textbf{Applications}  \end{tabular} }                 &  \begin{tabular}[c]{@{}l@{}}  PSIA \\Mandelbrot \\ PSIA\_TS (time-stepping) \\ Mandelbrot\_TS (time-stepping) \\Constant\\Uniform\\Normal\\Exponential\\Gamma \end{tabular} & 
\begin{tabular}[c]{@{}l@{}}  $[5.9 \cdot 10^7$ .. $6.6 \cdot 10^7]$ FLOP per iteration\\ $[5.9 \cdot 10^1$ .. $2.6 \cdot 10^8]$ FLOP per iteration\\  $[5.9 \cdot 10^7$ .. $6.5 \cdot 10^7]$ FLOP per iteration \\ $[5.9 \cdot 10^1$ .. $2.6 \cdot 10^8]$ FLOP per iteration \\ $2.3 \cdot 10^8$ FLOP per iteration\\ $[10^3$ .. $7 \cdot 10^8]$ FLOP per iteration\\ $\mu = 9.5 \cdot 10^8$ FLOP, $\sigma= 7 \cdot 10^7$ FLOP, $[6 \cdot 10^8$ .. $1.3 \cdot 10^9 ]$ FLOP per iteration\\ $\lambda = 1/3 \cdot 10^8$ FLOP, $[9.48 \cdot 10^2$ .. $4.5 \cdot 10^9]$ FLOP per iteration\\ $k = 2$, $\theta = 10^8$ FLOP, $[4.1 \cdot 10^6$ .. $2.7 \cdot 10^9]$ FLOP per iteration \end{tabular}  \\
\cline{2-3}  
&  Problem size  
& \begin{tabular}[c]{@{}l@{}} $N =$ 400,000 iterations, all applications except for \\ PSIA\_TS $N= 4,000$ iterations per time-step $\times 10$ time-steps\\Mandelbrot $N=262,144$ iterations\\Mandelbrot\_TS $N=16,384$ iterations per time-step $\times 10$ time-steps \end{tabular}
\\ \hline
				\textbf{Loop scheduling}                & \begin{tabular}[c]{@{}l@{}}STATIC\\ SS, FSC, mFSC, GSS, TSS, FAC, WF\\ AWF-B, -C, -D, -E, AF\end{tabular}                     & \begin{tabular}[c]{@{}l@{}}Static\\ Nonadaptive dynamic\\ Adaptive dynamic\end{tabular}                                                                    \\ \hline
				\textbf{Computing system}               & \begin{tabular}[c]{@{}l@{}}miniHPC \\ (heterogeneous HPC cluster)\end{tabular}                                     & \begin{tabular}[c]{@{}l@{}}22 Intel Broadwell nodes ($22 \cdot 20$ cores), relative core weight $= 0.817$\\ 4 Intel Xeon Phi KNL nodes ($4 \cdot 64$ cores), relative core weight $= 0.183$\\ $P= 128$ heterogeneous (64 Broadwell + 64 KNL) cores\\ $P= 416$ heterogeneous (352 Broadwell + 64 KNL) cores \end{tabular} \\ \hline
				\multirow{5}{*}{  \begin{tabular}[c]{@{}l@{}} \\ \\ \\ \\ \\  \\  \\ \textbf{Perturbations}  \end{tabular}}  & Nominal conditions & \texttt{np} (no perturbations)  \\ \cline{2-3} & PE availability                                                                                          & \begin{tabular}[c]{@{}l@{}} \texttt{pea-cm} (constant mild): \aliB{$\mu=75\%$, $\sigma=0\%$}\\  \texttt{pea-cs} (constant severe): \aliB{$\mu=25\%$, $\sigma=0\%$}\\ \texttt{pea-em} (exponential mild): \aliB{$\mu=78\%$, $\sigma=24 \cdot 10^{-3}\%$}\\ \texttt{pea-es} (exponential severe): \aliB{$\mu=31\%$, $\sigma=89 \cdot 10^{-3}\%$}\end{tabular}                                  \\ \cline{2-3}
				& Bandwidth                                                                                                       & \begin{tabular}[c]{@{}l@{}} \texttt{bw-cm} (constant mild): \aliB{$\mu=1 \cdot 10^{-5}\%$, $\sigma=0\%$}\\ \texttt{bw-cs} (constant severe): \aliB{$\mu=1 \cdot 10^{-7}\%$, $\sigma=0\%$}\\ \texttt{bw-em} (exponential mild): \aliB{$\mu=1.1 \cdot 10^{-1}\%$, $\sigma=9 \cdot 10^{-2}\%$}\\ \texttt{bw-es} (exponential severe): \aliB{$\mu=23 \cdot 10^{-2}\%$, $\sigma=19 \cdot 10^{-2}\%$}\end{tabular}                                                                          \\ \cline{2-3}
				& Latency                                                                                                         & \begin{tabular}[c]{@{}l@{}} \texttt{lat-cm} (constant mild): \aliB{$\mu=1 \cdot 10^{-5}\%$, $\sigma=0\%$}\\ \texttt{lat-cs} (constant severe): \aliB{$\mu=1 \cdot 10^{-7}\%$, $\sigma=0\%$}\\ \texttt{lat-em} (exponential mild): \aliB{$\mu=1.2 \cdot 10^{-5}\%$, $\sigma=1.5 \cdot 10^{-5}\%$}\\ \texttt{lat-es} (exponential severe): \aliB{$\mu=2.9 \cdot 10^{-7}\%$, $\sigma=1.8 \cdot 10^{-7}\%$}\end{tabular}   \\ \cline{2-3}
				& \aliB{Combined}                                                                                                             & \begin{tabular}[c]{@{}l@{}} \texttt{all-cm} (constant mild): \aliB{\texttt{pea-cm}, \texttt{bw-cm}, and \texttt{lat-cm}}\\ \texttt{all-cs} (constant severe): \aliB{\texttt{pea-cs}, \texttt{bw-cs}, and \texttt{lat-cs}}\\ \texttt{all-em} (exponential mild): \aliB{\texttt{pea-em}, \texttt{bw-em}, and \texttt{lat-em}}\\ \texttt{all-es} (exponential severe): \aliB{\texttt{pea-es}, \texttt{bw-es}, and \texttt{lat-es}} \end{tabular}         
				\\ \hline
				\multirow{2}{*}{  \begin{tabular}[c]{@{}l@{}} \textbf{Experimentation}  \end{tabular}} & Native\tnote{1} &  PSIA and Mandelbrot on $128$ and $416$ cores under targeted perturbations \\ \cline{2-3}
				& Simulative  &   \begin{tabular}[c]{@{}l@{}} PSIA and Mandelbrot on $128$ and $416$ cores under all perturbations \\ Synthetic applications on $128$ and $416$ cores under all perturbations  \end{tabular}  \\ \bottomrule
			\end{tabular}%
		}
			\begin{tablenotes}
				\item[1] Direct experiments on real HPC systems.
			\end{tablenotes}
		\end{threeparttable}
	}
	\end{table}

\subsection{Applications}
\label{subsec:app}
This work considers \aliC{two} real-world applications and five synthetic workloads.\\

\textbf{Real applications.}\\
\textit{1. PSIA.} The parallel spin-image algorithm~\cite{Eleliemy:2017b}~(PSIA), is a \aliC{computationally-intensive} application from computer vision. 
The PSIA is \aliD{embarrassingly parallel application and} algorithmically load imbalanced where the computational effort of a loop iteration depends on the input data. 
The performance of PSIA has been studied in prior work~\cite{Eleliemy:2017b} and \aliD{was} enhanced for \aliD{execution on} a heterogeneous cluster by using nonadaptive DLS techniques.
The total number of \aliC{parallel} loop iterations \aliD{in PSIA} is 400,000.

\noindent\textit{2. Mandelbrot.} \aliC{\aliD{This} application computes the Mandelbrot set~\cite{mandelbrot1980fractal} and generates its image. 
The program is based on one of the codes available online\footnote{https://github.com/CaptGreg/SenecaOOP345-attic/blob/master/parallel-pgm/mpi/mandelbrot-mpi-dynamic.c}. The application is parallelized such that the calculation of the value at every single pixel of a 2D image is a loop iteration, that is performed in parallel. The application is modified to compute the function $f_c(z) = z^4 + c$ \aliD{instead of $f_c(z) = z^2 + c$} to increase the number of computations per task. The size of the generated image is $512 \times 512$ pixels resulting in $2^{18}$ parallel loop iterations.}

\noindent\textit{3. PSIA\_TS.} This application is similar to PSIA. Unlike PSIA, PSIA\_TS is executed in \mbox{time-steps}. 
It simulates applying \mbox{spin-image} transformations to an object in motion~(a video), where at each \mbox{time-step} a certain number of \mbox{spin-images}~($4,000$) is created.
PSIA\_TS is executed for 10~\mbox{time-steps}.

\noindent\textit{4. Mandelbrot\_TS.} This is the \mbox{time-stepping} version of Mandelbrot application. 
At each \mbox{time-step}, the generated Mandelbrot set image at time $t$ is \mbox{zoomed-in} by $5\%$ on the center of the image to generate the image at $t+1$. 
Mandelbrot\_TS is executed for 10 \mbox{time-steps}.
The workload per \mbox{time-step} is reduced compared to Mandelbrot (\mbox{single-sweep}) such that the execution time of 10~\mbox{time-steps} of Mandelbrot\_TS is comparable to the execution time of the \mbox{single-sweep} version. 
This is desirable for the purpose of native experimentation given the large set of experiments performed~(see Table~\ref{tbl:ex}), to avoid extremely long execution times.

\noindent\textbf{Synthetic workloads.}\\
Five synthetic workloads are examined in this work.
Each of the five synthetic workloads contains 400,000~parallel loop iterations.
The \aliD{number of floating point operations (FLOP count)} in each loop iteration is assumed to follow five different probability distributions, namely: constant, uniform, normal, exponential, and gamma probability distributions. 
This assumption captures the characteristics of a wide range of applications.
The probability distribution parameters used to generate these FLOP counts are \aliD{also} given in Table~\ref{tbl:ex}.

\subsection{Loop scheduling}
\label{subsec:dls}
\aliC{Thirteen} loop scheduling techniques are used to assess the performance of the above \aliC{seven} applications under \aliD{various execution scenarios}. 
These techniques represent a wide range of \emph{static} and \emph{dynamic} loop scheduling approaches. 
The dynamic loop scheduling (DLS) \aliD{techniques} can further be distinguished into \aliD{five} adaptive and \aliD{seven} nonadaptive \aliD{techniques}.

\aliD{In general,} the DLS techniques can be implemented using centralized or decentralized execution and control approach. 
The decentralized control approach was found to scale better by eliminating a centralized master, and hence, the \mbox{master-level} contention~\cite{HPCS}.
\discuss{The decentralized control approach was used previously~\cite{Mohammed:2018c} using Intel MPI one-sided communications. 
The Intel implementation uses extra threads that run in the background to handle the \mbox{one-sided} communications. 
These threads introduce additional overhead during execution, and could \aliD{prevent} the application progress if these threads could not find enough computational power to execute.
\aliD{Therefore, it was found that the centralized \mbox{two-sided} communication implementation of DLS is more suitable for this work.}
}

\aliC{
The dynamic load balancing tool (\dlbtool{}~\cite{carino2007tool}) is extended and used to parallelize the applications with dynamic loop \mbox{self-scheduling} \aliD{and employs MPI \mbox{two-sided} communications for work distribution among processes}.
The \dlb{} implements a \mbox{master-worker} execution model, where the master is responsible for handling work requests from workers.
In addition, the master act \aliD{also} as a worker, and checks for outstanding work requests with a certain \aliD{adjustable} frequency.  
The \dlb{} is designed to parallelize an application with minimum changes.
Algorithm~\ref{algo:sil_code} shows, in blue \aliD{font color,} the lines needed to be added \aliD{to the} application code to parallelize it.  
The \dlbtool{} originally contained the implementation of nine loop scheduling techniques, \aliD{namely STATIC, mFSC, GSS, FAC, AWF-B, AWF-C, AWF-D, AWF-E, and AF.} 
In this work, the tool is extended into \dlb{} to support four more dynamic loop scheduling techniques, namely SS, FSC, TSS, and WF.
}
\subsection{\underline{Sim}ulation-assisted \underline{A}lgorithm \underline{S}election}
\label{subsec:sil}
\aliC{
In this work, the \dlb{} is extended to support the \sil{} \aliD{as the fourteenth option in the \dlb{}}.
Taking the same approach of the \dlb{} of minimal application code changes, an application can use the \sil{} by inserting only two function calls, shown in green \aliD{font color,} in Algorithm~\ref{algo:sil_code}.

The \texttt{\sil{}\_setup} function sets up the main data structure \texttt{\sil{}\_info} that holds important information, such as the number of PEs, the number of loop iterations, the path to the simulator, the FLOP file that contains the FLOP count per loop iteration, and the platform file that describes the computing system. 
In addition, \texttt{\sil{}\_setup} asynchronously starts the simulation of the application performance immediately with a portfolio of DLS techniques in parallel. 
The \texttt{\sil{}\_setup} sets the scheduling technique to a default DLS (AWF-B in this work), to allow the application to start and avoid delaying the application execution.

The \texttt{\sil{}\_update} checks (every 5 seconds in this work) if the simulation is finished, and selects the DLS technique allows the application to finish the largest number of tasks in the shortest time; otherwise it will keep the selected DLS technique unchanged.
The \texttt{\sil{}\_update} reruns the simulation again if 50 seconds (the \sil{} calling frequency) have passed since the simulator was previously called.
The \texttt{\sil{}\_update} prevents the start of a new instance of the simulator unless the earlier one is completed or the number of remaining unscheduled iterations is less than or equal the number of PEs. 

}

\begin{algorithm}[]
	\caption{Dynamic load balancing with \sil{} support using the extended  \textit{DLB\_tool}}
	\label{algo:sil_code}
	\KwData{\sil{}\_info, DLS\_info, h, $\sigma$, N, P}
    \#include \textless mpi.h\textgreater \\
    \#include ``DLB\_\sil{}''
	
	MPI\_Init(\&argc, \&argv);
	MPI\_Comm\_size(MPI\_COMM\_WORLD, \&nprocs);
    MPI\_Comm\_rank(MPI\_COMM\_WORLD, \&myid);
	
	\textcolor{ForestGreen}{scheduling\_method = \sil{}\_setup(\sil{}\_info, P, N, h, sigma, sim\_path, FLOP\_file, platform\_file);}
	
	\textcolor{blue}{DLS\_setup(MPI\_COMM\_WORLD, DLS\_info);}
	
	\textcolor{blue}{DLS\_startLoop (DLS\_info, N, scheduling\_method);}
	
	\While{not DLS\_terminated(DLS\_info)}{
		 \textcolor{ForestGreen}{\sil{}\_update(DLS\_info, \sil{}\_info);}
			
	      \textcolor{blue}{DLS\_startChunk(DLS\_info, Cstart, Csize);}
		  
		  Compute\_iterations(Cstart, size);
		   
		 \textcolor{blue}{DLS\_endChunk(DLS\_info);}
	}
      \textcolor{blue}{DLS\_endLoop(DLS\_info);}
\end{algorithm}

%
%

\subsection{Computing system}
\label{subsec:comp}
\aliD{The native experiments were conducted on} \emph{miniHPC}\footnote{miniHPC is a fully controlled non-production HPC cluster at the Department of Mathematics and Computer Science at the University of Basel, Switzerland.}, \aliD{a research and teaching cluster at the Department of Mathematics and Computer Science at the University of Basel, Switzerland. It} consists of 26~compute nodes: 22~nodes each with one dual socket Intel Xeon E5-2640~v4~(20~cores) configuration and 4 nodes each with one Intel Xeon Phi Knights Landing 7210~processor~(64~cores). 
All nodes are interconnected with Intel Omni-Path fabrics in a nonblocking two-level fat-tree topology.

\clearpage

\subsection{Simulation}
\label{subsec:sim}
\noindent\textbf{Applications.} 

\lsim{}\footnote{https://github.com/unibas-dmi-hpc/LoopSim}, an \simdag{}-based simulator, is used to simulate the applications of interest, where the loop iterations in the application code are represented as tasks~\cite{Mohammed:2018a}. 
To represent the computational effort \aliD{associated with} an application's loop iterations, the number of floating point operations (FLOP) of each loop iteration is counted using PAPI counters~\cite{papi}.
The FLOP count per iteration is then read by \lsim{} during execution to simulate the computation per iteration.
All DLS techniques supported by the \dlb{} are also implemented in \lsim{} and tasks are assigned to free and requesting simulated cores, similar to the native execution.
	
The pseudocode of \lsim{} is presented in Listing~\ref{simulator_code}.
\lsim{} reads in the number of iterations (tasks), \texttt{start task~ID}, the path to the file that contains the FLOP count per loop iteration, the path to the computing system representation (see below), the selected scheduling technique, and the maximum \emph{simulated time}.
The simulator reads the data and simulates the loop execution using the selected DLS technique.
It then outputs the simulated time and the number of tasks executed in this simulated time.
This information is read by the \sil{}, which compares different DLS techniques based on this information and selects the DLS technique that results in the shortest execution time and largest number of finished tasks.

\lstset{language=C,caption={\simdag{} loop simulator},label=simulator_code}
\lstset{escapeinside={<@}{@>}}
\begin{lstlisting}[frame=single] 
#include <simdag.h>
#include <DLS\_scheduling.h>

<@\textcolor{gray}{//read input}@>
read_input(num_tasks, FLOP_file, start_task_ID, \
platform_file, DLS_t, max_sim_t);
<@\textcolor{gray}{//create tasks that represent loop iterations}@>
Task_array = create_tasks(num_tasks, FLOP_file);
scheduled_tasks = 0;
<@\textbf{ while(executed\_tasks < num\_tasks) \&\& (get\_sim\_time() < max\_sim\_t)}@>
<@\textbf{ \{ }@>
idle_processes = get_idle_processes();
<@\textbf{ foreach(idle\_process in idle\_processes) }@>
<@\textbf{ \{ }@>
<@\textcolor{gray}{//read and update finsihed tasks}@>
executed_tasks += get_finished_tasks(idle_process);
<@\textcolor{gray}{// send work request to master}@>
send_work_request(idle_process, master);
chunk = calculate_chunk(Task_array, num_tasks, \
scheduled_tasks, DLS_t);
<@\textcolor{gray}{//assign work to worker}@>
send_work(master, idle_process);
scheduled_tasks += chunk;
<@\textbf{ \} }@>
<@\textcolor{gray}{//resume simulation untill a task is finished, i.e., a process is idle}@>
simulate_execution(platform_file);
<@\textbf{ \} }@>
print("simulated time: " + get_sim_time());
print("finished tasks: " + executed_tasks);

\end{lstlisting}

\noindent\textbf{Computing system.}
A computing system is represented in \simgrid{} via an XML file denoted as \texttt{platform file}.
\simgrid{} registers each processor core from their representation as a \texttt{host} in the \texttt{platform file}.
The computational speed of a processor core is estimated \ali{by measuring a loop execution time and dividing it by the total number of floating point operations included in the loop}~\cite{Mohammed:2018a}.
A Xeon core was found to be four times faster than a Xeon Phi core \ali{as indicated by the relative core weights~(cf. Table~\ref{tbl:ex})}.
The network bandwidth and latency represented in the \texttt{platform file} are calibrated with the \simgrid{} calibration procedure\footnote{http://simgrid.gforge.inria.fr/contrib/smpi-calibration-doc/}.

\subsection{Perturbations}
\label{subsec:perts}
Three different categories of perturbations are considered in this work, namely \emph{delivered computational speed}, \emph{available network bandwidth}, and \emph{available network latency}. 
Two intensities are considered, \texttt{mild} and \texttt{severe}, for each category. 
Two scenarios are considered for each intensity, where the value of the delivered computational speed is either \texttt{constant} or \texttt{exponentially} distributed. 
%

All perturbations (cf. Table~\ref{tbl:ex})  are considered to occur periodically, with a period of $100$~seconds where the perturbations affect the system only during $50\%$ of the perturbation period.
The network (bandwidth and latency) perturbations commence with the application execution, while the delivered computational speed perturbations begin 50~seconds after the start of the application.
%
%
Another perturbation scenario is created by combining all perturbations from the other individual categories.

\noindent\textbf{Perturbations in simulative experiments.} All perturbations are enacted in \simgrid{} during simulation via the \texttt{availability}, \texttt{bandwidth}, \texttt{latency}, and \texttt{platform} files to represent perturbations in delivered computational speed, network available bandwidth, and network latency, respectively.

\noindent\textbf{Perturbations in native experiments.} \aliC{A program (CPU burner) is \aliD{launched in parallel} and pinned on the same processor cores as the application to induce perturbations on the PE availability in native execution. 
The program is executed periodically every 100 seconds and is only active during a fraction of this period that corresponds to the required PE availability perturbation (75\% or 25\%).

For \aliD{injecting perturbations in the link latency}, the MPI communication functions are intercepted using the MPI profiling interface (PMPI), and certain delays are inserted to simulate longer communication latencies. 
Given that the applications of interest are computationally-intensive and the communicated data size between application's processes is minimal, perturbations in the network bandwidth does not have a significant effect on the application performance, as can be seen from the simulative experiments below. 
Therefore, perturbations in the network bandwidth are excluded from native experimentation. 

A combined perturbations scenario is created for the native execution by combining PE availability perturbations and network latency perturbations.
As both perturbation distributions (constant and exponential) have a comparable effect on the performance, where the impact of constantly distributed perturbations is more evident, only the constant distribution of perturbations is considered in the native experiments. 
}
\section{Evaluation and Analysis}
\label{sec:evalaution}
\aliC{An analysis of the load imbalance of the real applications considered in this work is presented \aliD{in this section}.} 
The performance results of the execution of the applications with different loop scheduling techniques under different execution scenarios are \aliD{illustrated} and discussed.

\subsection{Load imbalance in PSIA and Mandelbrot}
\label{subsec:load_imabalance}
Both, PSIA and Mandelbrot, applications suffer from load imbalance that stems from the variation in the number of computational operations per loop iteration.
The number of computations varies \aliD{in both applications} due to a conditional statement in their code that can increase or decrease the number of computations per loop iteration based on the input data.
As a measure of the variation of the loop iteration execution times for both applications, the standard deviation of loop iterations execution times $\sigma$ is \aliD{derived} for both applications \aliD{by means of their} sequential execution on a single processor core (to avoid any parallelization overheads). 
The median of 20 measurements of $\sigma$ for PSIA was found to be $0.00327$, whereas it was one order of magnitude higher  for the Mandelbrot, namely $0.06056$. 

\begin{figure}[]
		\centering
		\centering
		\subfloat[PSIA performance on 416 cores]{%
			\includegraphics[clip, trim=0cm 0cm 0cm 0cm, scale=0.36]{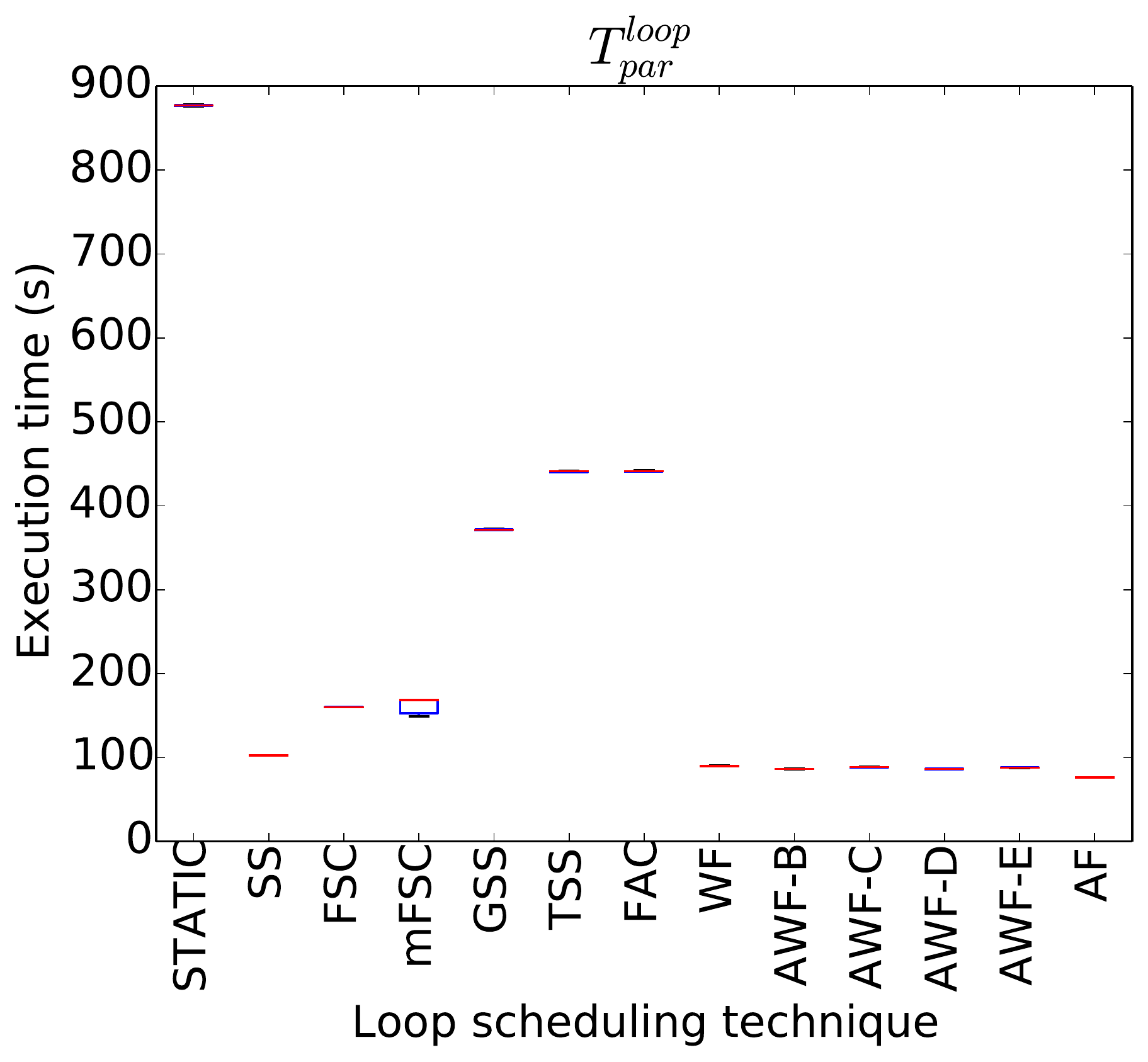}%
			\label{subfig:PSIA_native_np_416}%
		}  
		\subfloat[Load imbalance of PSIA on 416 cores]{%
			\includegraphics[clip, trim=0cm 0cm 0cm 0cm,scale=0.36]{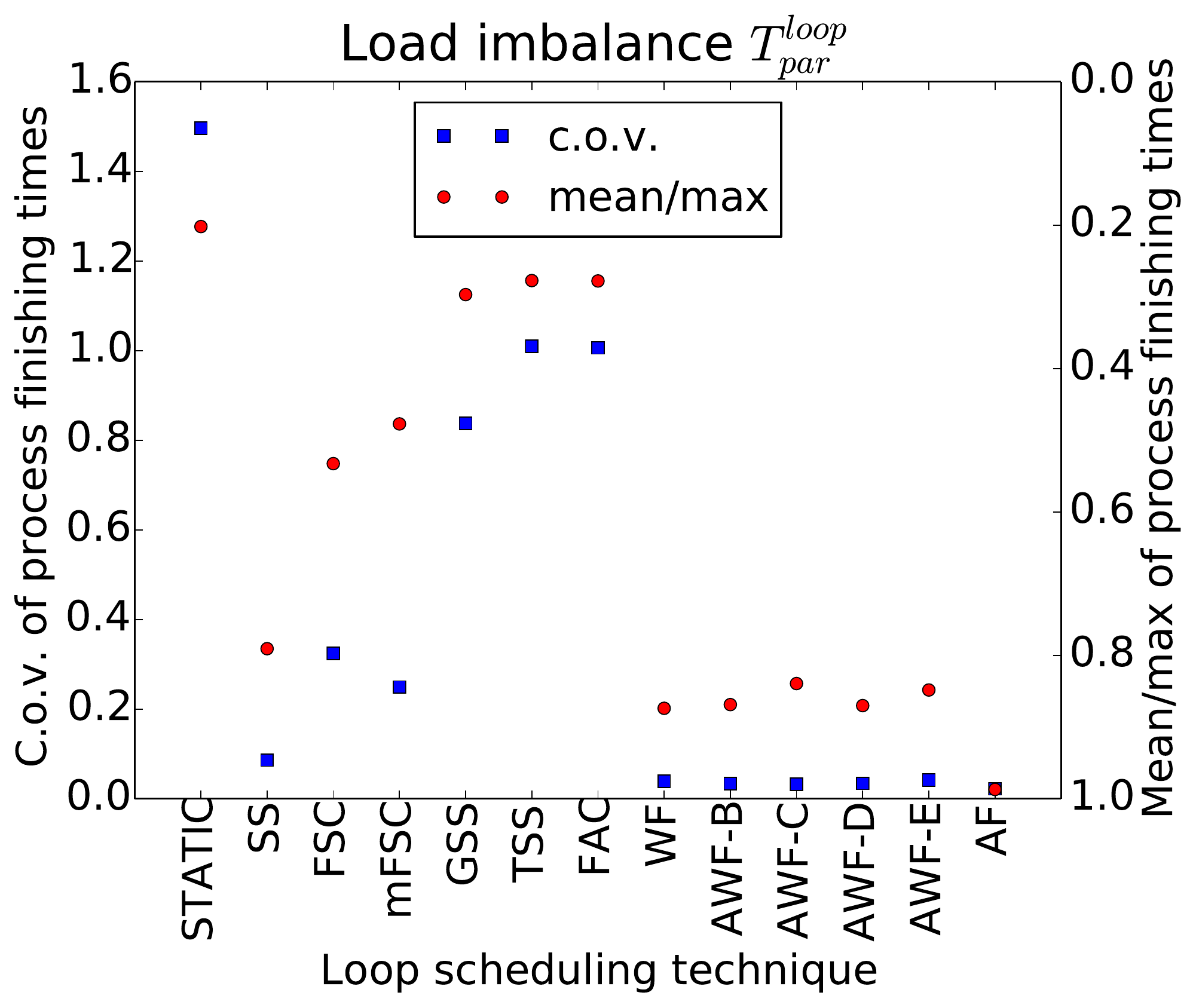}
			\label{subfig:PSIA_imbalance_416}%
		} 
		\\	
		\subfloat[Mandelbrot performance on 416 cores]{%
			\includegraphics[clip, trim=0cm 0cm 0cm 0cm,scale=0.36]{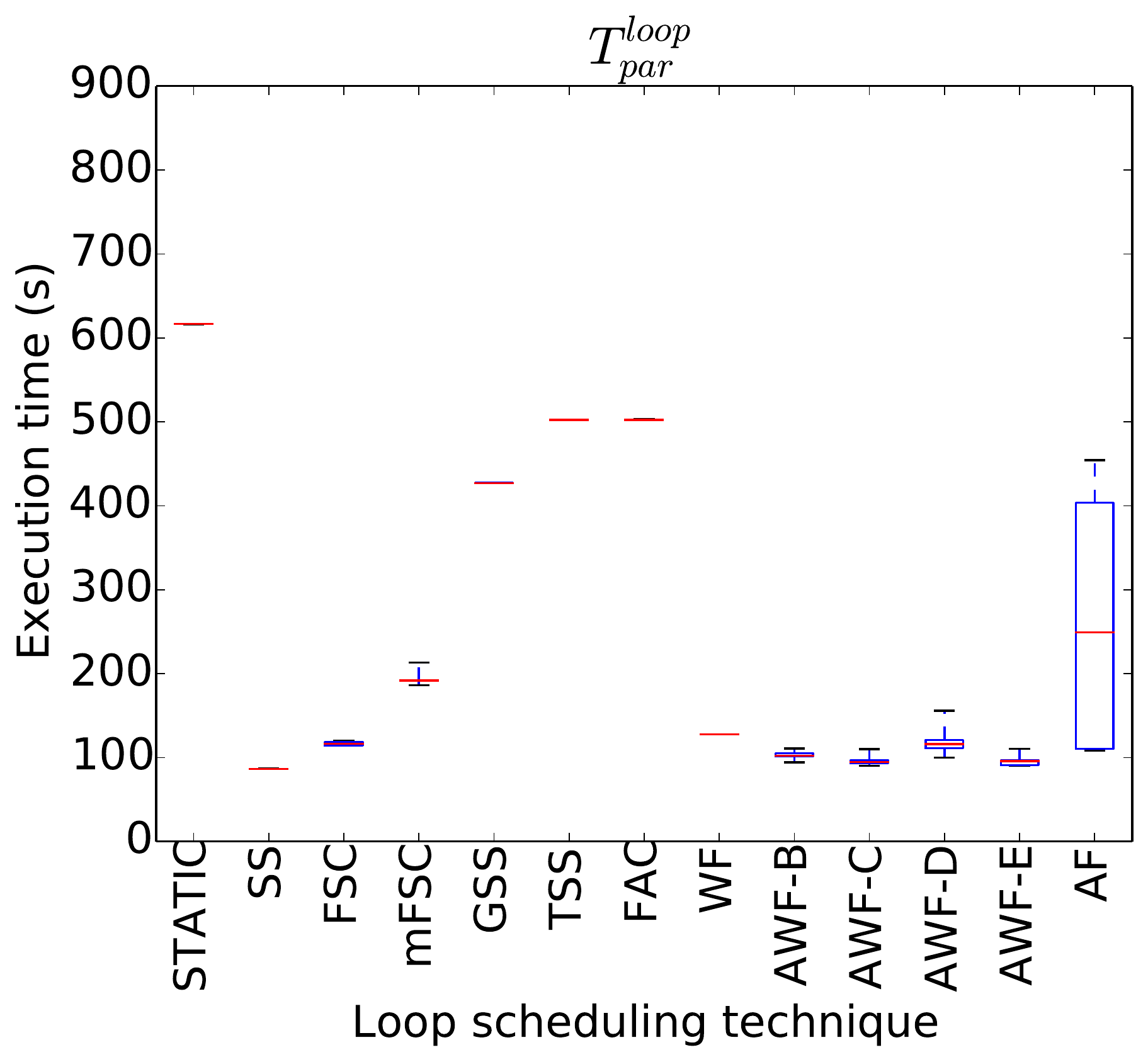}%
			\label{subfig:Mandelbrot_native_np_416}%
		} 
		\subfloat[Load imbalance of Mandelbrot on 416 cores ]{%
			\includegraphics[clip, trim=0cm 0cm 0cm 0cm,scale=0.36]{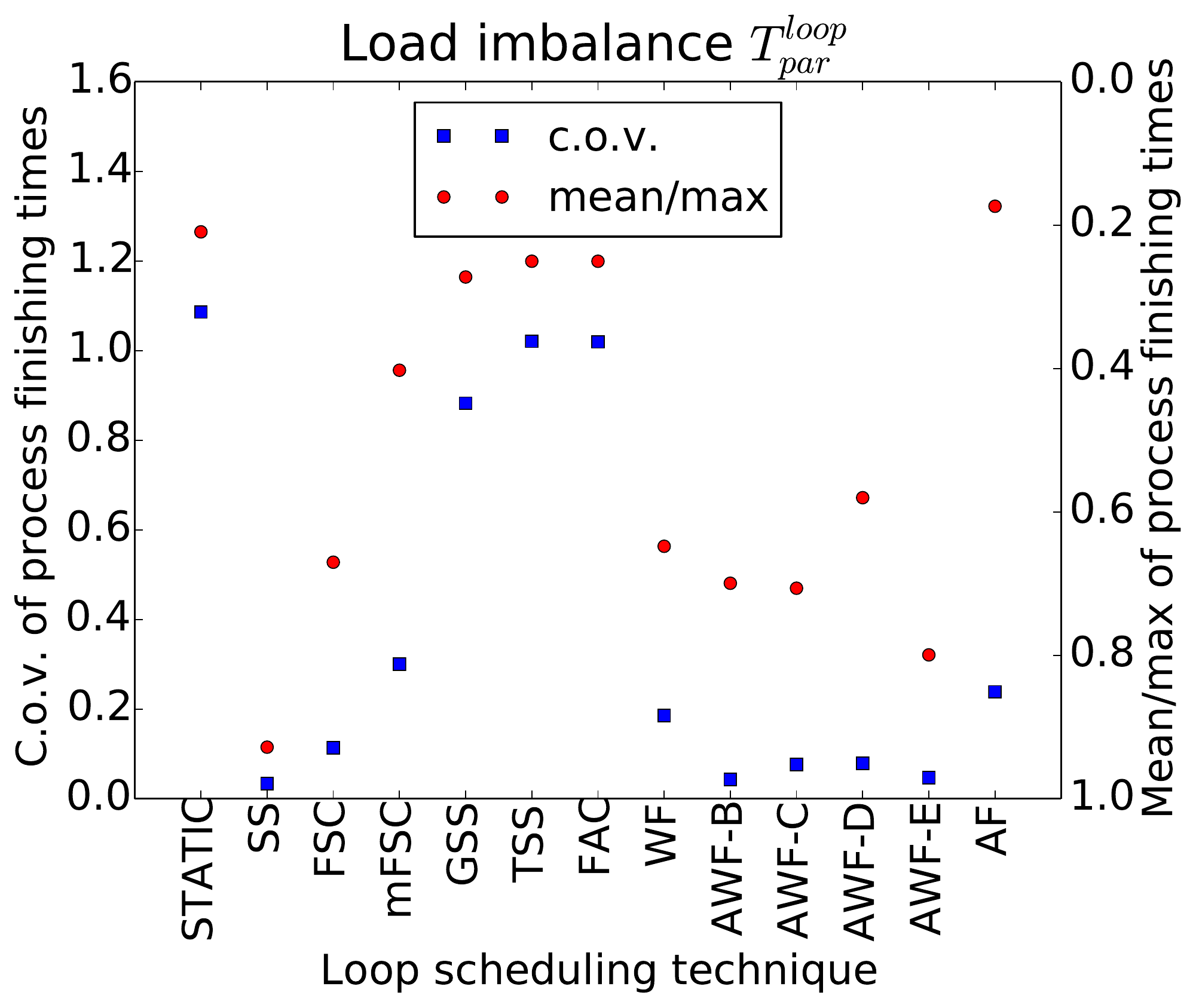}
			\label{subfig:Mandelbrot_imbalance_416}%
		} 
		\caption{Execution load imbalance of the native execution of PSIA and Mandelbrot on 416 heterogeneous cores of the miniHPC system under no perturbations. The parallel loop execution time $T^{loop}_{par}$ and the median of the load imbalance metrics over five executions are reported.}
		\label{fig:imbalance_416}
\end{figure}

\begin{figure}[]
		\centering
		\centering
		\subfloat[PSIA performance on 128 cores]{%
			\includegraphics[clip, trim=0cm 0cm 0cm 0cm, scale=0.36]{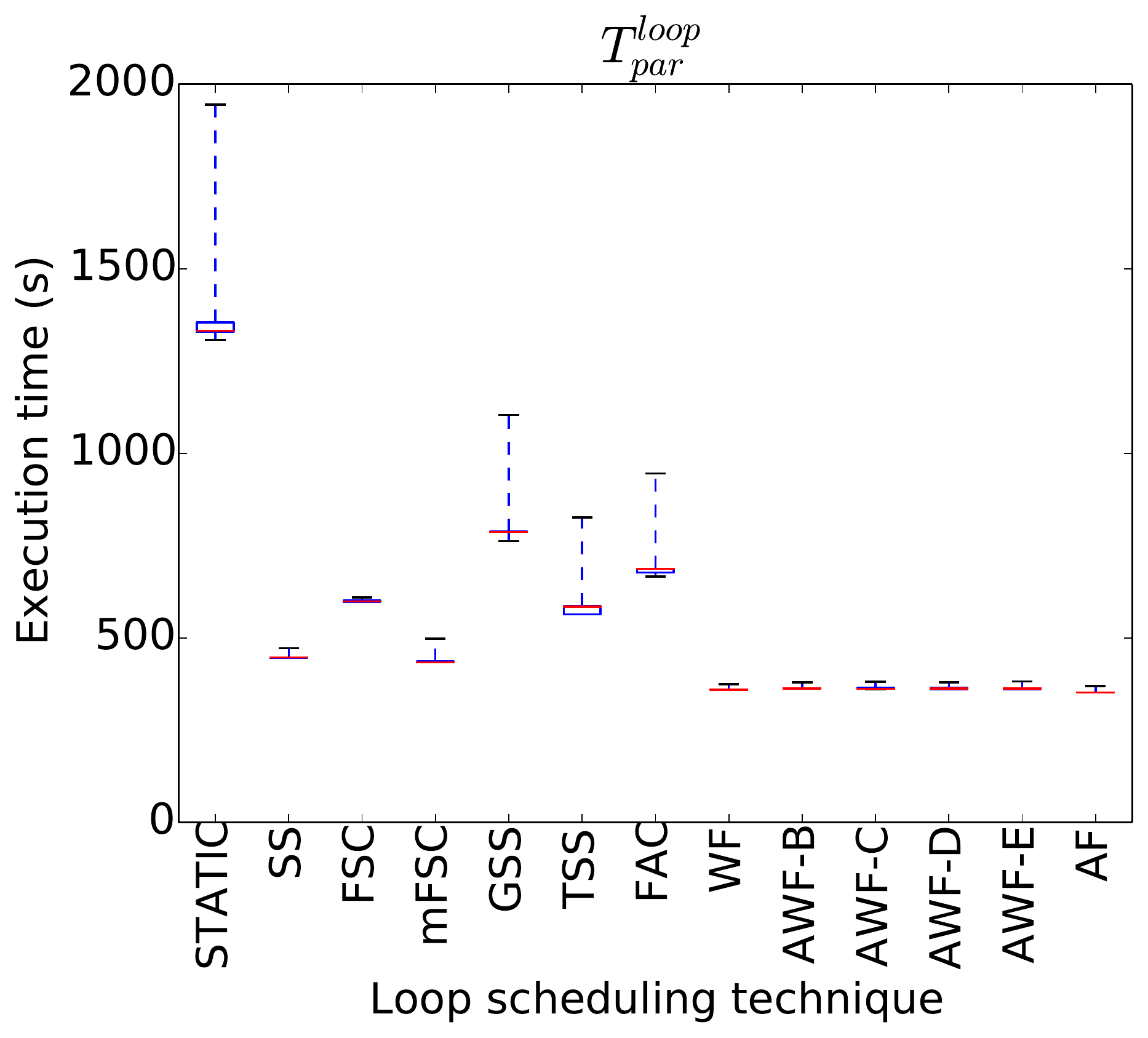}%
			\label{subfig:PSIA_native_np_128}%
		} 
		\subfloat[Load imbalance of PSIA on 128 cores]{%
			\includegraphics[clip, trim=0cm 0cm 0cm 0cm,scale=0.36]{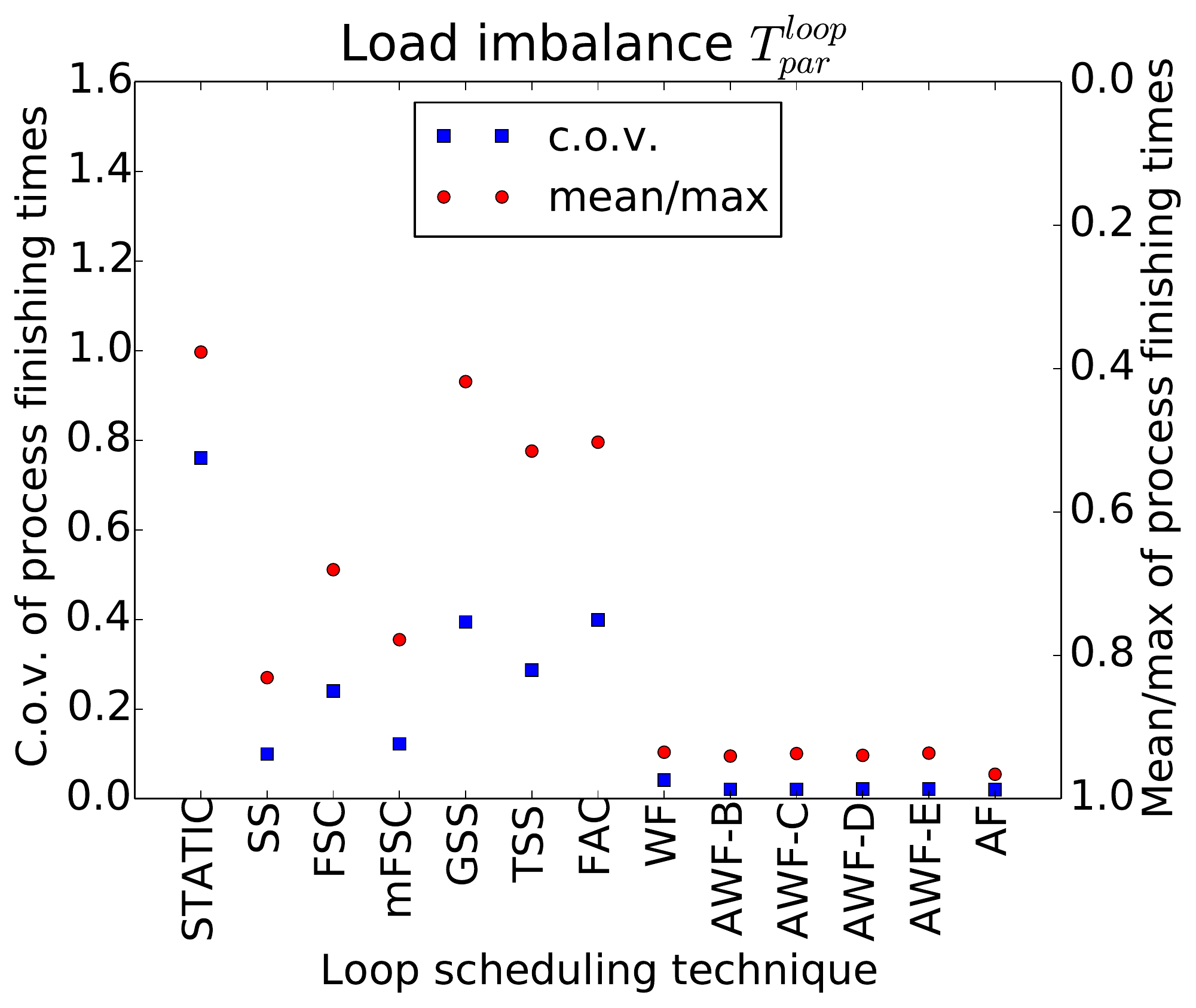}
			\label{subfig:PSIA_imbalance_128}%
		} 
		\\	
		\subfloat[Mandelbrot performance on 128 cores]{%
			\includegraphics[clip, trim=0cm 0cm 0cm 0cm,scale=0.36]{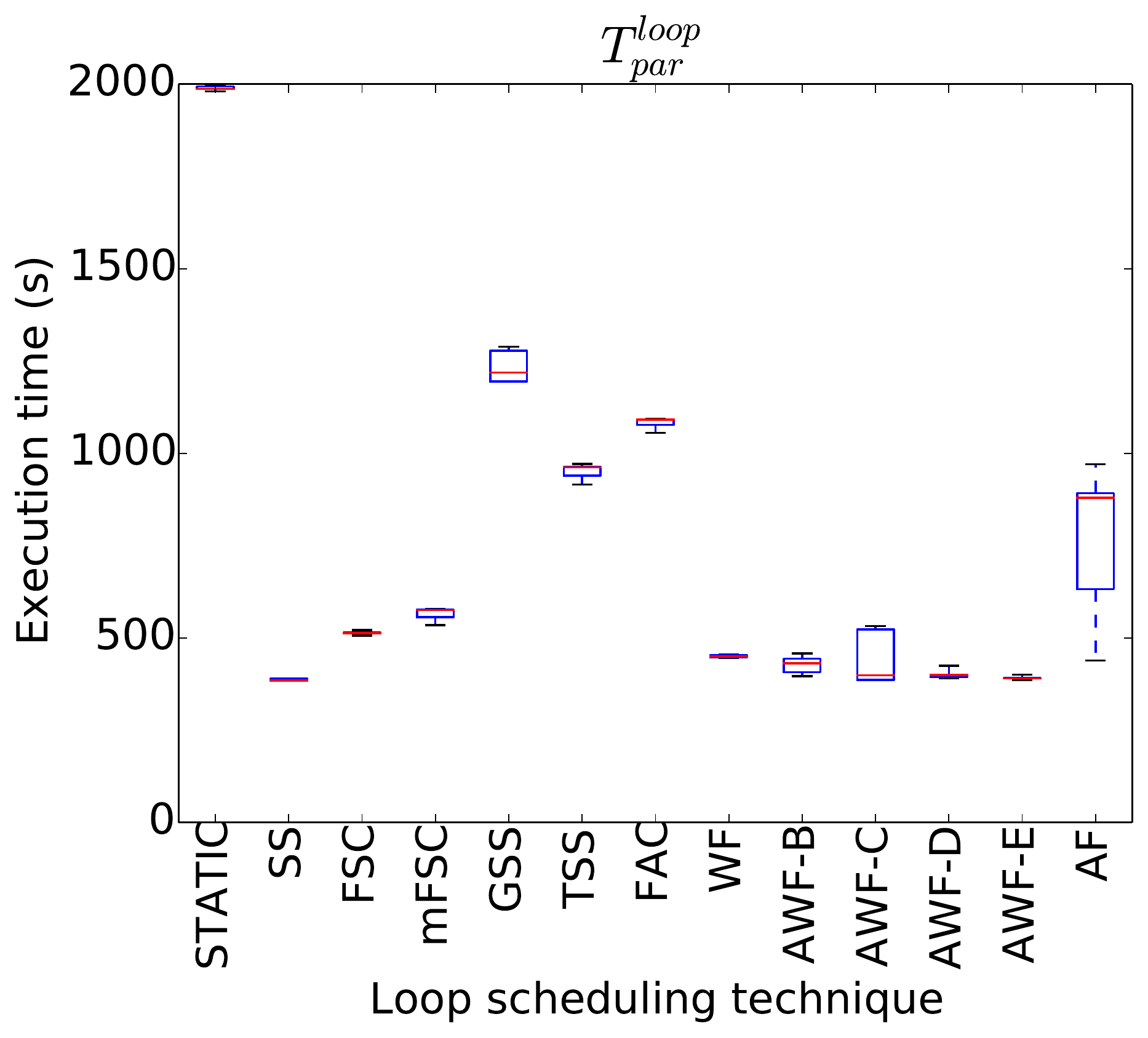}%
			\label{subfig:Mandelbrot_native_np_128}%
		} 
		\subfloat[Load imbalance of Mandelbrot on 128 cores ]{%
			\includegraphics[clip, trim=0cm 0cm 0cm 0cm,scale=0.36]{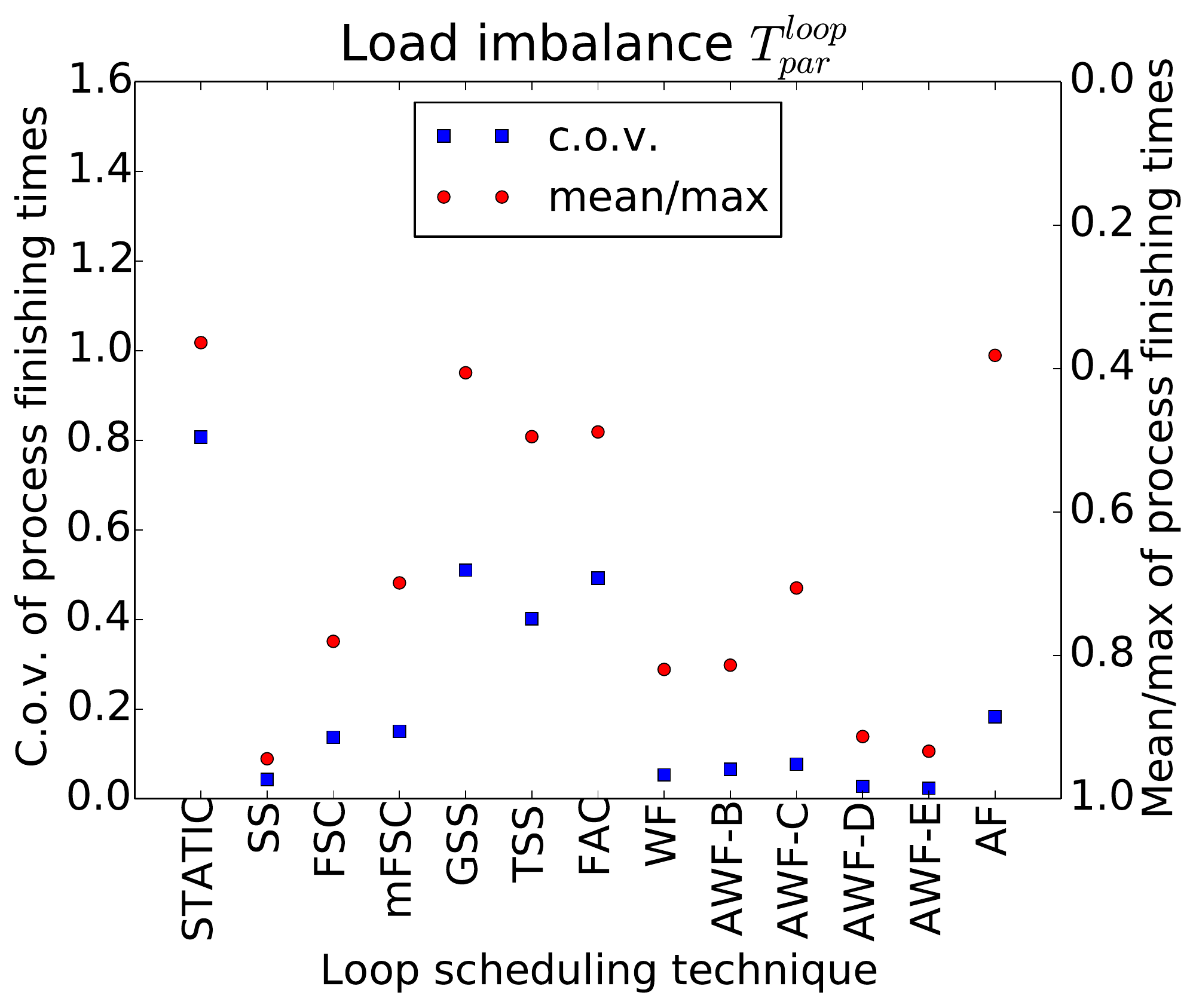}
			\label{subfig:Mandelbrot_imbalance_128}%
		} 
		\caption{Execution load imbalance of the native execution of PSIA and Mandelbrot on 128 heterogeneous cores of the miniHPC system under no perturbations. The parallel loop execution time $T^{loop}_{par}$ and the median of the load imbalance metrics over five executions are reported.}
		\label{fig:imbalance_128}

\end{figure}

Two metrics are considered to measure the load imbalance of the parallel execution of the applications on the miniHPC, namely the \emph{coefficient of variation} (c.o.v.) of the parallel processes finishing times~\cite{FAC} and the \emph{ratio of the mean process finishing times to the maximum process finishing time} (mean/max).
The c.o.v. is calculated as the ratio between the standard deviation of processes finishing times to their mean value. 
A severe \aliD{execution} load imbalance corresponds to a high value of c.o.v. and a low value of mean/max. 
\figurename{~\ref{subfig:PSIA_imbalance_416}} and \figurename{~\ref{subfig:Mandelbrot_imbalance_416}} show the median of the two load imbalance metrics over five executions. 
Both metrics indicate high load imbalance \aliD{for} the applications executed with STATIC, FSC, mFSC, GSS, TSS, and FAC, which correspond to longer parallel loop execution times in \figurename{~\ref{subfig:PSIA_native_np_416}} and \figurename{~\ref{subfig:Mandelbrot_native_np_416}}.
\aliD{A} similar performance can also be observed in \figurename{~\ref{fig:imbalance_128}} \aliD{for} PSIA and Mandelbrot \aliD{executed} on 128 heterogeneous cores of the miniHPC system.  
\aliD{An} inspection of the Mandelbrot execution times with AF, for both system sizes, reveals that the high variation in the execution time with 416 cores is due to the small number of loop iterations per PE. 
The small number of loop iterations per PE  and the high variation of loop iterations execution times of Mandelbrot did not \aliD{offer sufficient opportunity to} AF to \aliD{accurately} learn the PE weights. 
In the execution of Mandelbrot on 128 cores, the number of loop iterations per core is higher than \aliD{on} 416 cores as the problem size is fixed. 
This allowed the improved and more stable performance of Mandelbrot with AF \aliD{on} 128 cores.

\subsection{Performance of Scientific Applications under Perturbations}
\label{subsec:perf}
%
%
\noindent\textbf{Simulative experiments.}
The simulative performance results of the two real applications, PSIA and Mandelbrot, under perturbations are shown in Figures~5 - 8. 
One can \aliD{note} that STATIC, GSS, TSS, and FAC perform poorly on heterogeneous systems. 
Also, WF can not accommodate the variability in the system due to perturbations, especially \aliD{to} perturbations in the delivered computational speed \aliD{of the PEs}.
\aliD{The performance of} FSC and mFSC is, in general, higher than \aliD{that of} STATIC, GSS, TSS, FAC, and WF.
However, FSC and mFSC are highly affected by the perturbations in the PE availability.
SS is resilient to perturbations in the delivered computational speed of the PEs. 
However, it is significantly influenced by the network latency variations, as can be seen in Figures~5 - 8 \texttt{lat-cs} and \texttt{lat-es}.

\begin{figure}[]
	\centering
	\subfloat[PSIA simulative performance on 128 cores]{%
		\includegraphics[clip, trim=0cm 0cm 0cm 0cm, width = 0.8\textwidth]{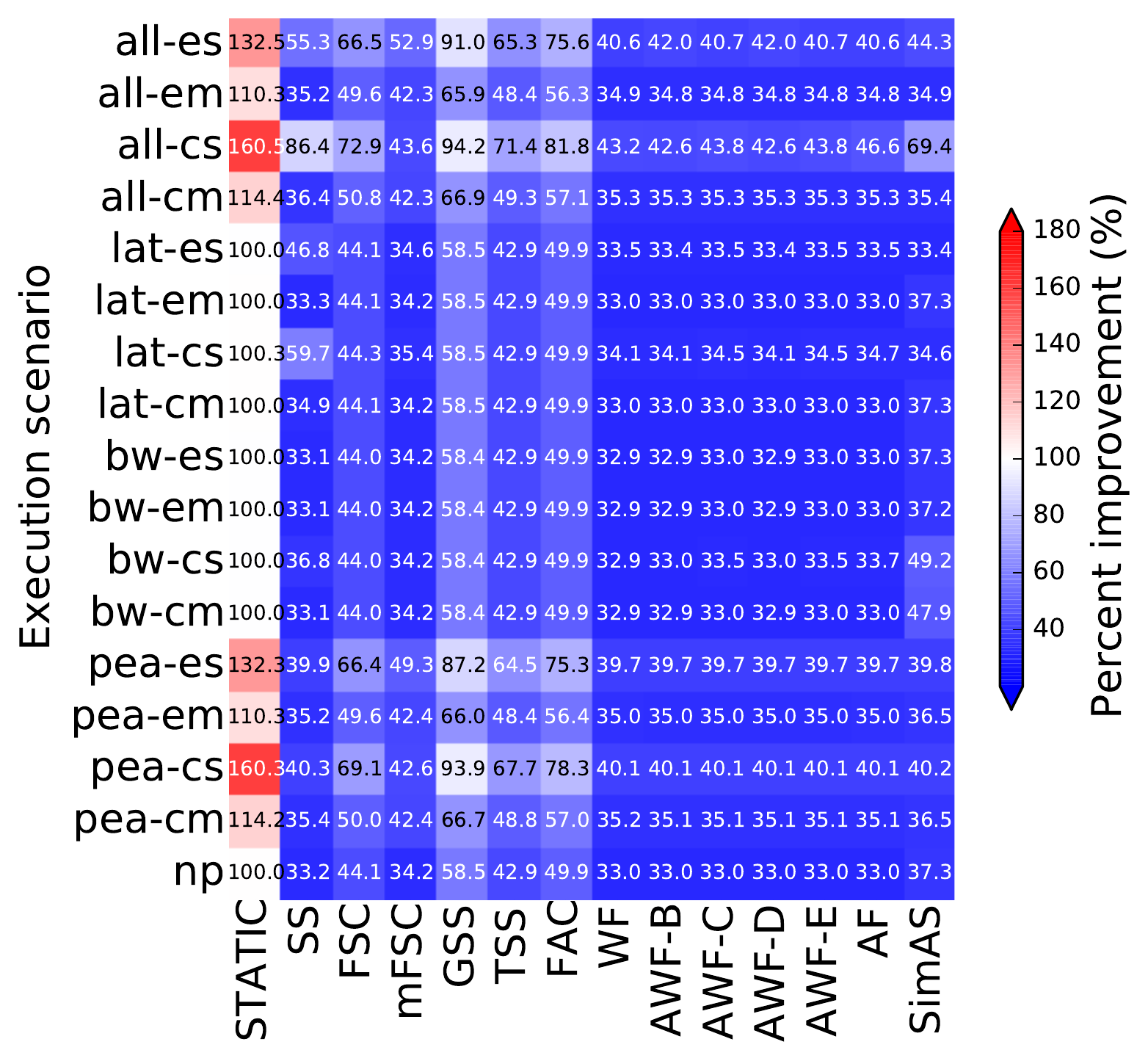}%
		\label{subfig:PSIA_128_sim_heatmap}%
	} \\
	\subfloat[Percentage of counts DLS techniques are selected by \sil{} ]{%
		\includegraphics[clip, trim=0cm 0cm 0cm 0cm, width = 0.8\textwidth]{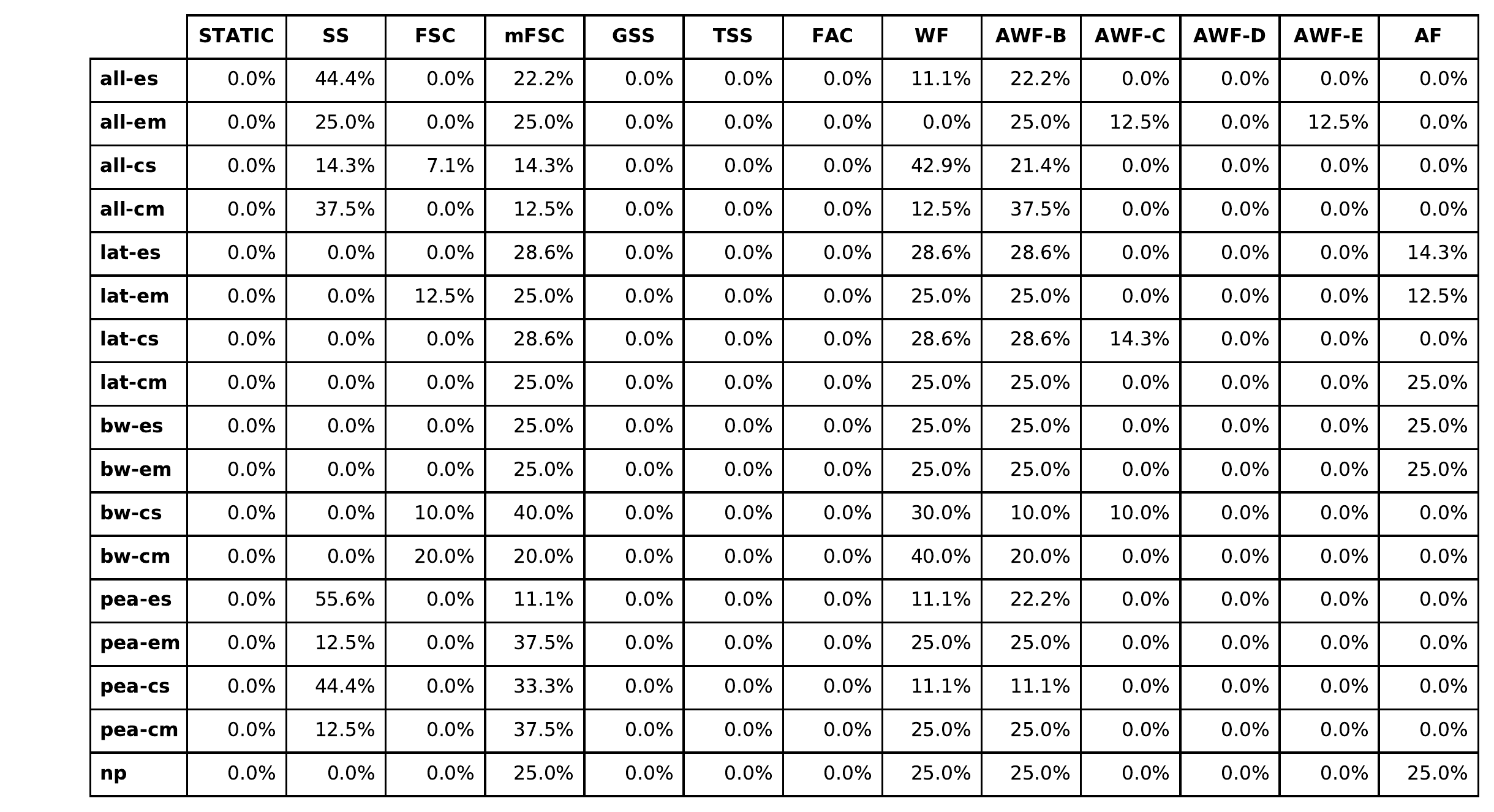}%
		\label{subfig:PSIA_128_sim_table}}
	\\
	\caption{ \textbf{Simulative} performance results of PSIA without (denoted with np) and with (the rest) perturbations using \sil{} and other thirteen loop scheduling techniques on 128 cores of miniHPC. Percent performance improvement normalized to STATIC in np scenario (baseline case without any perturbations and baseline load balancing method). White, red, and blue denote baseline ($=100\%$), degraded ($>100\%$), and improved performance ($<100\%$), respectively.
		The table shows the DLS techniques dynamically selected by \sil{} during execution.} 
	\label{fig:SimAS_psia_sim}
\end{figure}

\begin{figure}[]
	\centering
	\subfloat[PSIA simulative performance on 416 cores]{%
		\includegraphics[clip, trim=0cm 0cm 0cm 0cm, width = 0.8\textwidth]{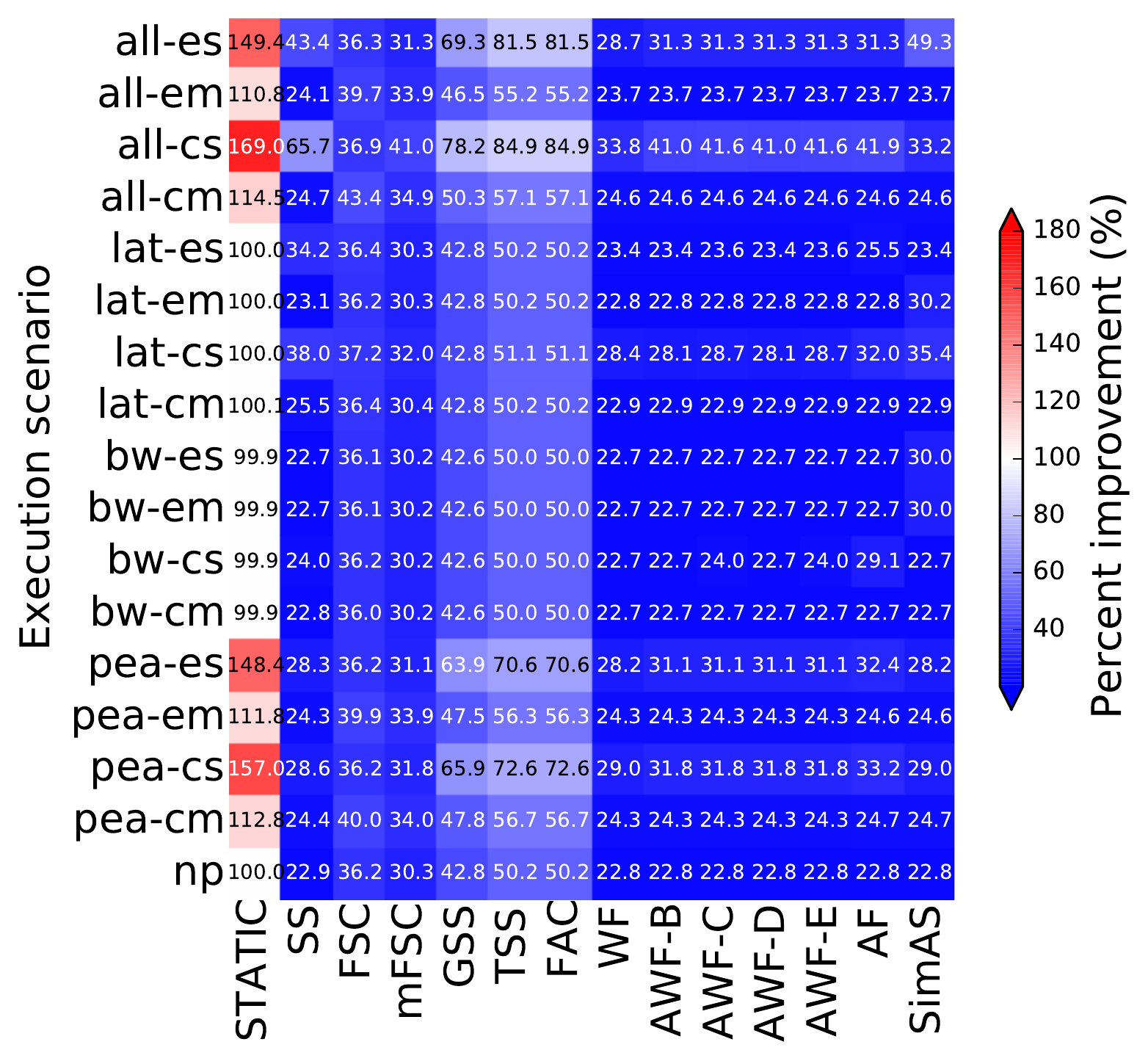}%
		\label{subfig:PSIA_416_sim_heatmap}%
	} \\
	\subfloat[Percentage of counts DLS techniques are selected by \sil{} ]{%
		\includegraphics[clip, trim=0cm 0cm 0cm 0cm, width = 0.8\textwidth]{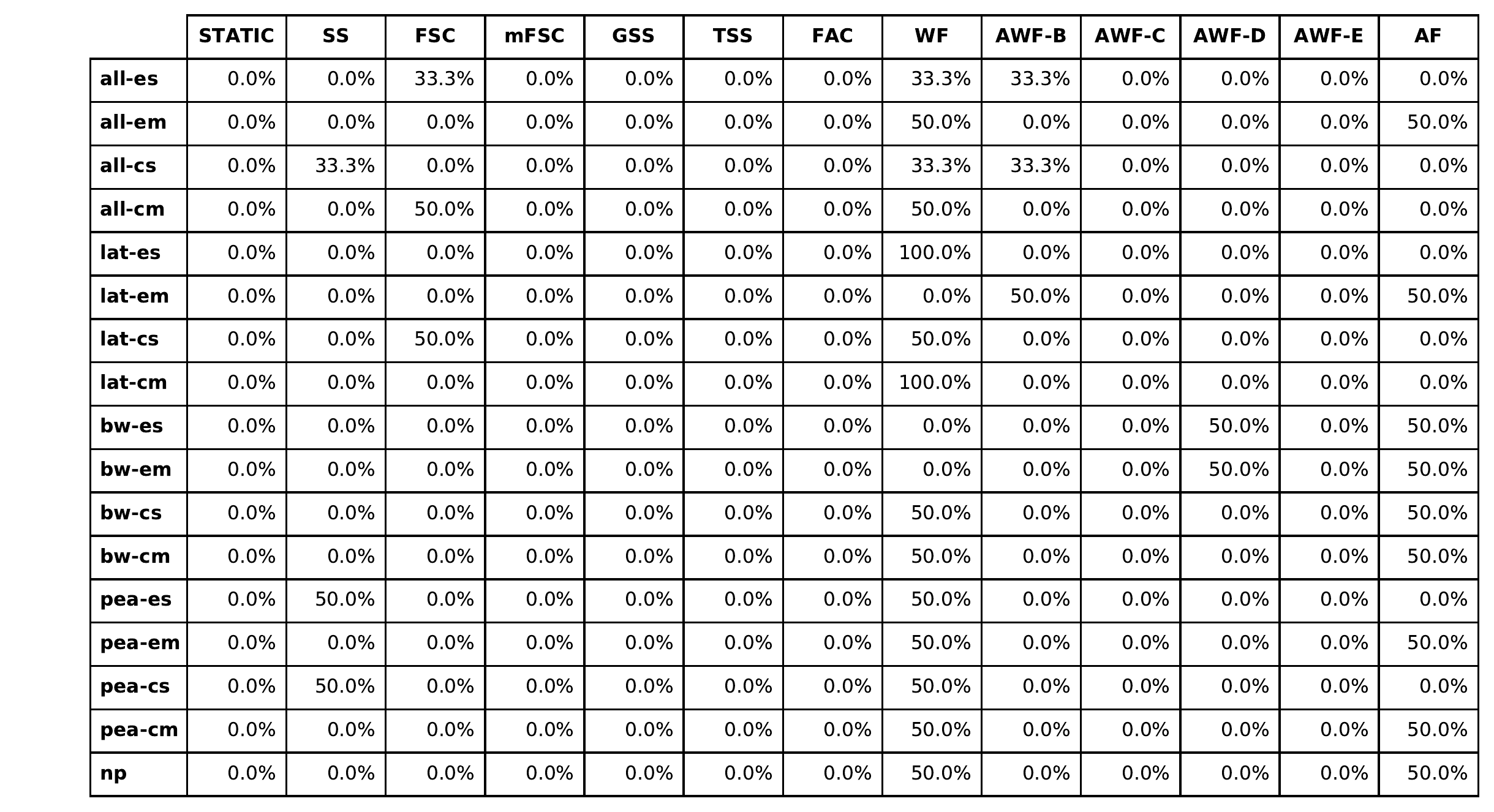}%
		\label{subfig:PSIA_416_sim_table}}
	\\
	\caption{ \textbf{Simulative} performance results of PSIA without (denoted with np) and with (the rest) perturbations using \sil{} and other thirteen loop scheduling techniques on 416 cores of miniHPC. Percent performance improvement normalized to STATIC in np scenario (baseline case without any perturbations and baseline load balancing method). White, red, and blue denote baseline ($=100\%$), degraded ($>100\%$), and improved performance ($<100\%$), respectively.
		The table shows the DLS techniques dynamically selected by \sil{} during execution.} 
	\label{fig:SimAS_psia_sim_416}
\end{figure}

\begin{figure}[]
	\centering
	\subfloat[Mandelbrot simulative performance on 128 cores]{%
		\includegraphics[clip, trim=0cm 0cm 0cm 0cm,width = 0.8\textwidth]{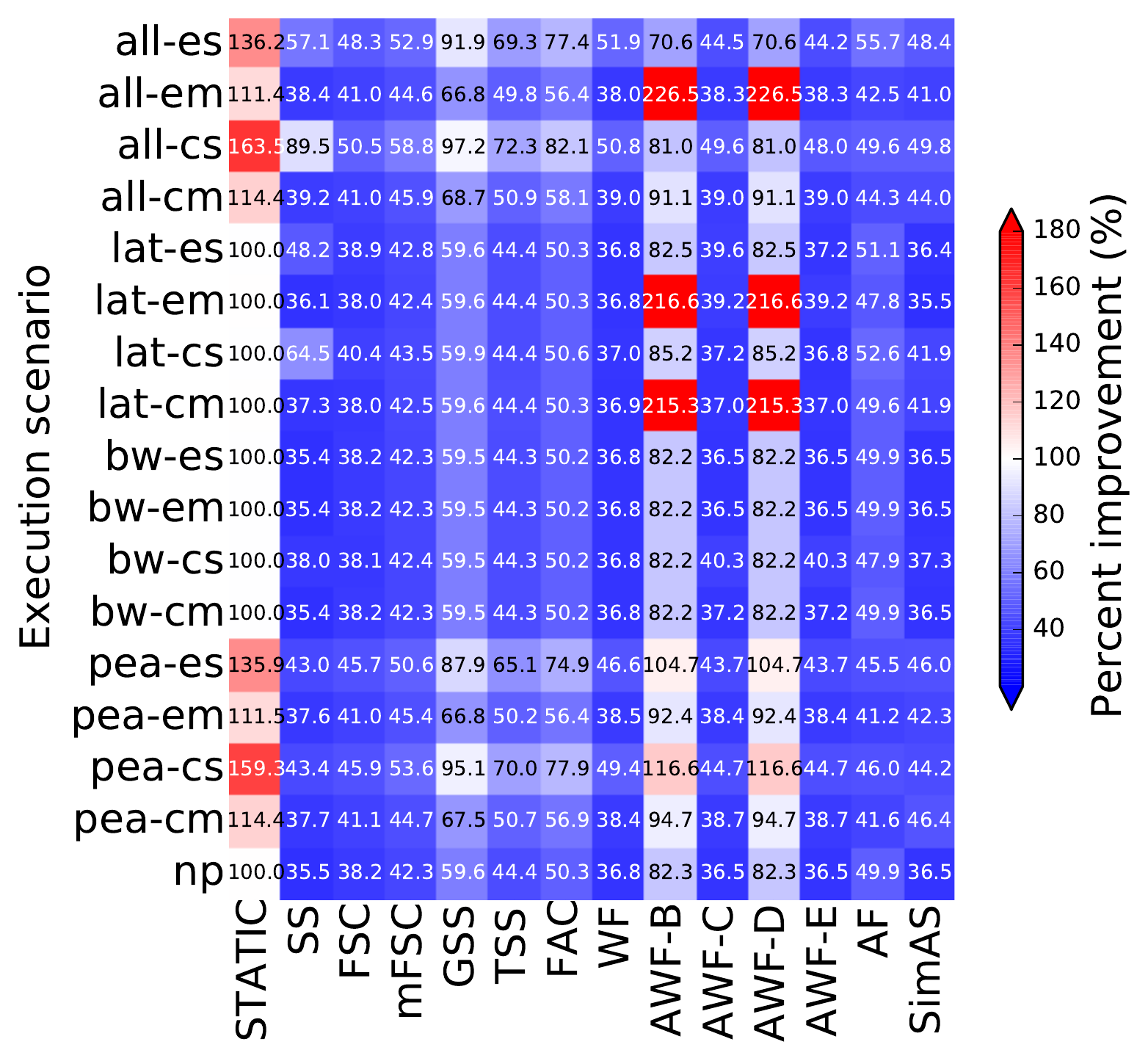}%
		\label{subfig:Mandelbrot_128_sim_heatmap}
	} \\
	\subfloat[Percentage of counts DLS techniques are selected by \sil{}]{%
		\includegraphics[clip, trim=0cm 0cm 0cm 0cm,width = 0.8\textwidth]{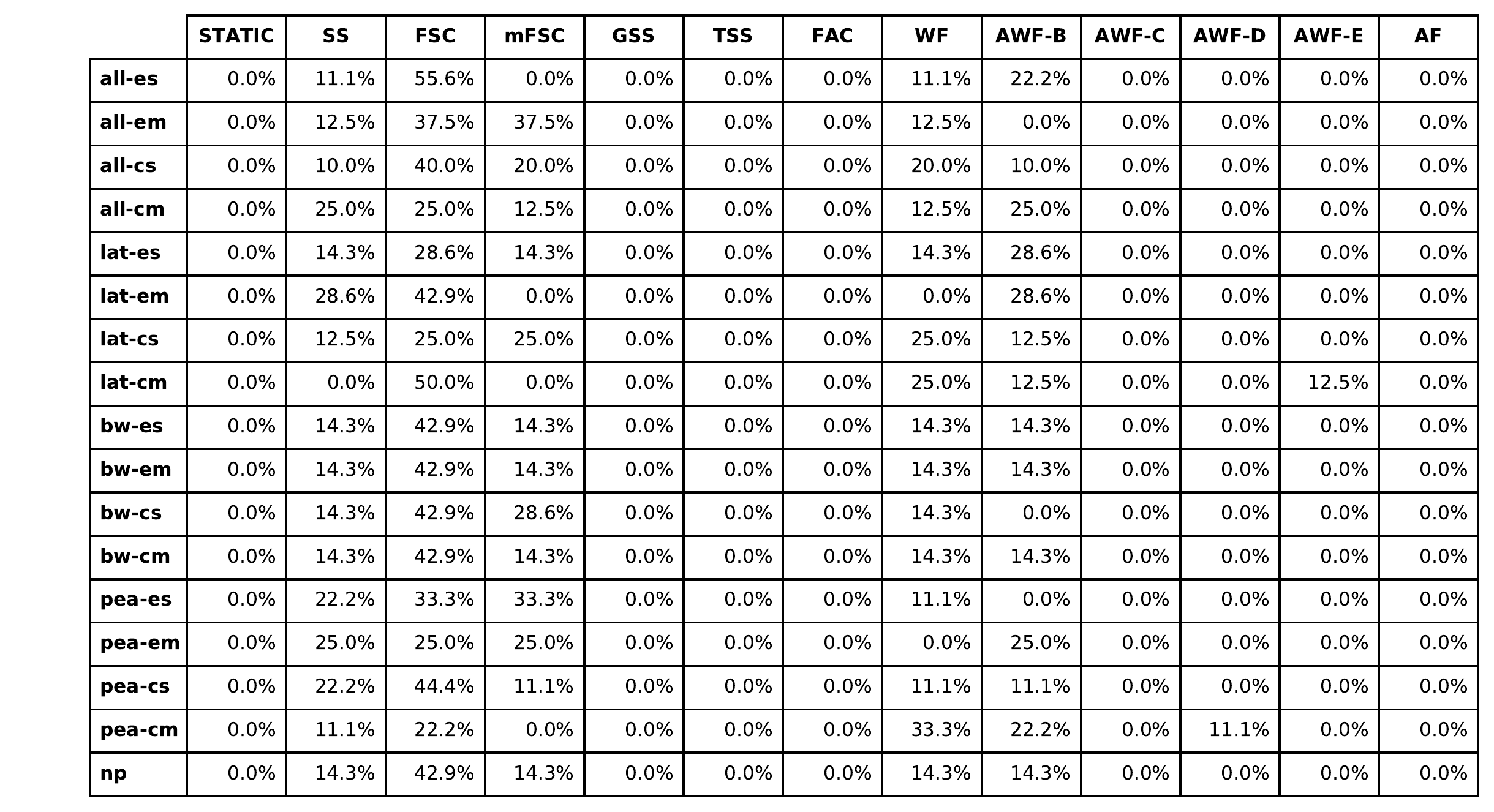}%
		\label{subfig:Mandelbrot_128_sim_table}%
	} 
	\caption{ \textbf{Simulative} performance results of Mandelbrot without (denoted with np) and with (the rest) perturbations using \sil{} and other thirteen loop scheduling techniques on 128 cores of miniHPC. Percent performance improvement normalized to STATIC in np scenario (baseline case without any perturbations and baseline load balancing method). White, red, and blue denote baseline ($=100\%$), degraded ($>100\%$), and improved performance ($<100\%$), respectively.
		The table shows the DLS techniques dynamically selected by \sil{} during execution.} 
	\label{fig:SimAS_mandel_sim}
\end{figure}

\begin{figure}[]
	\centering
	\subfloat[Mandelbrot simulative performance on 416 cores]{%
		\includegraphics[clip, trim=0cm 0cm 0cm 0cm,width = 0.8\textwidth]{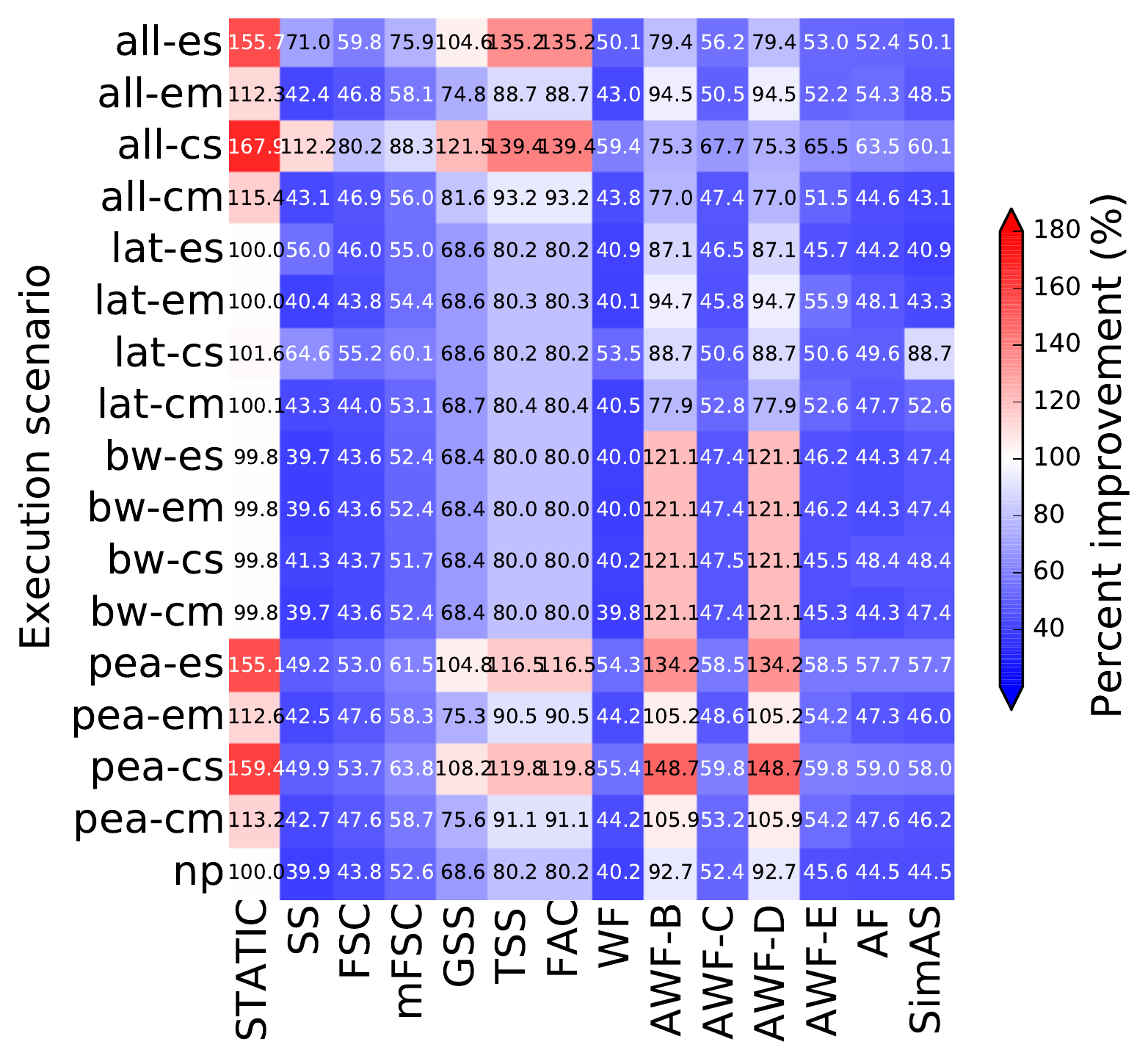}%
		\label{subfig:Mandelbrot_416_sim_heatmap}
	} \\
	\subfloat[Percentage of counts DLS techniques are selected by \sil{}]{%
		\includegraphics[clip, trim=0cm 0cm 0cm 0cm,width = 0.8\textwidth]{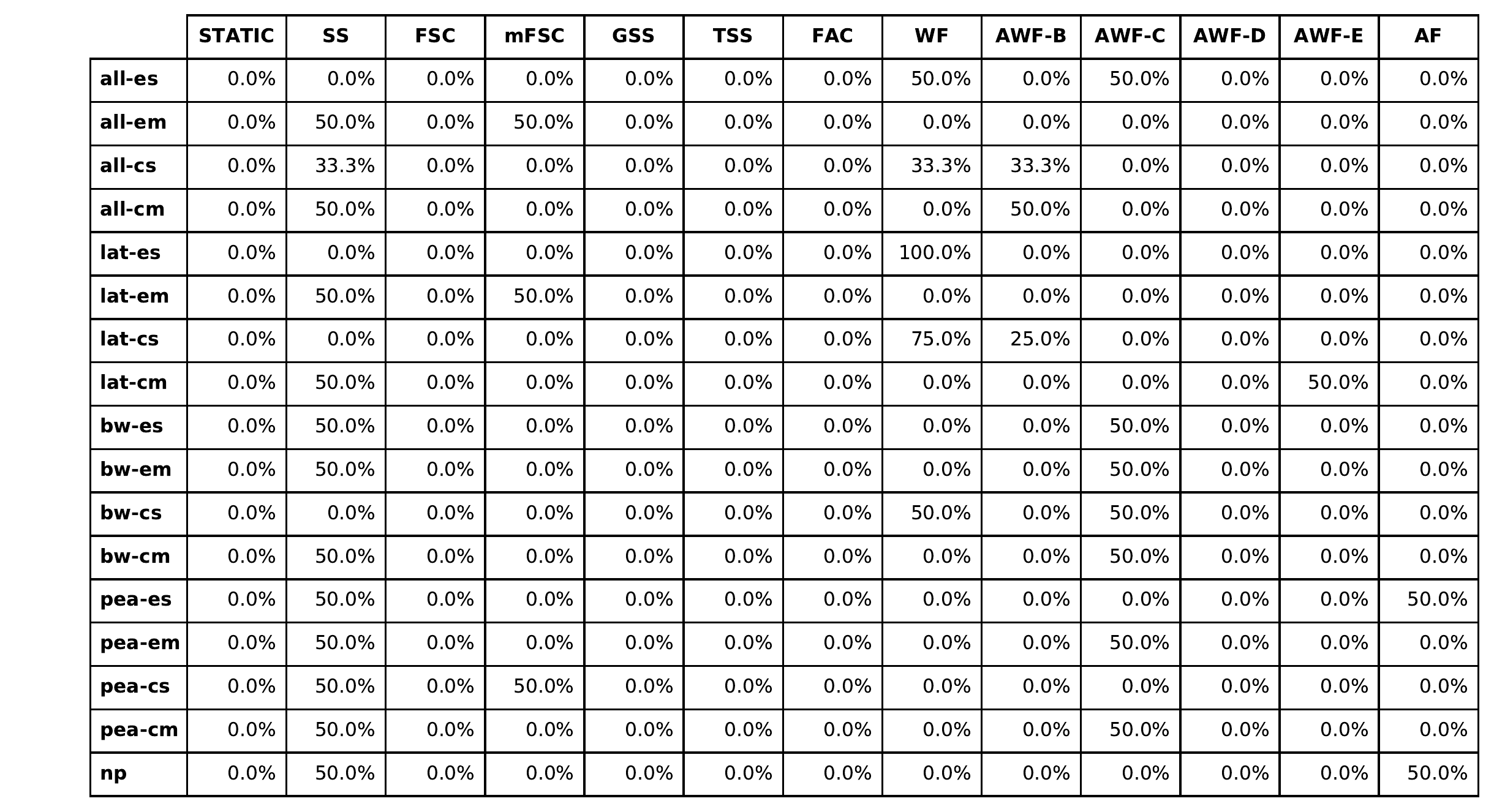}%
		\label{subfig:Mandelbrot_416_sim_table}%
	} 
	\caption{ \textbf{Simulative} performance results of Mandelbrot without (denoted with np) and with (the rest) perturbations using \sil{} and other thirteen loop scheduling techniques on 416 cores of miniHPC. Percent performance improvement normalized to STATIC in np scenario (baseline case without any perturbations and baseline load balancing method). White, red, and blue denote baseline ($=100\%$), degraded ($>100\%$), and improved performance ($<100\%$), respectively.
		The table shows the DLS techniques dynamically selected by \sil{} during execution.} 
	\label{fig:SimAS_mandel_sim_416}
\end{figure}

\begin{figure}[]
	\centering
	\subfloat[Constant workload simulative performance on 128 cores]{%
		\includegraphics[clip, trim=0cm 0cm 0cm 0cm, width = 0.8\textwidth]{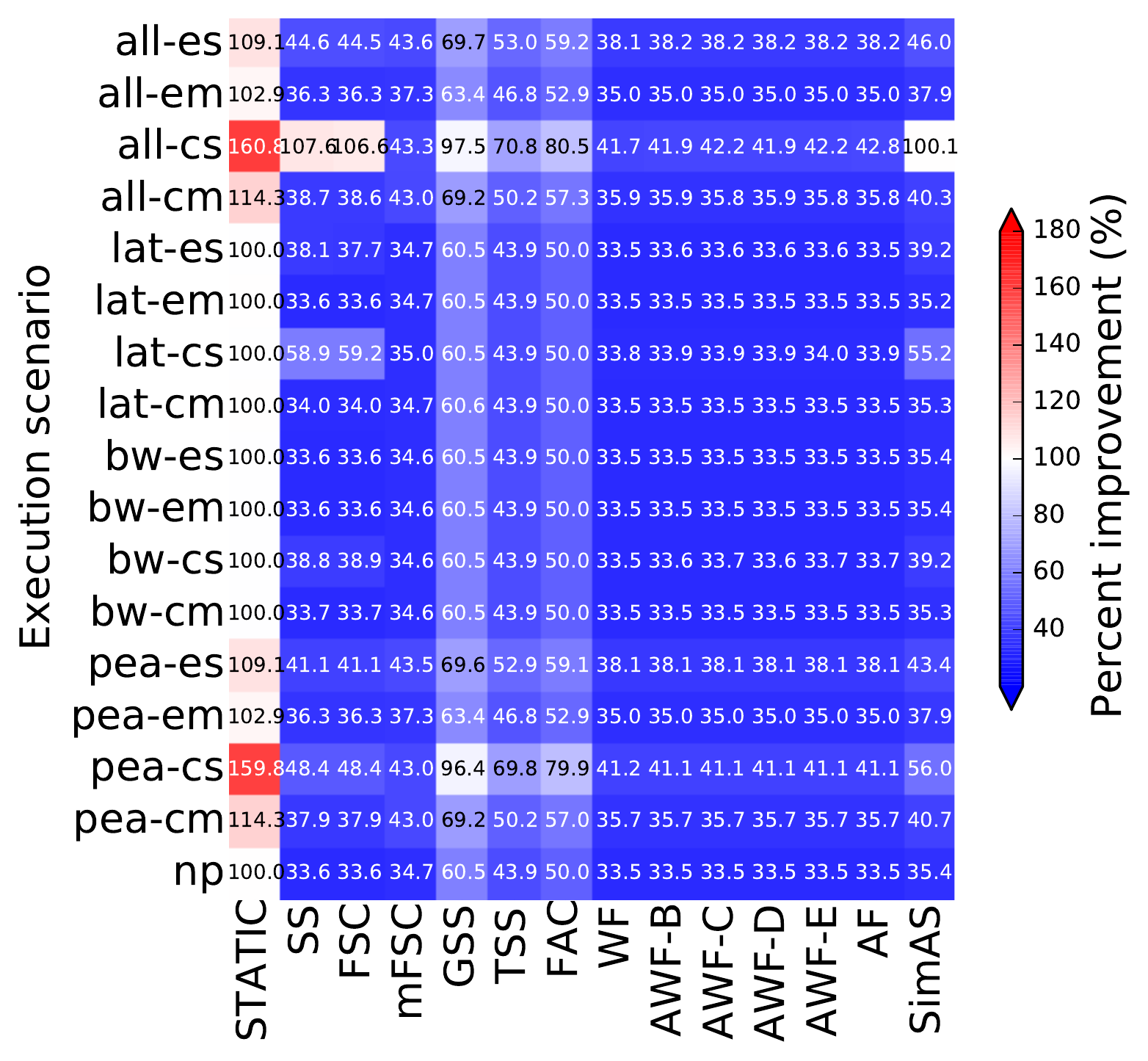}%
		\label{subfig:constant_128_sim_heatmap}%
	} \\
	\subfloat[Percentage of counts DLS techniques are selected by \sil{} ]{%
		\includegraphics[clip, trim=0cm 0cm 0cm 0cm, width = 0.8\textwidth]{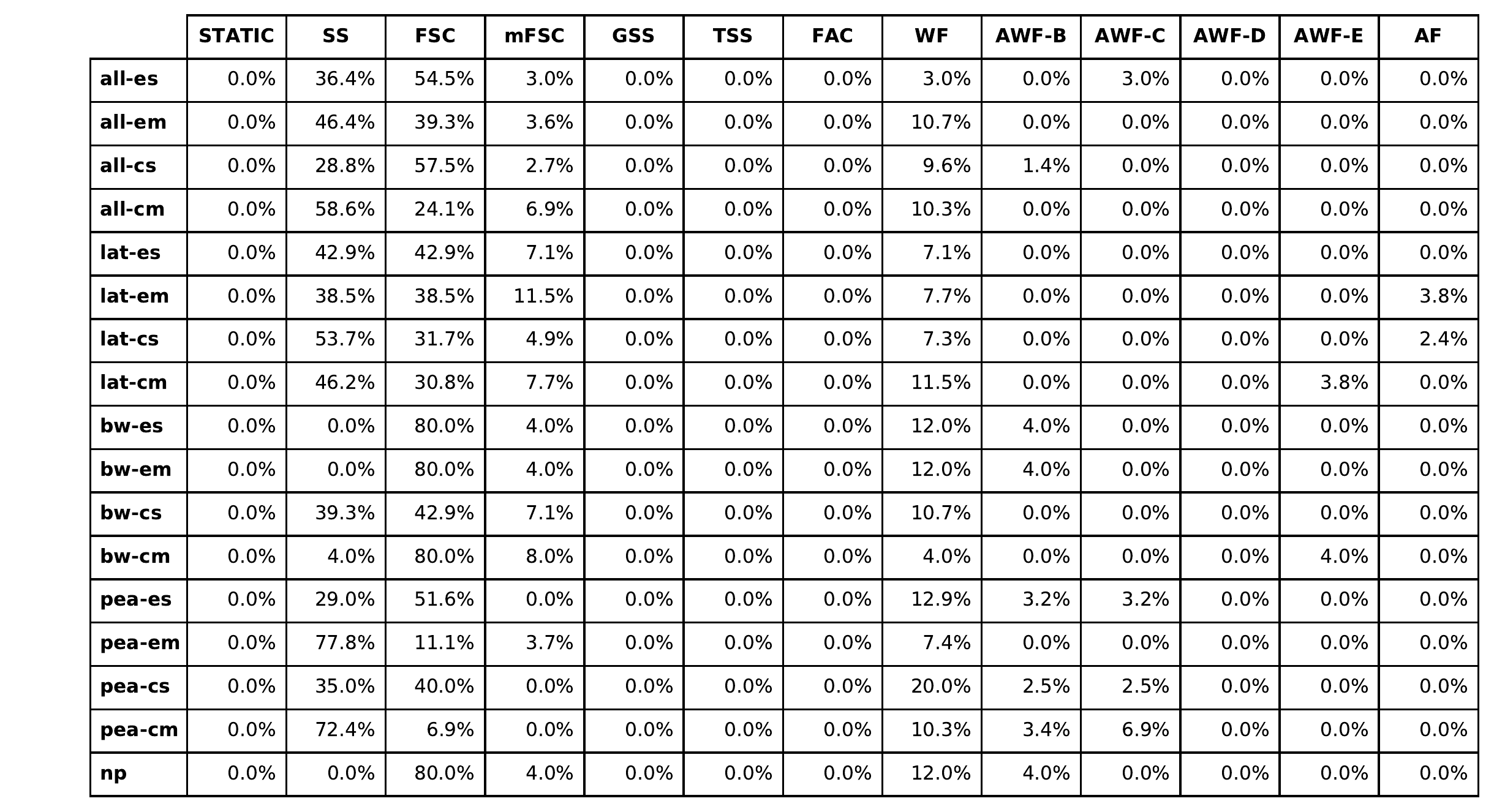}%
		\label{subfig:constant_128_sim_table}}
	\\
	\caption{ \textbf{Simulative} performance results of Constant synthetic workload without (denoted with np) and with (the rest) perturbations using \sil{} and other thirteen loop scheduling techniques on 128 cores of miniHPC. Percent performance improvement normalized to STATIC in np scenario (baseline case without any perturbations and baseline load balancing method). White, red, and blue denote baseline ($=100\%$), degraded ($>100\%$), and improved performance ($<100\%$), respectively.
		The table shows the DLS techniques dynamically selected by \sil{} during execution.} 
	\label{fig:SimAS_Constant_sim}
\end{figure}

\begin{figure}[]
	\centering
	\subfloat[Constant workload simulative performance on 416 cores]{%
		\includegraphics[clip, trim=0cm 0cm 0cm 0cm, width = 0.8\textwidth]{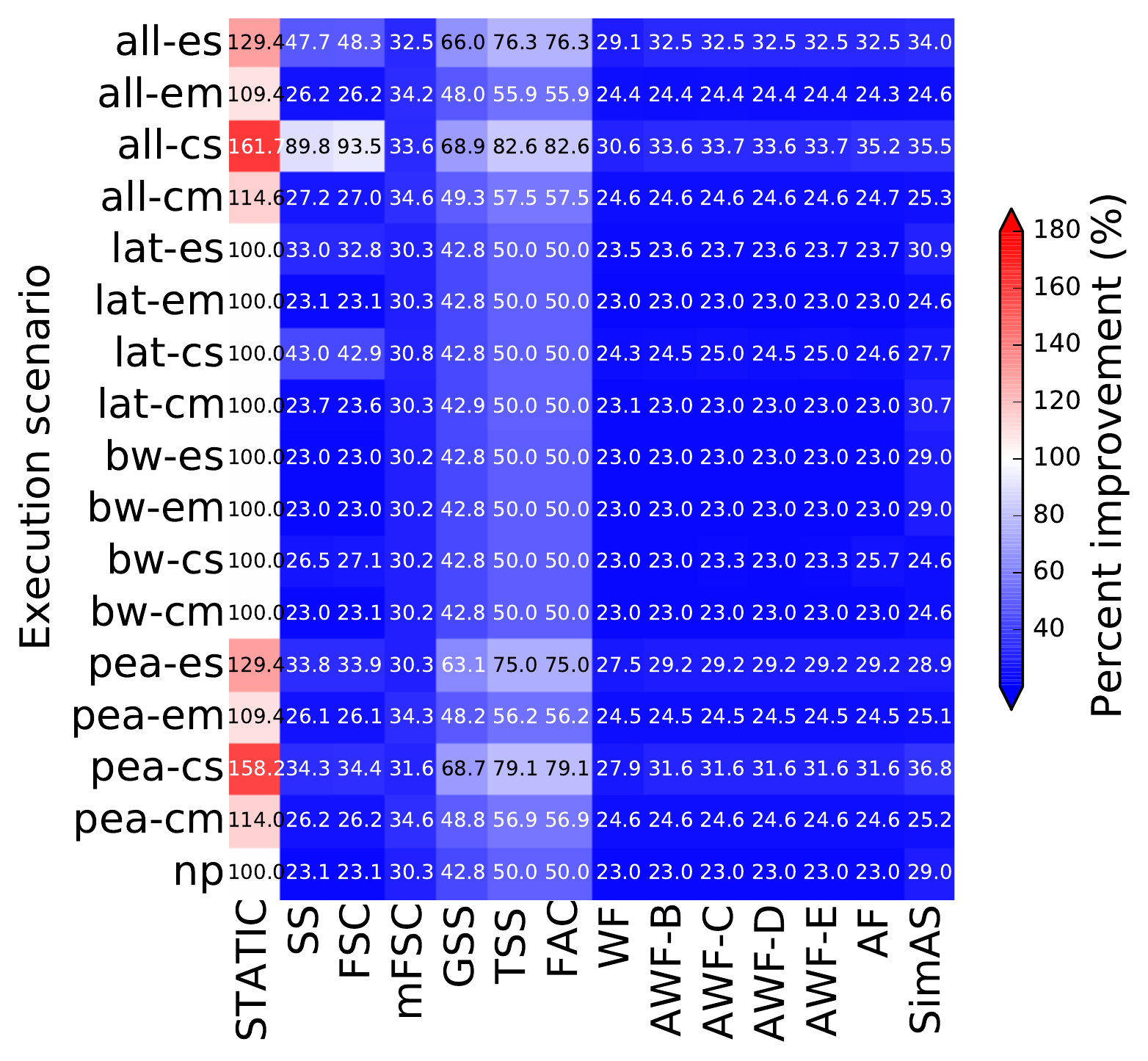}%
		\label{subfig:constant_416_sim_heatmap}%
	} \\
	\subfloat[Percentage of counts DLS techniques are selected by \sil{} ]{%
		\includegraphics[clip, trim=0cm 0cm 0cm 0cm, width = 0.8\textwidth]{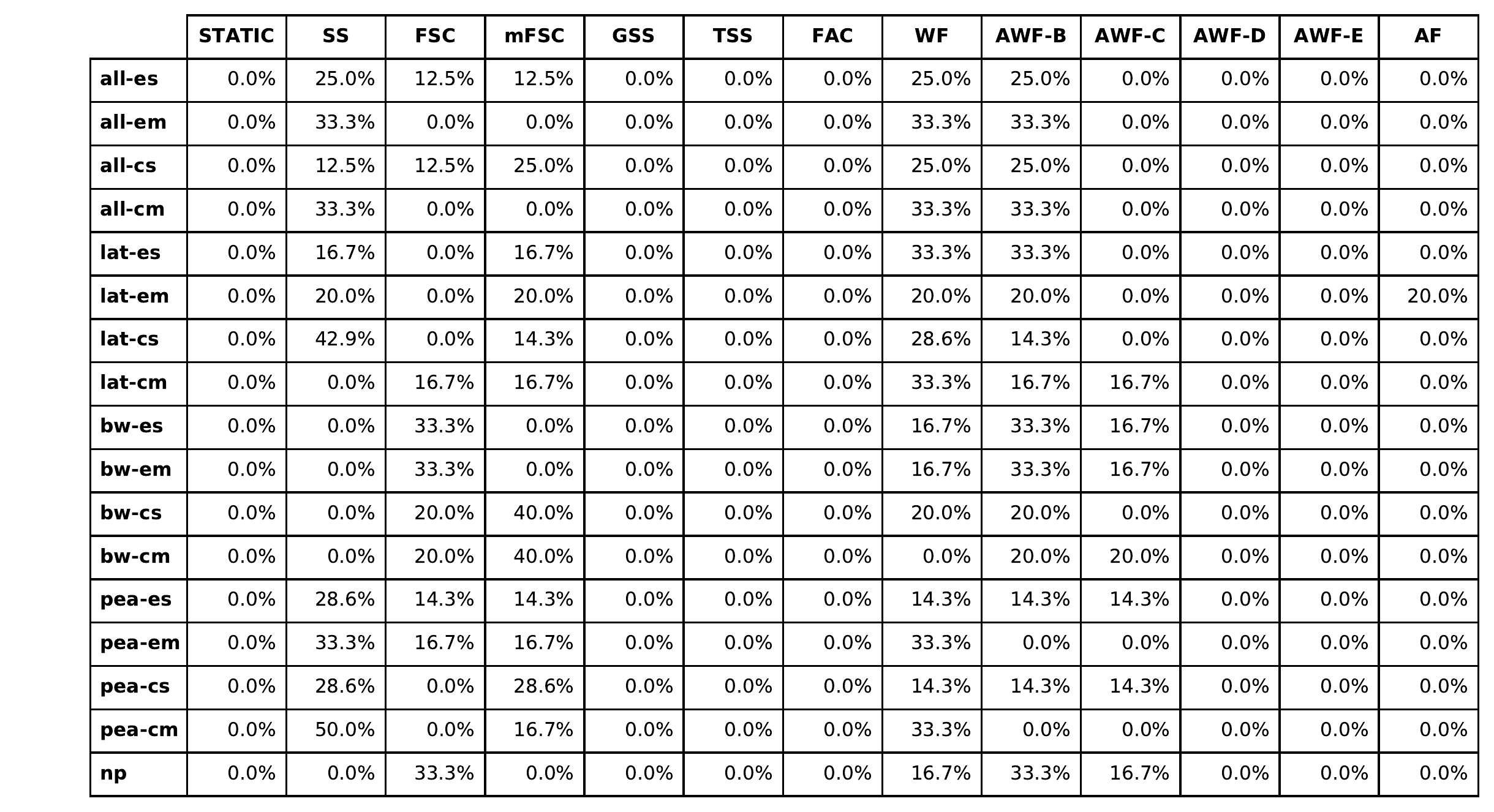}%
		\label{subfig:constant_416_sim_table}}
	\\
	\caption{ \textbf{Simulative} performance results of Constant synthetic workload without (denoted with np) and with (the rest) perturbations using \sil{} and other thirteen loop scheduling techniques on 416 cores of miniHPC. Percent performance improvement normalized to STATIC in np scenario (baseline case without any perturbations and baseline load balancing method). White, red, and blue denote baseline ($=100\%$), degraded ($>100\%$), and improved performance ($<100\%$), respectively.
		The table shows the DLS techniques dynamically selected by \sil{} during execution.} 
	\label{fig:SimAS_Constant_sim_416}
\end{figure}

\begin{figure}[]
	\centering
	\subfloat[Uniform workload simulative performance on 128 cores]{%
		\includegraphics[clip, trim=0cm 0cm 0cm 0cm, width = 0.8\textwidth]{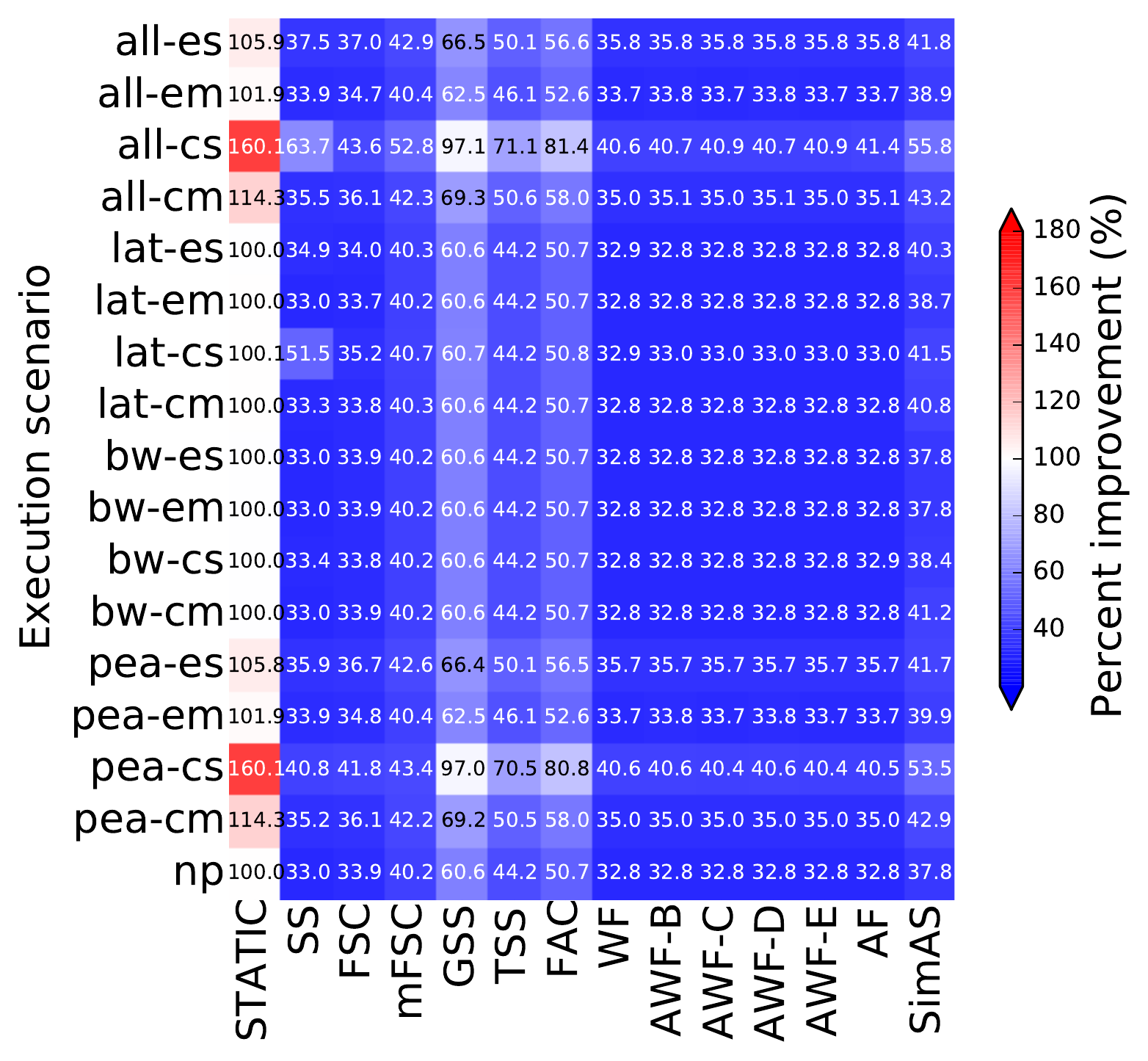}%
		\label{subfig:Uniform_128_sim_heatmap}%
	} \\
	\subfloat[Percentage of counts DLS techniques are selected by \sil{} ]{%
		\includegraphics[clip, trim=0cm 0cm 0cm 0cm, width = 0.8\textwidth]{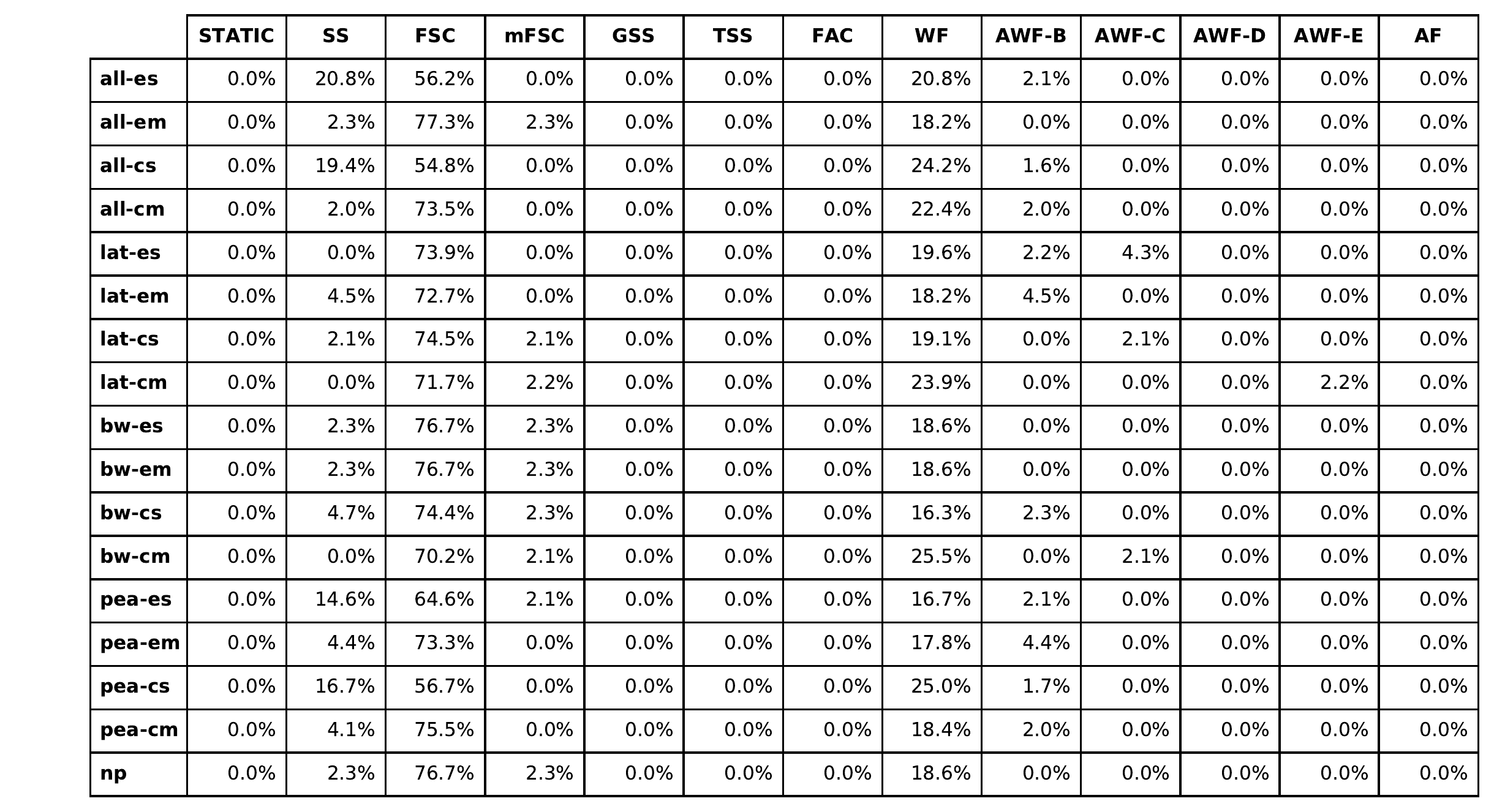}%
		\label{subfig:Uniform_128_sim_table}}
	\\
	\caption{ \textbf{Simulative} performance results of Uniform synthetic workload without (denoted with np) and with (the rest) perturbations using \sil{} and other thirteen loop scheduling techniques on 128 cores of miniHPC. Percent performance improvement normalized to STATIC in np scenario (baseline case without any perturbations and baseline load balancing method). White, red, and blue denote baseline ($=100\%$), degraded ($>100\%$), and improved performance ($<100\%$), respectively.
		The table shows the DLS techniques dynamically selected by \sil{} during execution.} 
	\label{fig:SimAS_Uniform_sim}
\end{figure}

\begin{figure}[]
	\centering
	\subfloat[Uniform workload simulative performance on 416 cores]{%
		\includegraphics[clip, trim=0cm 0cm 0cm 0cm, width = 0.8\textwidth]{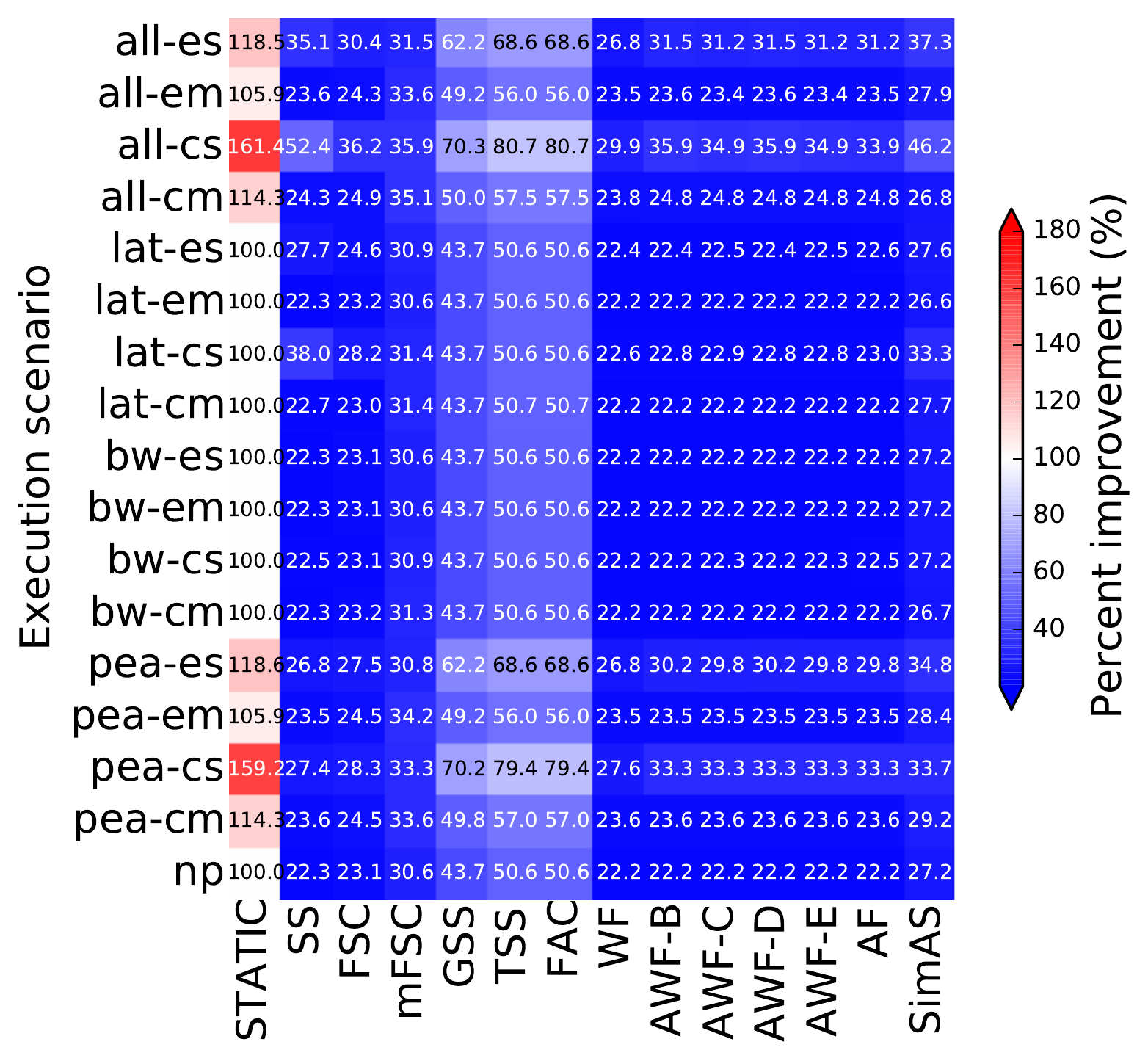}%
		\label{subfig:Uniform_416_sim_heatmap}%
	} \\
	\subfloat[Percentage of counts DLS techniques are selected by \sil{} ]{%
		\includegraphics[clip, trim=0cm 0cm 0cm 0cm, width = 0.8\textwidth]{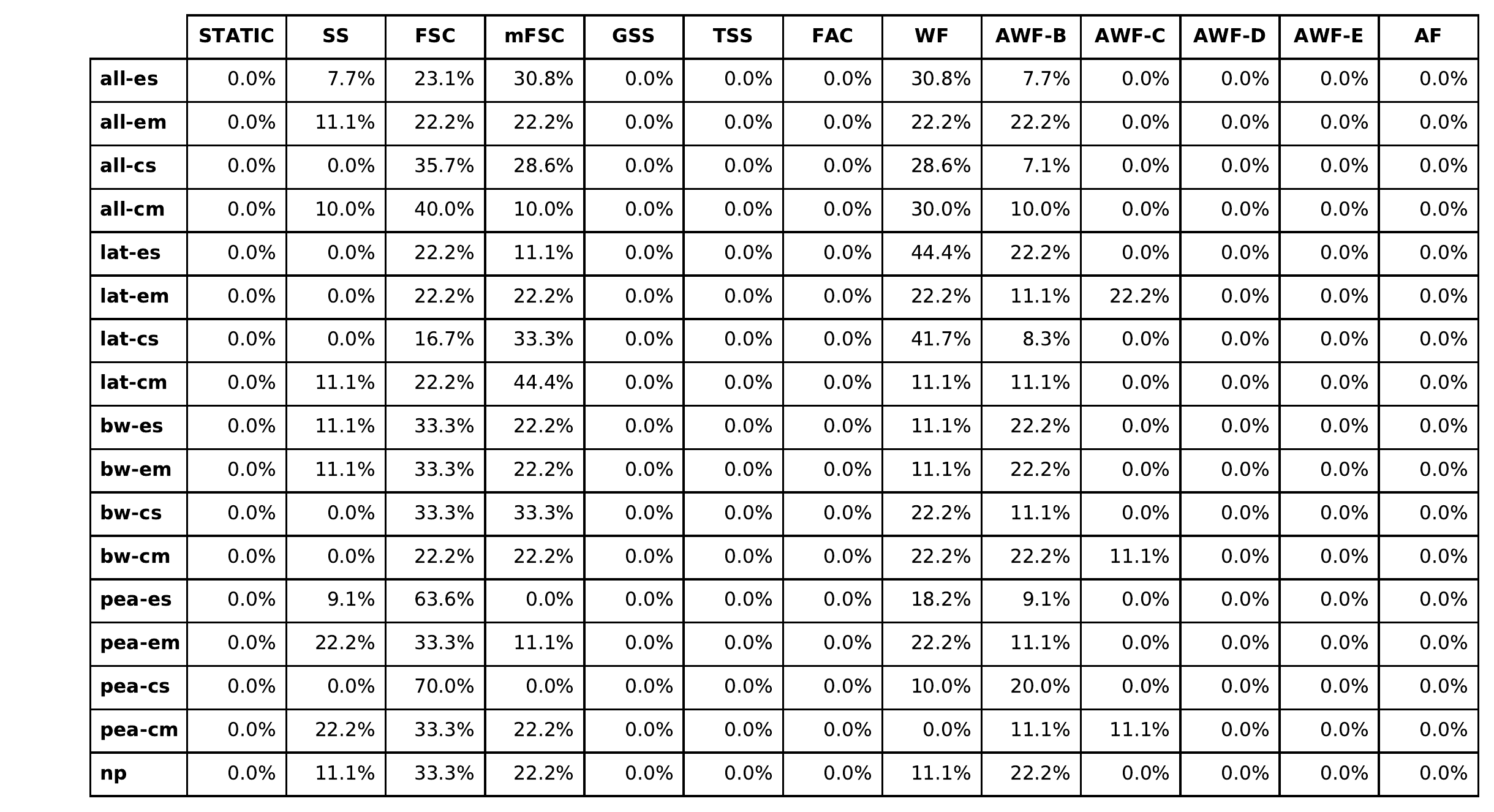}%
		\label{subfig:Uniform_416_sim_table}}
	\\
	\caption{ \textbf{Simulative} performance results of Uniform synthetic workload without (denoted with np) and with (the rest) perturbations using \sil{} and other thirteen loop scheduling techniques on 416 cores of miniHPC. Percent performance improvement normalized to STATIC in np scenario (baseline case without any perturbations and baseline load balancing method). White, red, and blue denote baseline ($=100\%$), degraded ($>100\%$), and improved performance ($<100\%$), respectively.
		The table shows the DLS techniques dynamically selected by \sil{} during execution.} 
	\label{fig:SimAS_Uniform_sim_416}
\end{figure}

\begin{figure}[]
	\centering
	\subfloat[Normal workload simulative performance on 128 cores]{%
		\includegraphics[clip, trim=0cm 0cm 0cm 0cm, width = 0.8\textwidth]{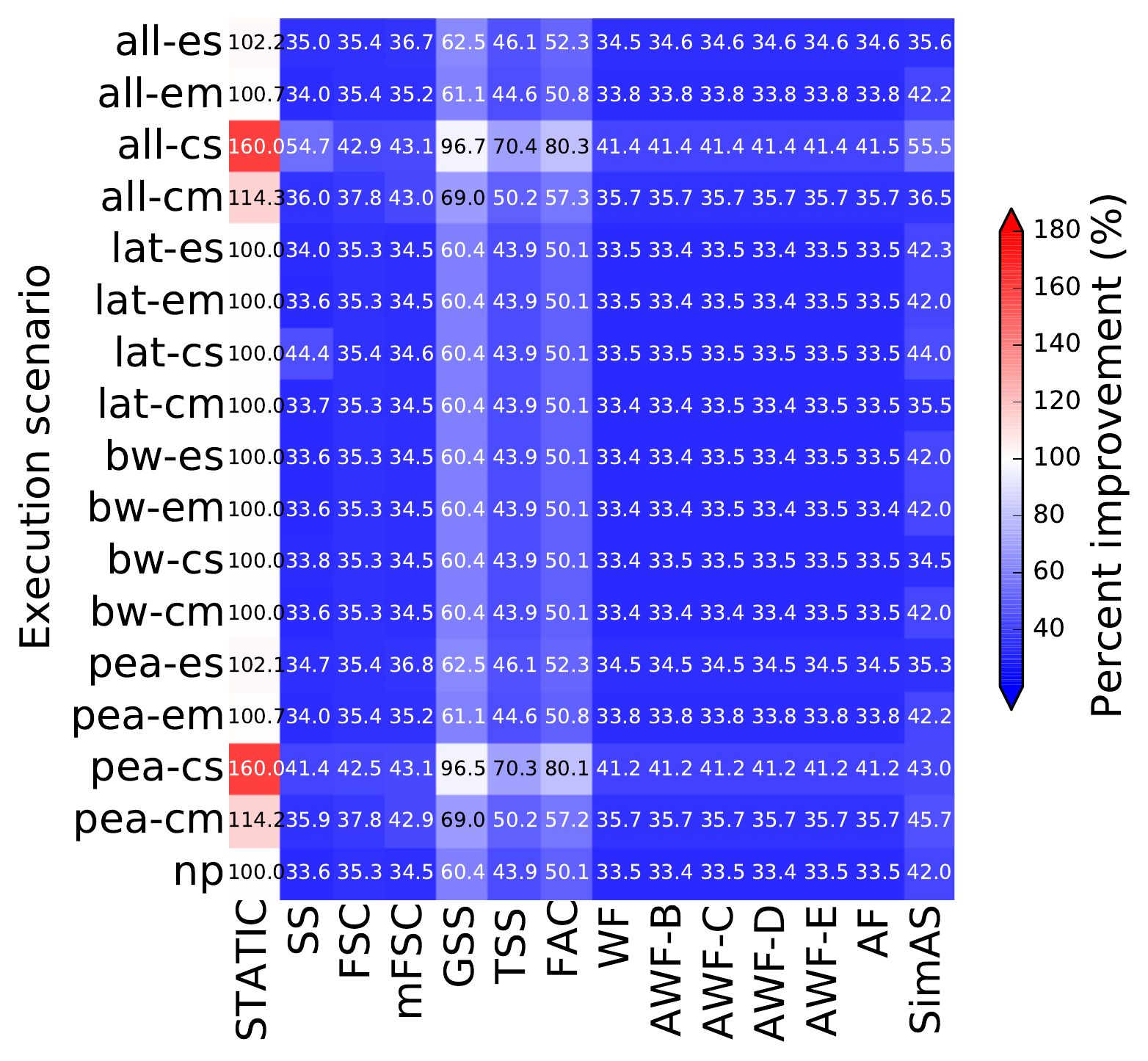}%
		\label{subfig:Normal_128_sim_heatmap}%
	} \\
	\subfloat[Percentage of counts DLS techniques are selected by \sil{} ]{%
		\includegraphics[clip, trim=0cm 0cm 0cm 0cm, width = 0.8\textwidth]{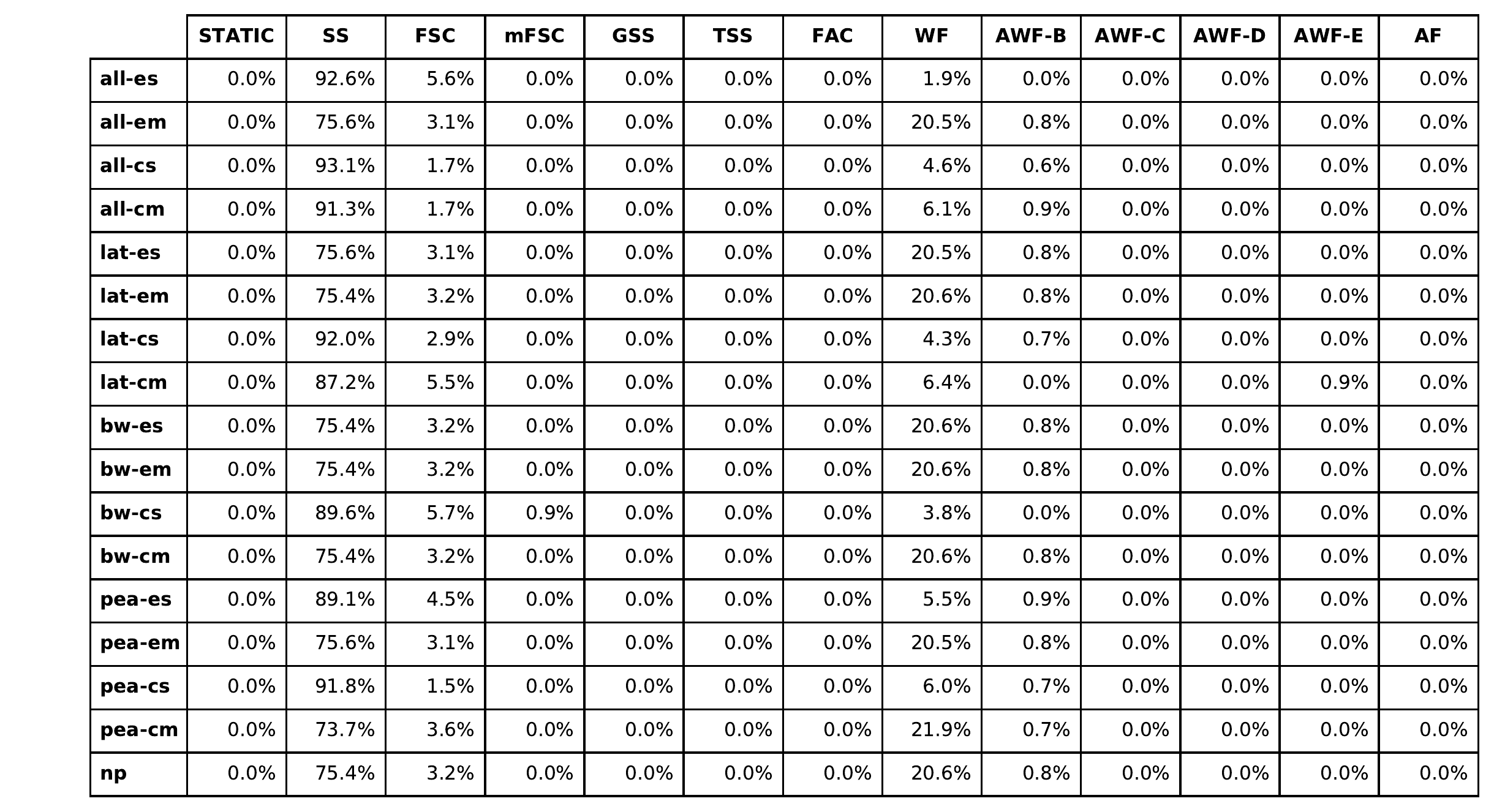}%
		\label{subfig:Normal_128_sim_table}}
	\\
	\caption{ \textbf{Simulative} performance results of Normal synthetic workload without (denoted with np) and with (the rest) perturbations using \sil{} and other thirteen loop scheduling techniques on 128 cores of miniHPC. Percent performance improvement normalized to STATIC in np scenario (baseline case without any perturbations and baseline load balancing method). White, red, and blue denote baseline ($=100\%$), degraded ($>100\%$), and improved performance ($<100\%$), respectively.
		The table shows the DLS techniques dynamically selected by \sil{} during execution.} 
	\label{fig:SimAS_Normal_sim}
\end{figure}

\begin{figure}[]
	\centering
	\subfloat[Normal workload simulative performance on 416 cores]{%
		\includegraphics[clip, trim=0cm 0cm 0cm 0cm, width = 0.8\textwidth]{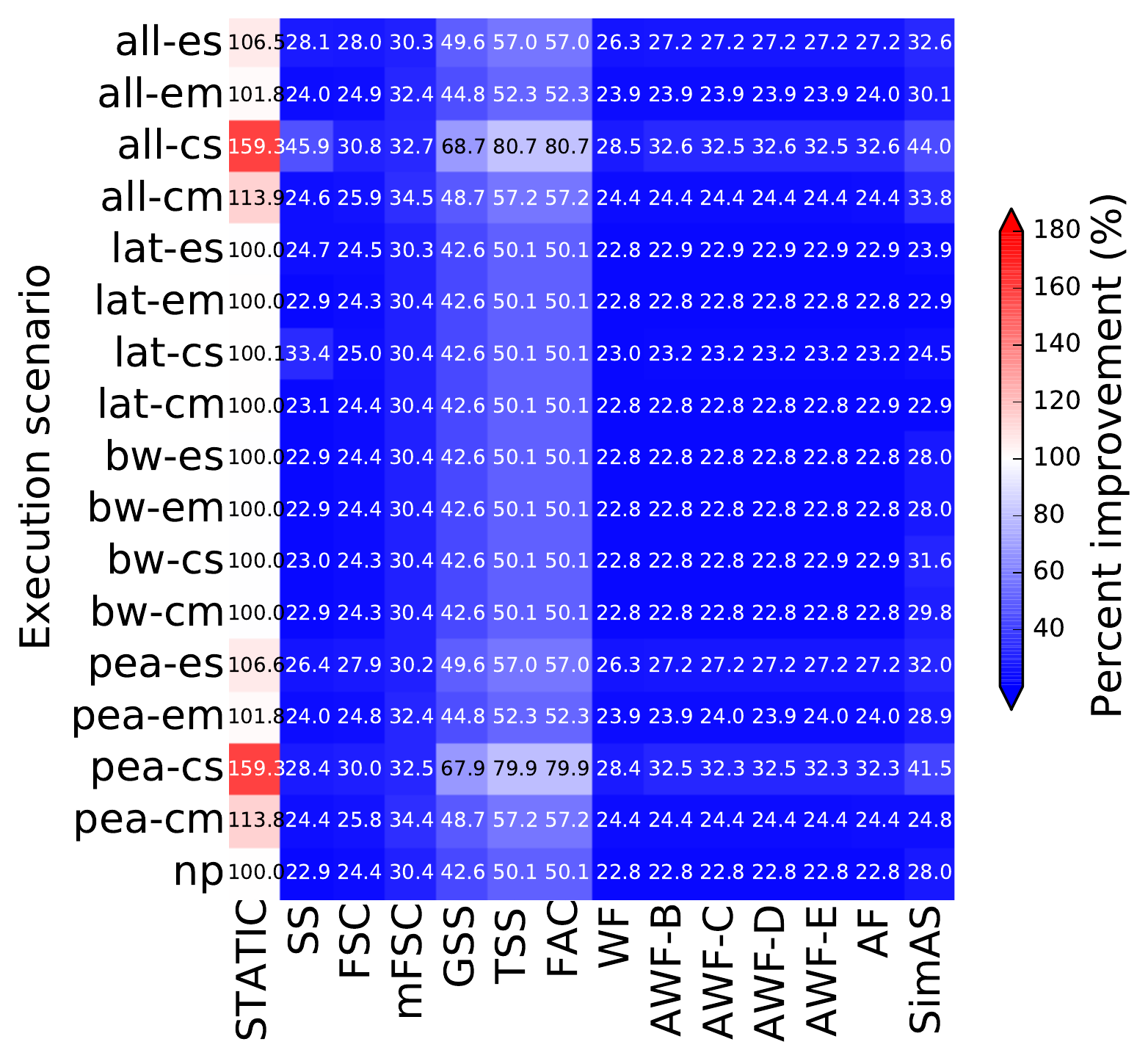}%
		\label{subfig:Normal_416_sim_heatmap}%
	} \\
	\subfloat[Percentage of counts DLS techniques are selected by \sil{} ]{%
		\includegraphics[clip, trim=0cm 0cm 0cm 0cm, width = 0.8\textwidth]{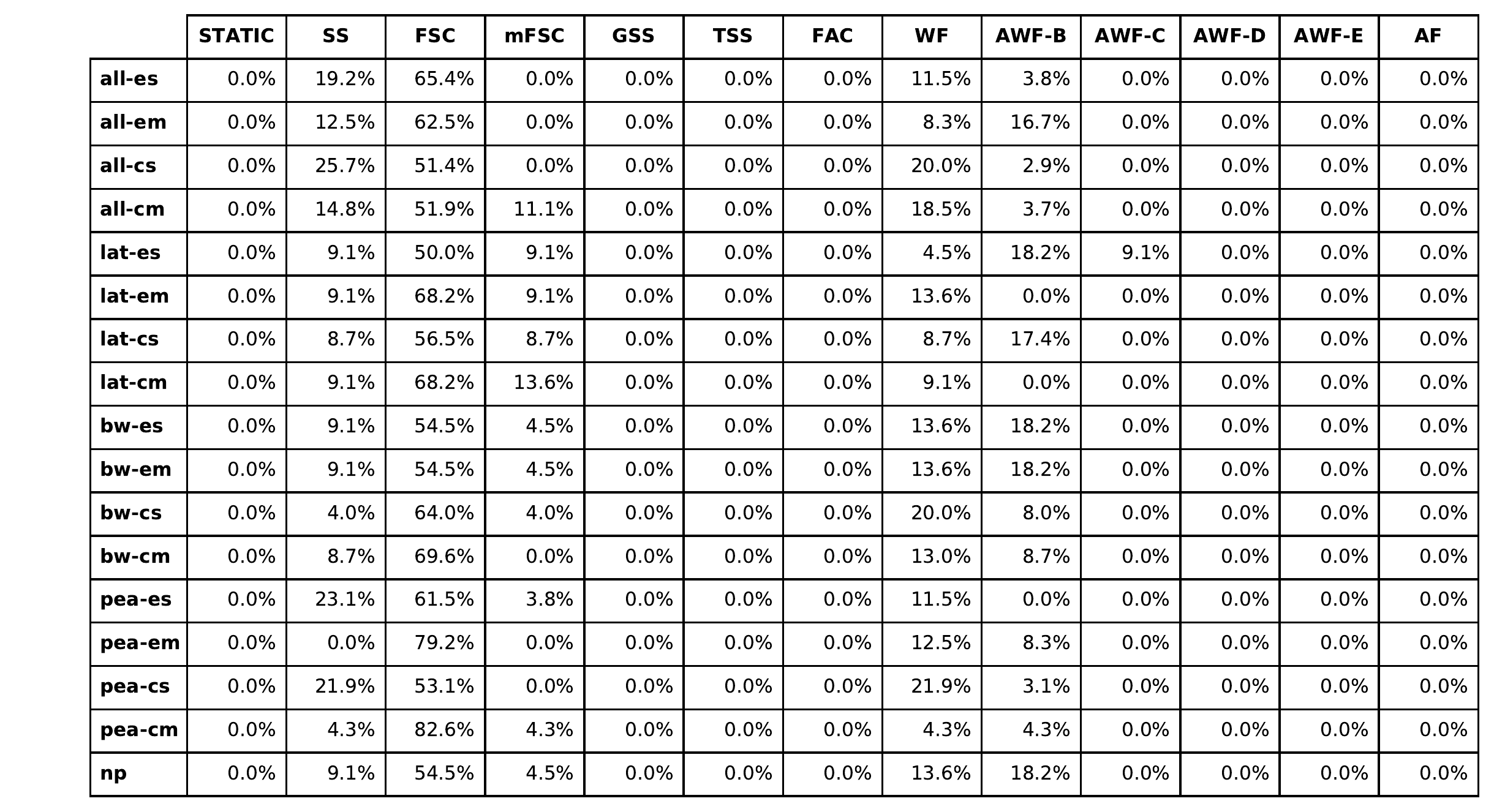}%
		\label{subfig:Normal_416_sim_table}}
	\\
	\caption{ \textbf{Simulative} performance results of Normal synthetic workload without (denoted with np) and with (the rest) perturbations using \sil{} and other thirteen loop scheduling techniques on 416 cores of miniHPC. Percent performance improvement normalized to STATIC in np scenario (baseline case without any perturbations and baseline load balancing method). White, red, and blue denote baseline ($=100\%$), degraded ($>100\%$), and improved performance ($<100\%$), respectively.
		The table shows the DLS techniques dynamically selected by \sil{} during execution.} 
	\label{fig:SimAS_Normal_sim_416}
\end{figure}

\begin{figure}[]
	\centering
	\subfloat[Exponential workload simulative performance on 128 cores]{%
		\includegraphics[clip, trim=0cm 0cm 0cm 0cm, width = 0.8\textwidth]{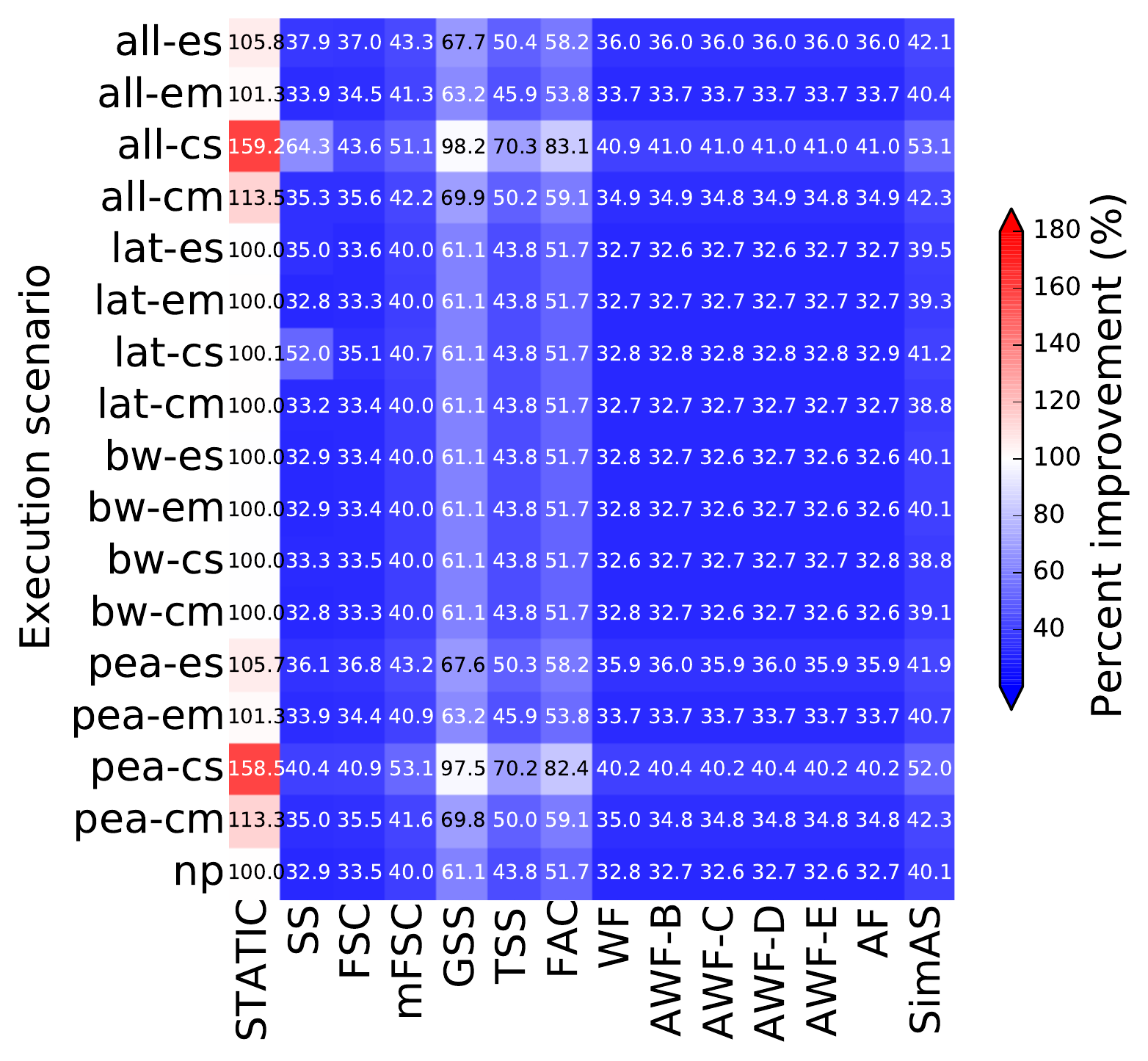}%
		\label{subfig:Exponential_128_sim_heatmap}%
	} \\
	\subfloat[Percentage of counts DLS techniques are selected by \sil{} ]{%
		\includegraphics[clip, trim=0cm 0cm 0cm 0cm, width = 0.8\textwidth]{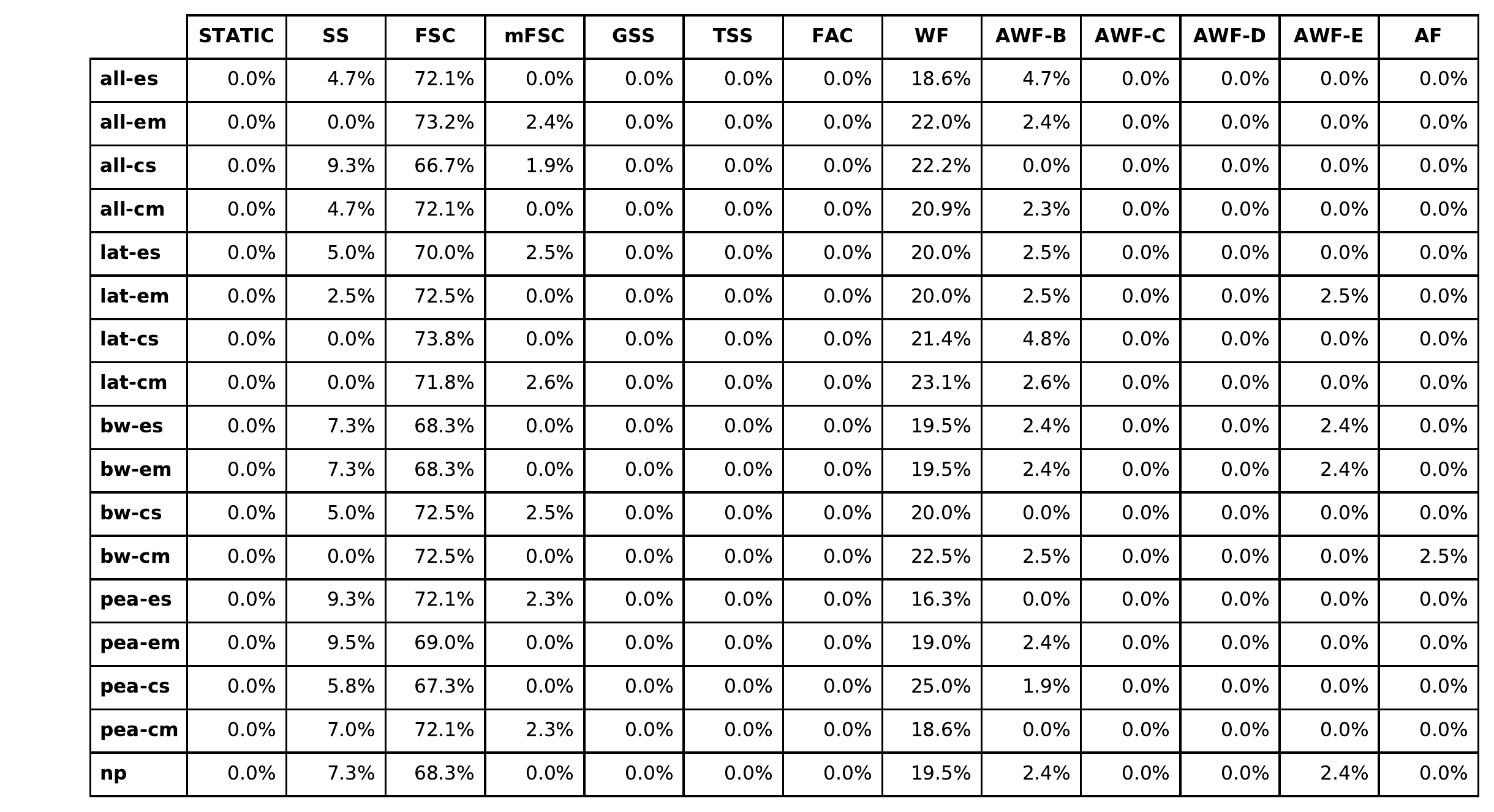}%
		\label{subfig:Exponential_128_sim_table}}
	\\
	\caption{ \textbf{Simulative} performance results of Exponential synthetic workload without (denoted with np) and with (the rest) perturbations using \sil{} and other thirteen loop scheduling techniques on 128 cores of miniHPC. Percent performance improvement normalized to STATIC in np scenario (baseline case without any perturbations and baseline load balancing method). White, red, and blue denote baseline ($=100\%$), degraded ($>100\%$), and improved performance ($<100\%$), respectively.
		The table shows the DLS techniques dynamically selected by \sil{} during execution.} 
	\label{fig:SimAS_Exponential_sim}
\end{figure}

\begin{figure}[]
	\centering
	\subfloat[Exponential workload simulative performance on 416 cores]{%
		\includegraphics[clip, trim=0cm 0cm 0cm 0cm, width = 0.8\textwidth]{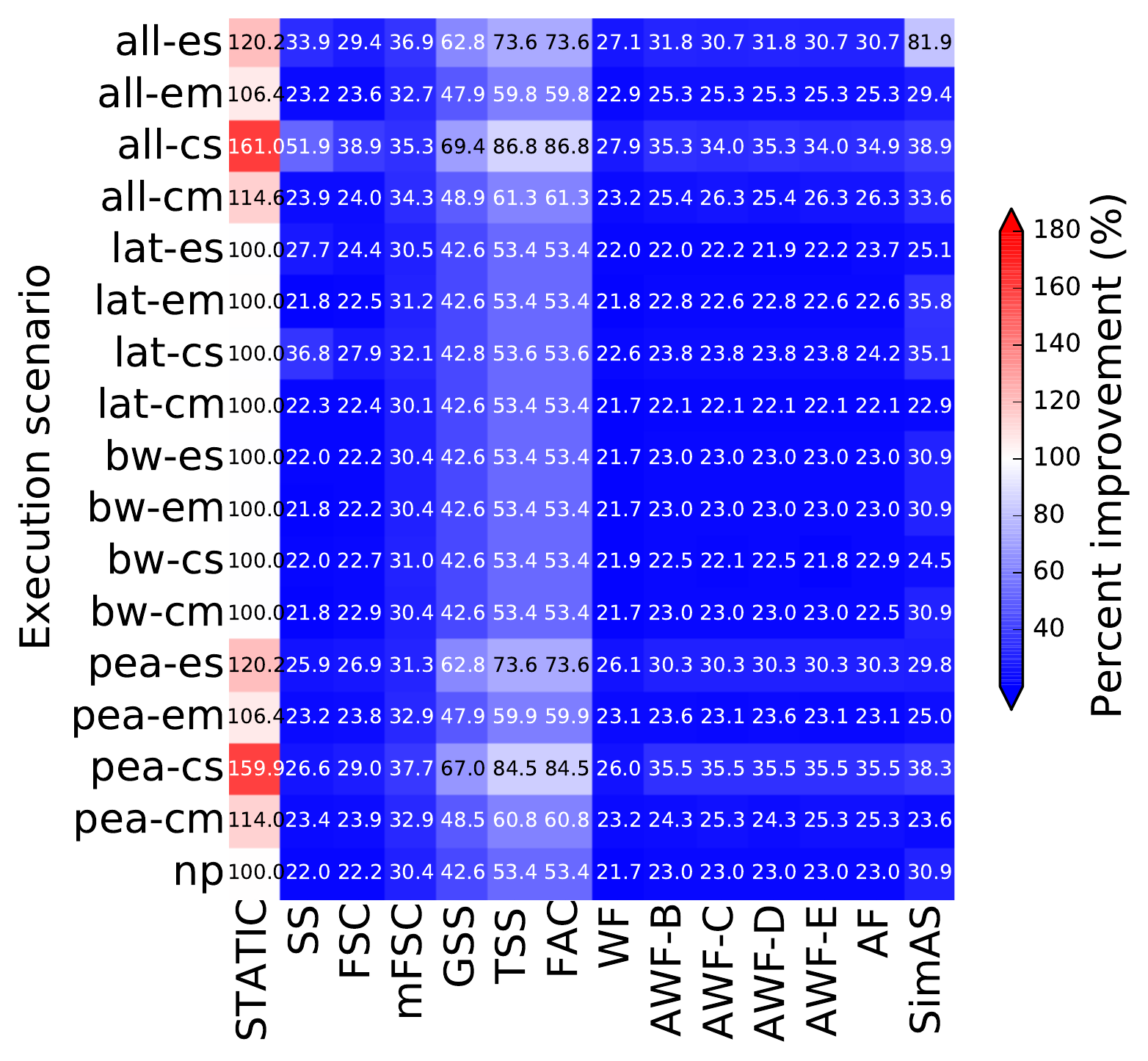}%
		\label{subfig:Exponential_416_sim_heatmap}%
	} \\
	\subfloat[Percentage of counts DLS techniques are selected by \sil{} ]{%
		\includegraphics[clip, trim=0cm 0cm 0cm 0cm, width = 0.8\textwidth]{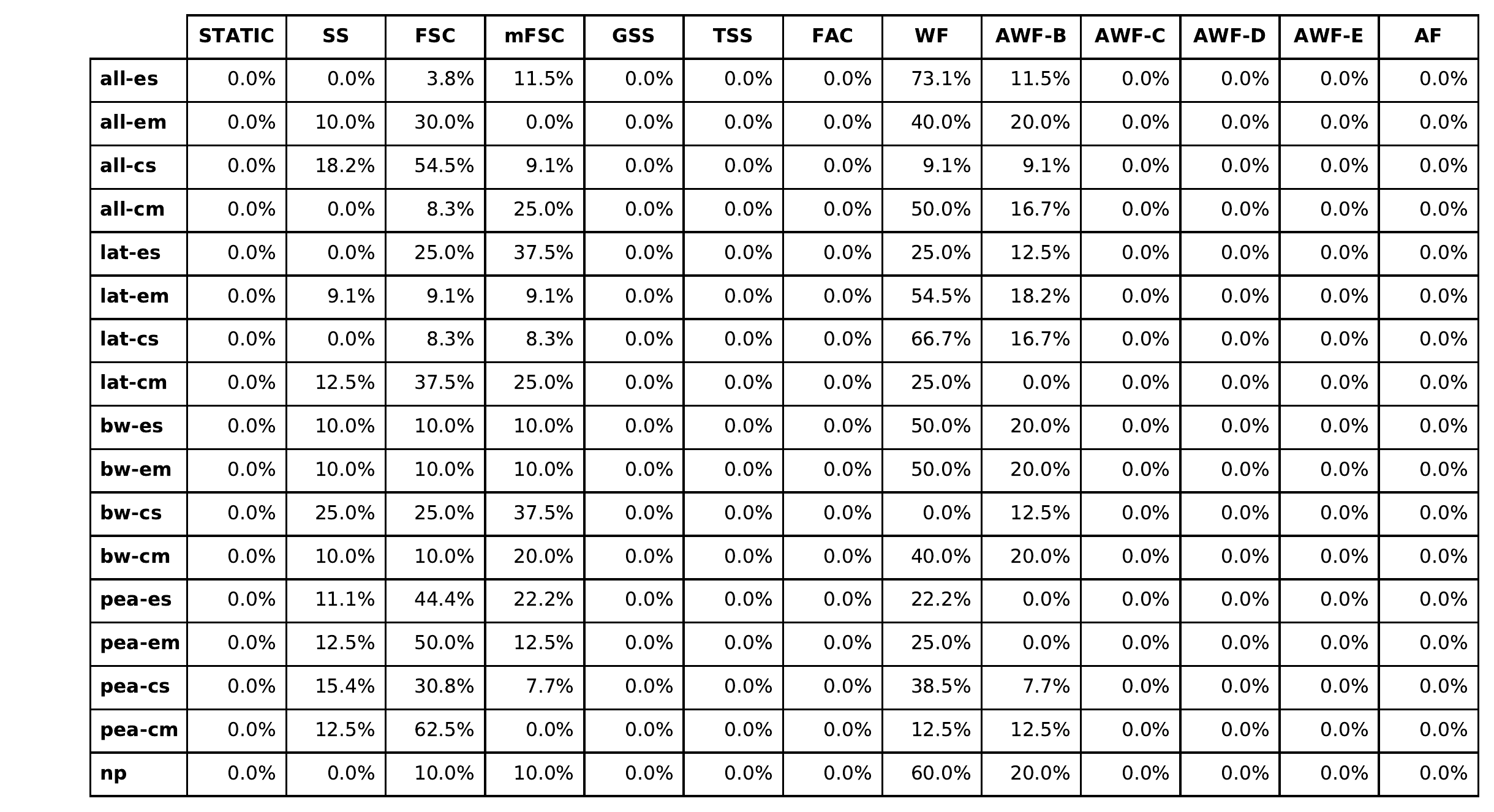}%
		\label{subfig:Exponential_416_sim_table}}
	\\
	\caption{ \textbf{Simulative} performance results of Exponential synthetic workload without (denoted with np) and with (the rest) perturbations using \sil{} and other thirteen loop scheduling techniques on 416 cores of miniHPC. Percent performance improvement normalized to STATIC in np scenario (baseline case without any perturbations and baseline load balancing method). White, red, and blue denote baseline ($=100\%$), degraded ($>100\%$), and improved performance ($<100\%$), respectively.
		The table shows the DLS techniques dynamically selected by \sil{} during execution.} 
	\label{fig:SimAS_Exponential_sim_416}
\end{figure}

\begin{figure}[]
	\centering
	\subfloat[Gamma workload simulative performance on 128 cores]{%
		\includegraphics[clip, trim=0cm 0cm 0cm 0cm, width = 0.8\textwidth]{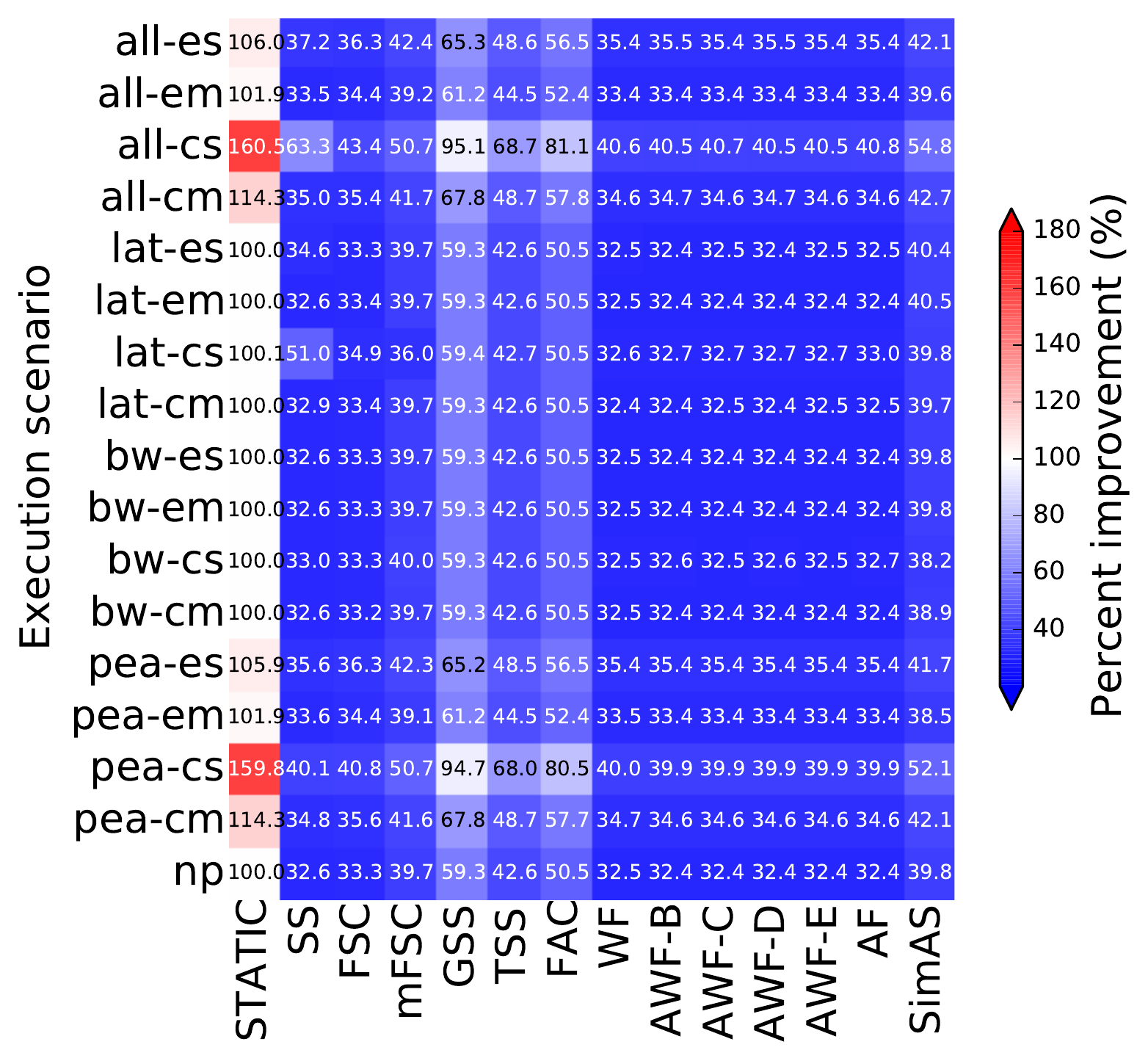}%
		\label{subfig:Gamma_128_sim_heatmap}%
	} \\
	\subfloat[Percentage of counts DLS techniques are selected by \sil{} ]{%
		\includegraphics[clip, trim=0cm 0cm 0cm 0cm, width = 0.8\textwidth]{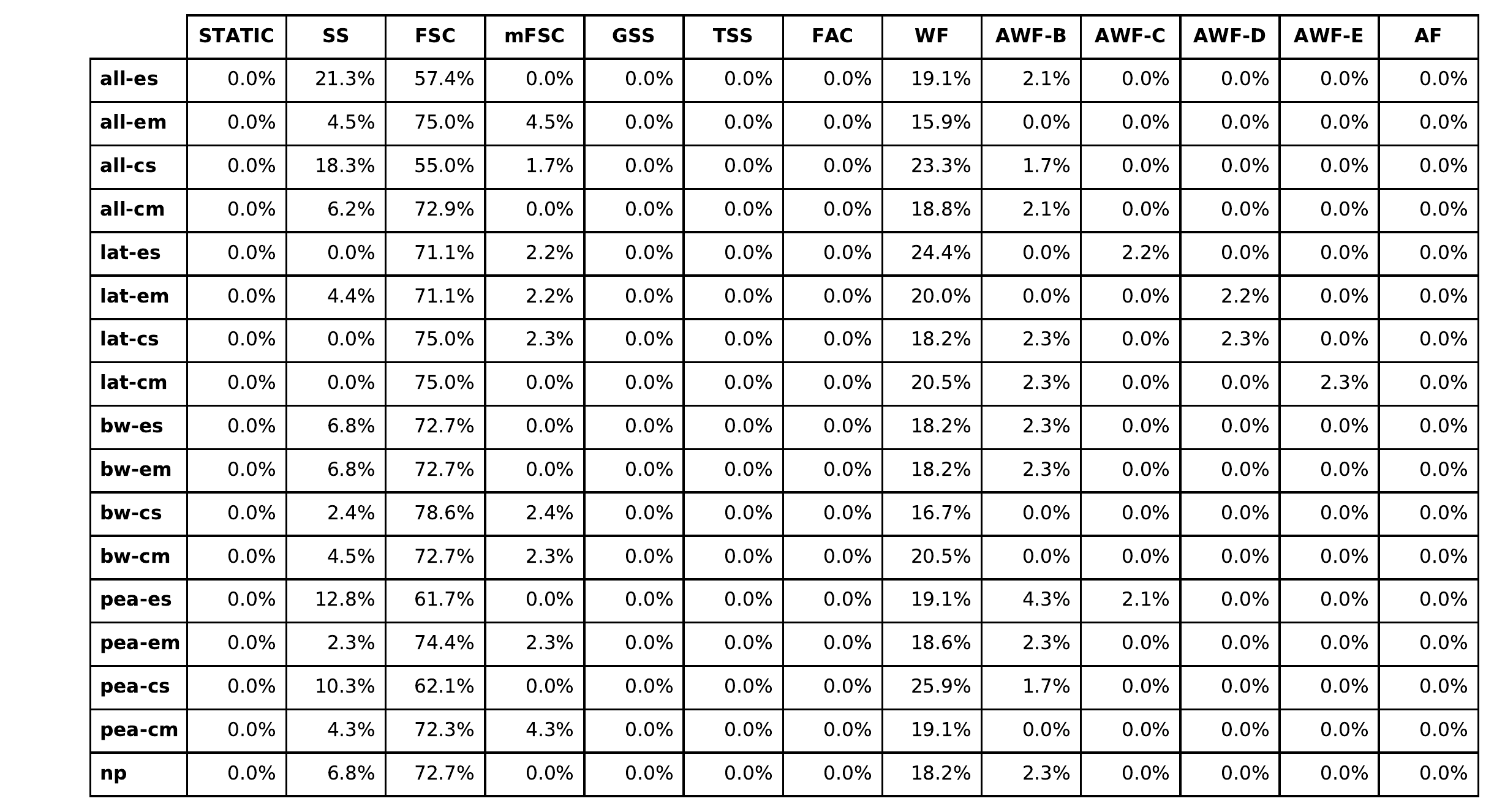}%
		\label{subfig:Gamma_128_sim_table}}
	\\
	\caption{ \textbf{Simulative} performance results of Gamma synthetic workload without (denoted with np) and with (the rest) perturbations using \sil{} and other thirteen loop scheduling techniques on 128 cores of miniHPC. Percent performance improvement normalized to STATIC in np scenario (baseline case without any perturbations and baseline load balancing method). White, red, and blue denote baseline ($=100\%$), degraded ($>100\%$), and improved performance ($<100\%$), respectively.
		The table shows the DLS techniques dynamically selected by \sil{} during execution.} 
	\label{fig:SimAS_Gamma_sim}
\end{figure}

\begin{figure}[]
	\centering
	\subfloat[Gamma workload simulative performance on 416 cores]{%
		\includegraphics[clip, trim=0cm 0cm 0cm 0cm, width = 0.8\textwidth]{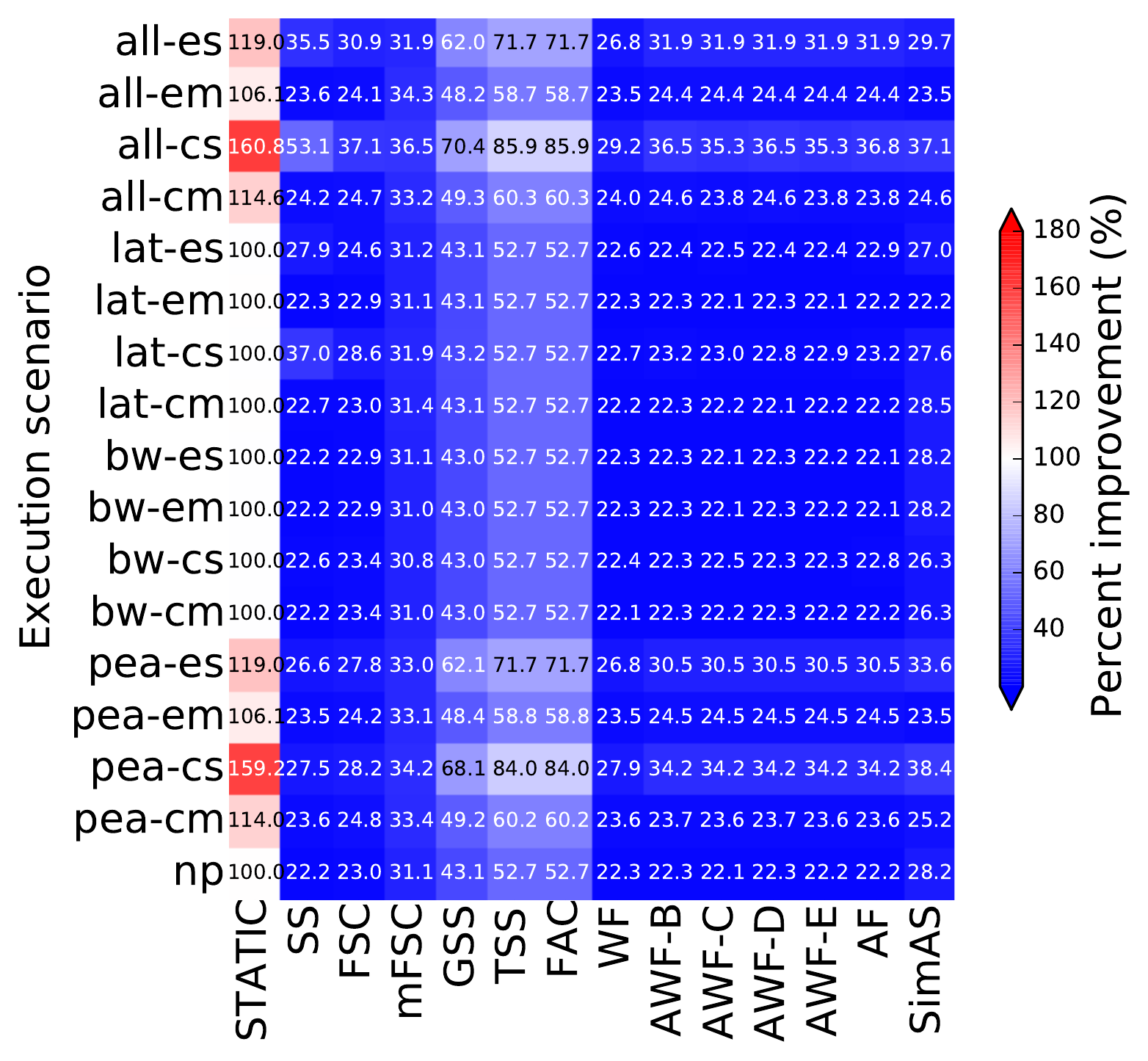}%
		\label{subfig:Gamma_416_sim_heatmap}%
	} \\
	\subfloat[Percentage of counts DLS techniques are selected by \sil{} ]{%
		\includegraphics[clip, trim=0cm 0cm 0cm 0cm, width = 0.8\textwidth]{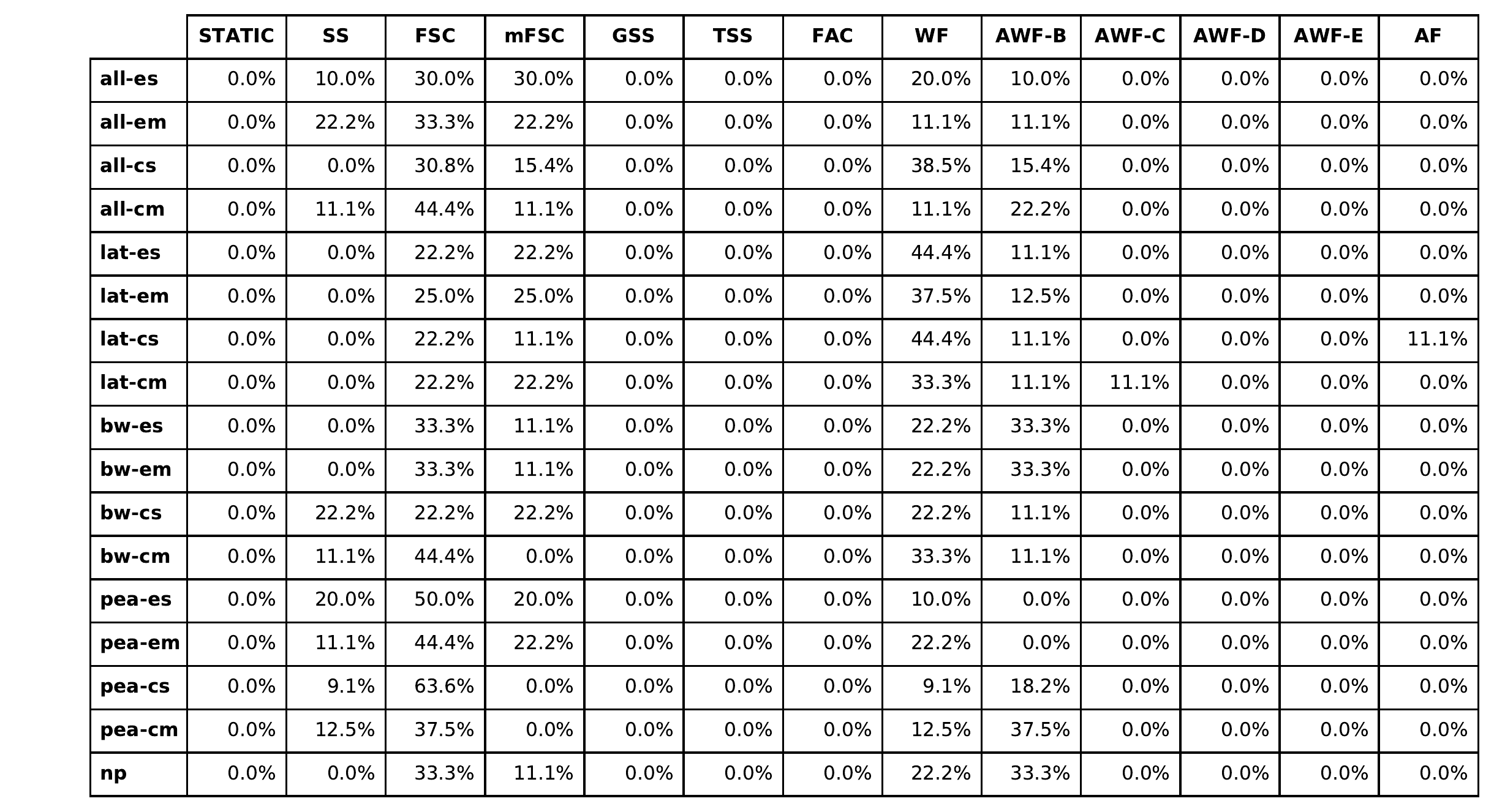}%
		\label{subfig:Gamma_416_sim_table}}
	\\
	\caption{ \textbf{Simulative} performance results of Gamma synthetic workload without (denoted with np) and with (the rest) perturbations using \sil{} and other thirteen loop scheduling techniques on 416 cores of miniHPC. Percent performance improvement normalized to STATIC in np scenario (baseline case without any perturbations and baseline load balancing method). White, red, and blue denote baseline ($=100\%$), degraded ($>100\%$), and improved performance ($<100\%$), respectively.
		The table shows the DLS techniques dynamically selected by \sil{} during execution.} 
	\label{fig:SimAS_Gamma_sim_416}
\end{figure}

Perturbations in the network bandwidth show a minimal influence on performance, as the PEs only communicate loop iteration indices to calculate the start index of the next chunk. 
Therefore, the communicated messages are small.
The bandwidth perturbations are, thus, not selected for \aliD{subsequent more} targeted native experiments under perturbations.

The adaptive techniques perform comparably except for AWF-E in Mandelbrot on 128 cores (\figurename{~\ref{subfig:Mandelbrot_128_sim_heatmap}}), with a slight advantage for AWF-B as can be seen in \figurename{~\ref{subfig:Mandelbrot_416_sim_heatmap}} \texttt{all-cs} and~\texttt{all-es}.
However, in certain cases, other techniques outperform the adaptive techniques. 
Specifically, SS outperforms AWF-B in \figurename{~\ref{subfig:Mandelbrot_128_sim_heatmap}} and \figurename{~\ref{subfig:Mandelbrot_416_sim_heatmap}} \texttt{pea-cs} and \texttt{pea-es}.

\emph{These results suggest that no single DLS outperforms all other techniques in all execution scenarios}. 
Therefore, the best strategy is to dynamically select a DLS based on the current application and system states.
The \sil{} is called every 50~seconds, when there is a work request, to select the best performing DLS. 
\aliD{The} DLS techniques with poor performance on heterogeneous systems, i.e., GSS, TSS, and FAC, are excluded from the DLS portfolio provided to the \sil{} to speed up the simulation.
A closer analysis of the \sil{}-based results reveals that it resulted in the \aliD{shortest} execution time in most execution scenarios, especially for Mandelbrot, as shown in~\figurename{~\ref{subfig:Mandelbrot_416_sim_heatmap}}~\texttt{lat-cs} and~\texttt{lat-es}, and for PSIA~\texttt{pea-cm} in \figurename{~\ref{subfig:PSIA_128_sim_heatmap}} and~\figurename{~\ref{subfig:PSIA_416_sim_heatmap}}. 
In other cases, the application performance with \sil{} was slightly \aliD{poorer} than the \aliD{best} execution time achieved by other DLS \aliD{techniques}. 
This is due to the fact that loop scheduling is, by definition, non-preemptive and the execution of already scheduled loop iterations can not be preempted to be resumed with the newly (expected more suitable) selected DLS.
Inspecting the simulation results of the synthetic workloads in Figures 9 - 18, one can see that the same observations from the real applications are also confirmed by the results of synthetic workloads.

\noindent\textbf{Native experiments.}
\begin{figure}
	\centering
	\subfloat[PSIA native performance on 128 cores]{%
		\includegraphics[clip, trim=0cm 0cm 0cm 0cm, width = 0.9\textwidth]{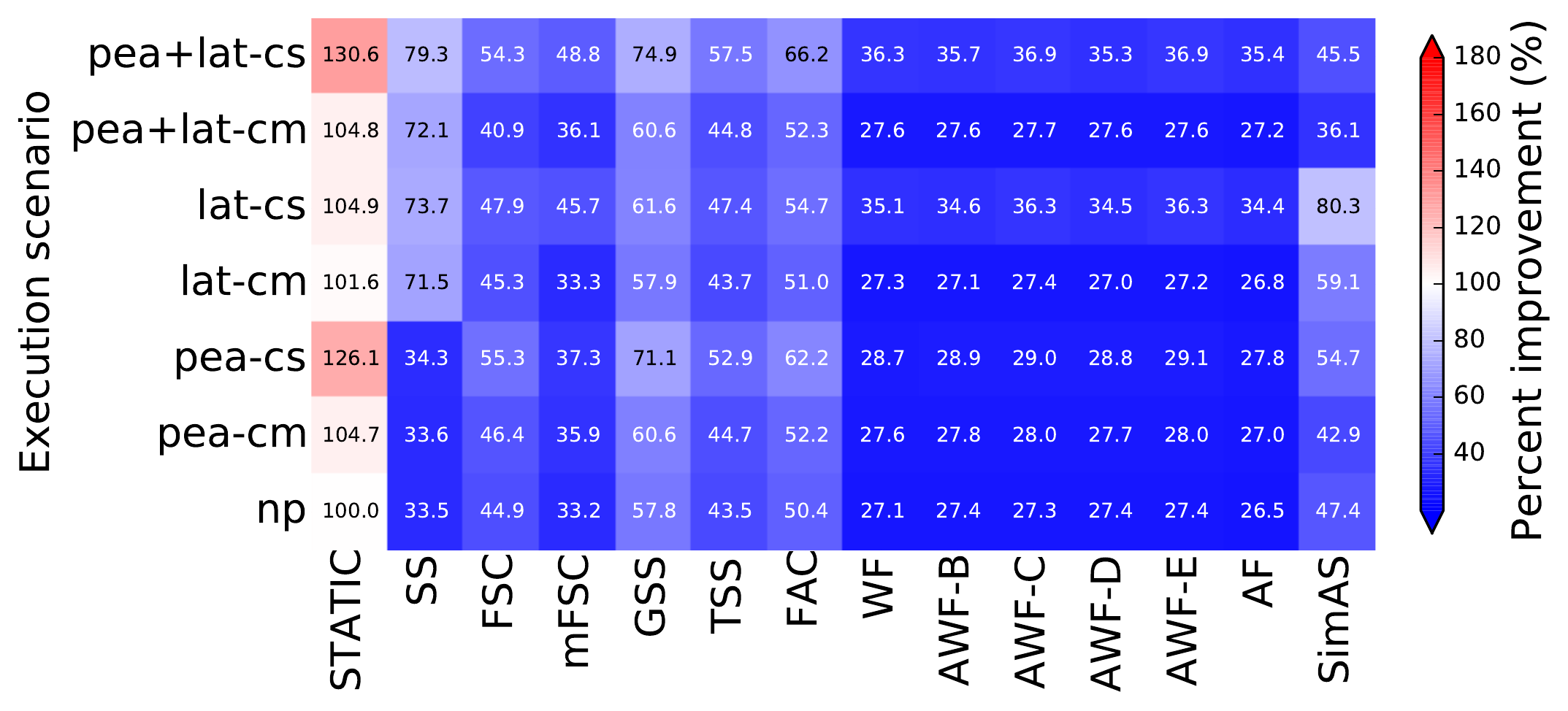}%
		\label{subfig:PSIA_128_native_heatmap}%
	} \\
	\subfloat[Percentage of counts DLS techniques are selected by \sil{} ]{%
		\includegraphics[clip, trim=0cm 0cm 0cm 0cm, width = 0.9\textwidth]{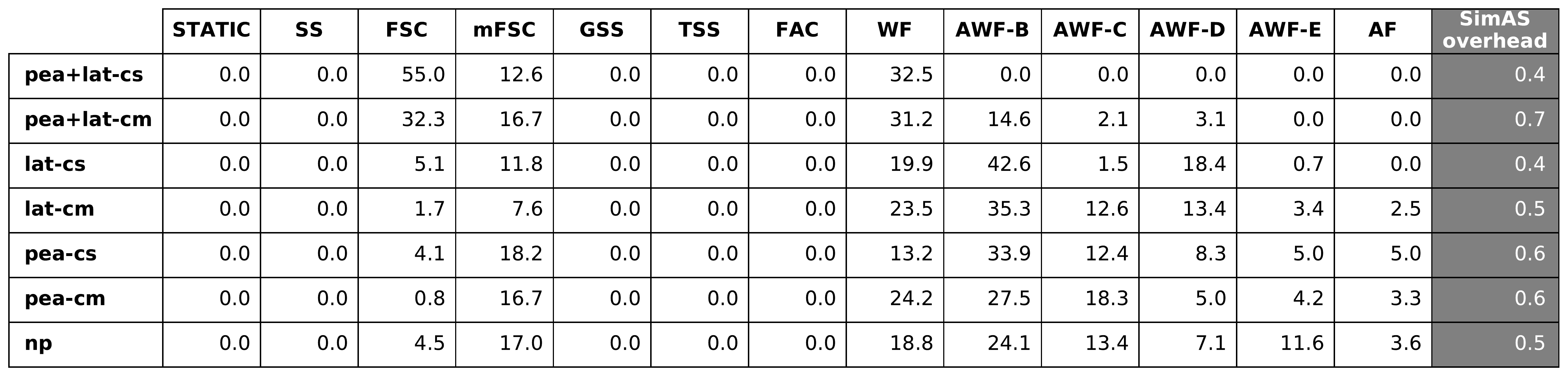}%
		\label{subfig:PSIA_128_native_table}}
	\\
	\caption{ \textbf{Native} performance results of PSIA without (denoted with np) and with (the rest) perturbations using \sil{} and other thirteen loop scheduling techniques on miniHPC. Percent performance improvement normalized to STATIC in np scenario (baseline case without any perturbations and baseline load balancing method). White, red, and blue denote baseline ($=100\%$), degraded ($>100\%$), and improved performance ($<100\%$), respectively.
		The table shows the DLS techniques dynamically selected by \sil{} and the percent of execution time spent in \sil{} calls.} 
	\label{fig:SimAS_PSIA_native}
\end{figure}

\begin{figure}
	\centering
	\subfloat[PSIA native performance on 416 cores]{%
		\includegraphics[clip, trim=0cm 0cm 0cm 0cm, width = 0.9\textwidth]{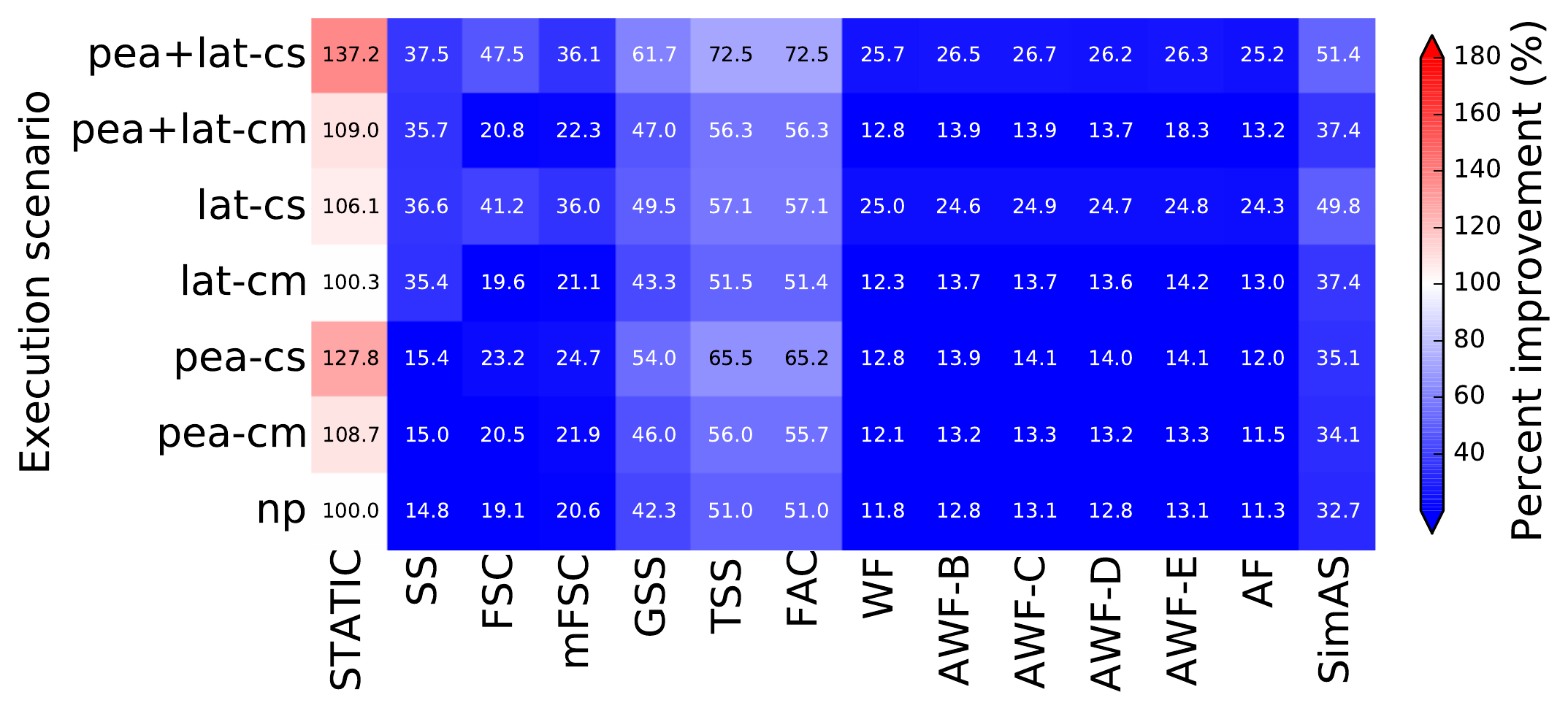}%
		\label{subfig:PSIA_416_native_heatmap}%
	} \\
	\subfloat[Percentage of counts DLS techniques are selected by \sil{} ]{%
		\includegraphics[clip, trim=0cm 0cm 0cm 0cm, width = 0.9\textwidth]{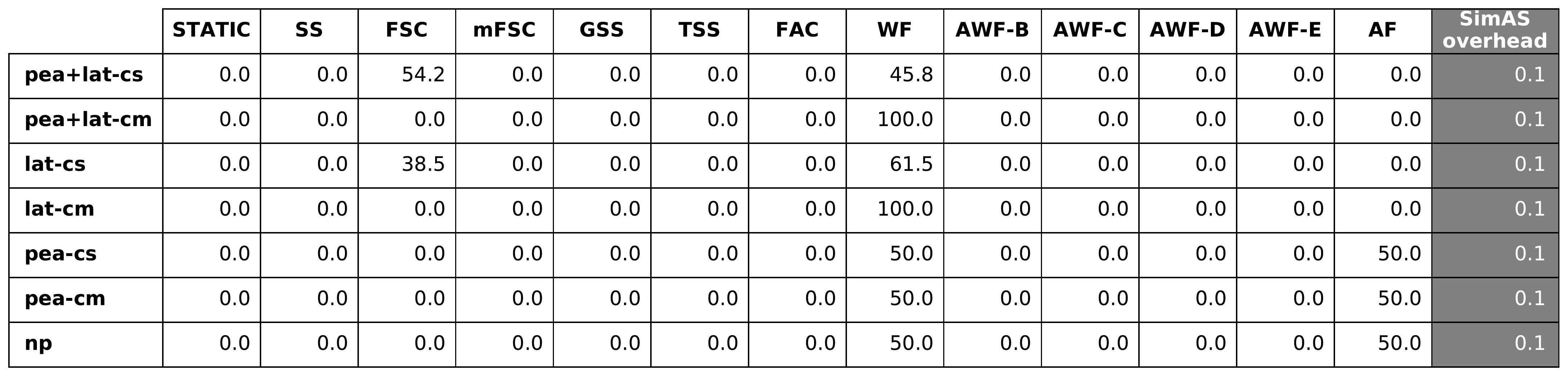}%
		\label{subfig:PSIA_416_native_table}}
	\\
	\caption{ \textbf{Native} performance results of PSIA without (denoted with np) and with (the rest) perturbations using \sil{} and other thirteen loop scheduling techniques on 416 cores of miniHPC. Percent performance improvement normalized to STATIC in np scenario (baseline case without any perturbations and baseline load balancing method). White, red, and blue denote baseline ($=100\%$), degraded ($>100\%$), and improved performance ($<100\%$), respectively.
		The table shows the DLS techniques dynamically selected by \sil{} and the percent of execution time spent in \sil{} calls.} 
	\label{fig:SimAS_PSIA_native_416}
\end{figure}

\begin{figure}
	\centering

	\subfloat[Mandelbrot native performance on 128 cores]{%
		\includegraphics[clip, trim=0cm 0cm 0cm 0cm, width = 0.9\textwidth]{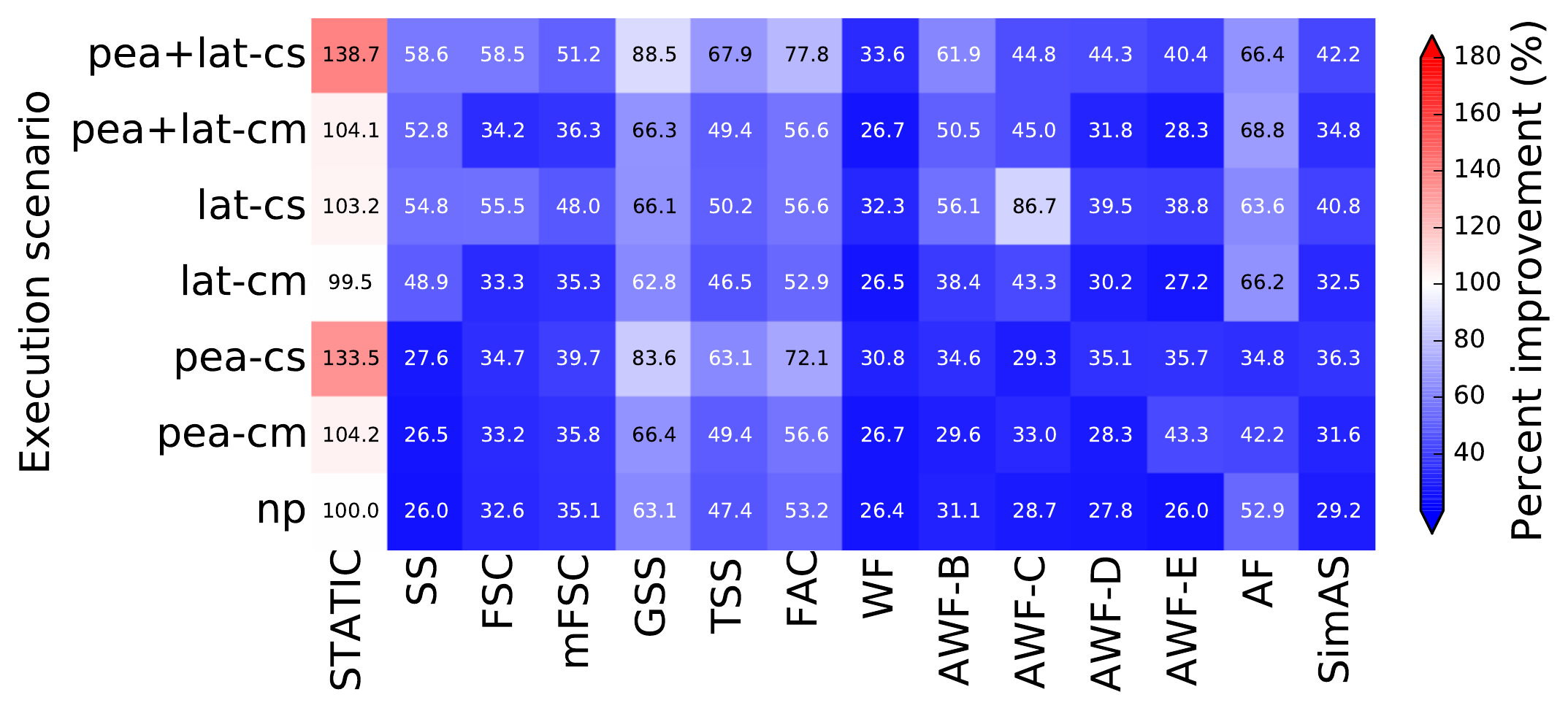}%
		\label{subfig:Mandelbrot_128_native_heatmap}%
	} \\
	\subfloat[Percentage of counts DLS techniques are selected by \sil{} ]{%
		\includegraphics[clip, trim=0cm 0cm 0cm 0cm, width = 0.9\textwidth]{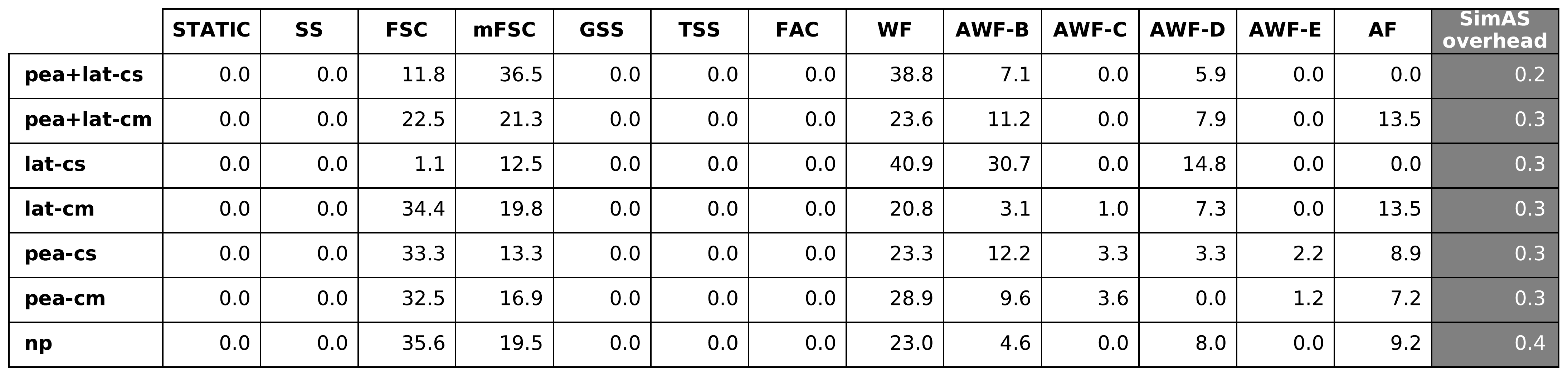}%
		\label{subfig:Mandelbrot_128_native_table}}
	\\
	\caption{ \textbf{Native} performance results of Mandelbrot without (denoted with np) and with (the rest) perturbations using \sil{} and other thirteen loop scheduling techniques on 128 cores of miniHPC. Percent performance improvement normalized to STATIC in np scenario (baseline case without any perturbations and baseline load balancing method). White, red, and blue denote baseline ($=100\%$), degraded ($>100\%$), and improved performance ($<100\%$), respectively.
		Each table shows the DLS techniques dynamically selected by \sil{} and the percent of execution time spent in \sil{} calls.} 
	\label{fig:SimAS_Mandelbrot_native}
\end{figure}
\clearpage

\begin{figure}
	\centering
	\subfloat[Mandelbrot native performance on 416 cores]{%
		\includegraphics[clip, trim=0cm 0cm 0cm 0cm, width = 0.9\textwidth]{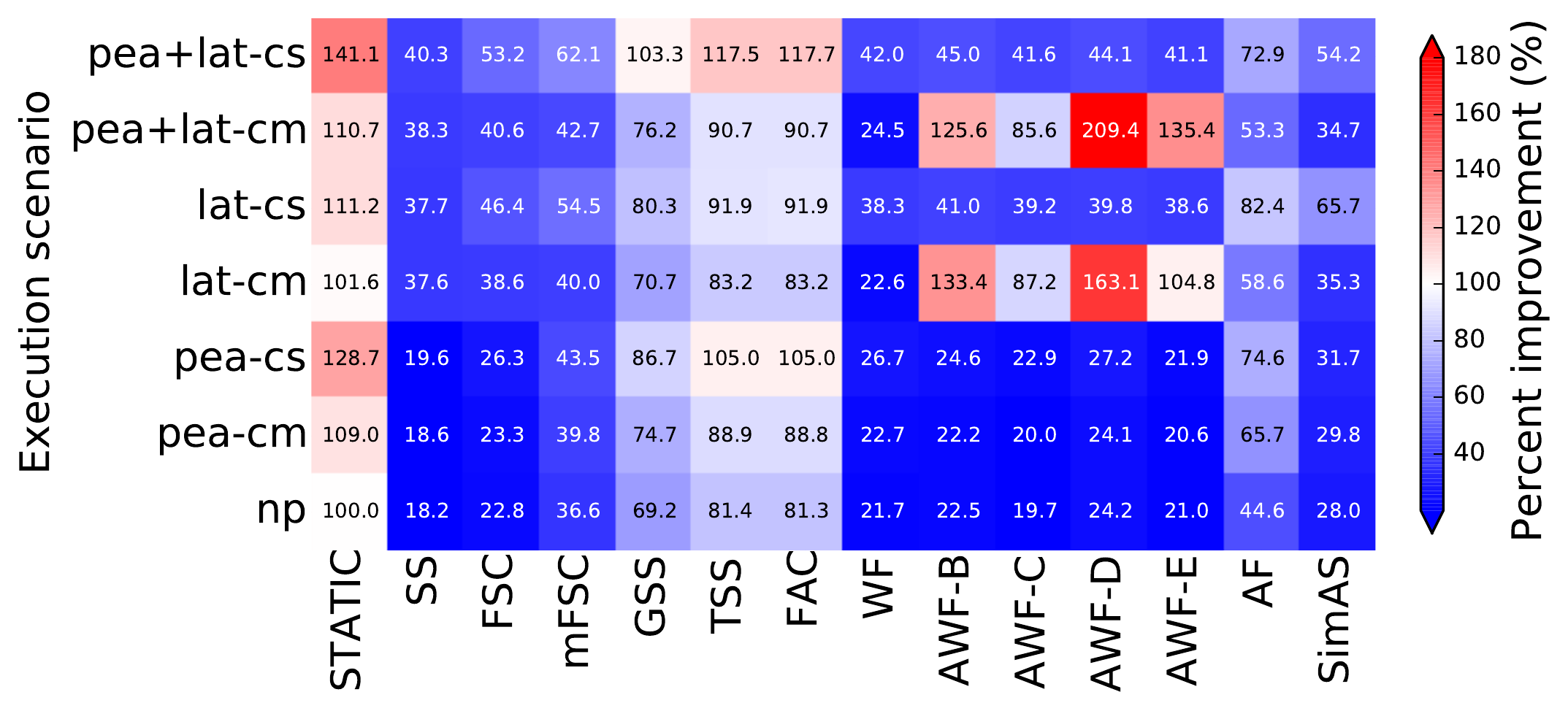}%
		\label{subfig:Mandelbrot_416_native_heatmap}%
	} \\
	\subfloat[Percentage of counts DLS techniques are selected by \sil{} ]{%
		\includegraphics[clip, trim=0cm 0cm 0cm 0cm, width = 0.9\textwidth]{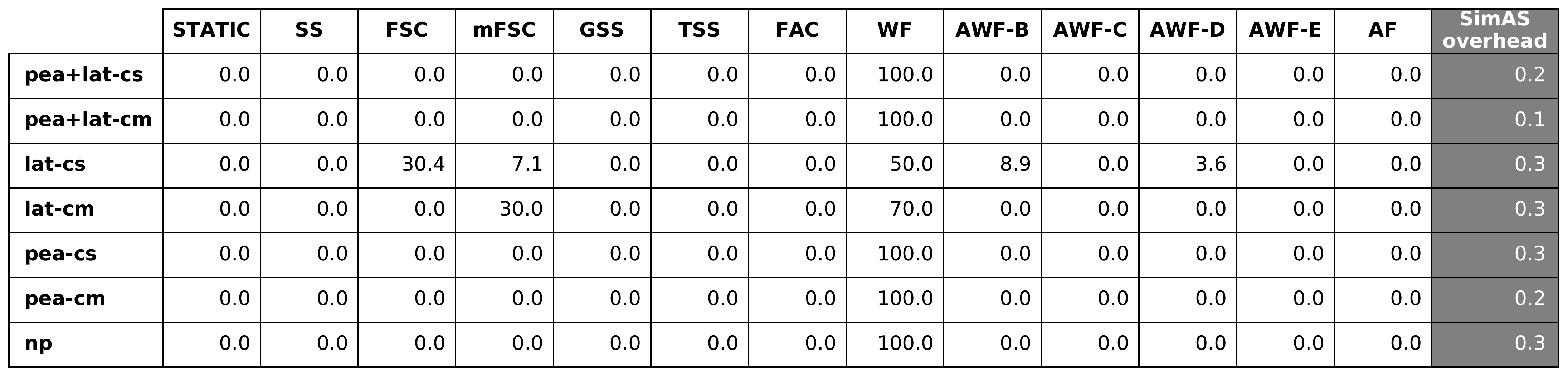}%
		\label{subfig:Mandelbrot_416_native_table}}
	\\
	\caption{ \textbf{Native} performance results of Mandelbrot without (denoted with np) and with (the rest) perturbations using \sil{} and other thirteen loop scheduling techniques on 416 cores of miniHPC. Percent performance improvement normalized to STATIC in np scenario (baseline case without any perturbations and baseline load balancing method). White, red, and blue denote baseline ($=100\%$), degraded ($>100\%$), and improved performance ($<100\%$), respectively.
		Each table shows the DLS techniques dynamically selected by \sil{} and the percent of execution time spent in \sil{} calls.} 
	\label{fig:SimAS_Mandelbrot_native_416}
\end{figure}

\aliD{A targeted selection of native experiments have been conducted for PSIA and Mandelbrot. 
The constant distribution of perturbation values was selected, as it significantly impacts the applications performance. 
Perturbations in the network bandwidth were excluded from native experimentation due to their minimal impact on performance (as shown above).} The performance results of PSIA and Mandelbrot with the thirteen DLS techniques under perturbations is \aliD{shown} in~Figures 19 -22. 
Similar to the \mbox{simulation-based predictions}, the nonadaptive DLS techniques perform poorly on \aliD{the perturbed heterogeneous system}. 
\aliD{In particular}, STATIC, GSS, TSS, and FAC are highly affected by all considered perturbations. 
Unlike \aliD{in the \mbox{simulation-based} predictions}, STATIC is also slightly affected by latency perturbations. 
This is due to the fact \aliD{that} STATIC is implemented in the \dlb{} in a \mbox{self-scheduling} manner, i.e., workers obtain chunks of loop iterations during execution when they become free.
The chunk size of STATIC is \aliD{equal to} the total number of loop iterations divided by the number of worker \aliD{processes}.
Therefore, each worker obtains exactly one chunk.
The adaptive techniques resulted in comparable performance. 
However,  in certain cases, AWF-E performed poorly in latency perturbations scenarios.
Similar to the \mbox{simulation-based} predictions, the AWF-B outperforms all other techniques in most the execution scenarios. 
The \sil{} results in the \aliD{shortest} execution time in most of the cases, especially for PSIA. 
The application performance with \sil{} \aliD{degraded} in certain cases due to the \mbox{non-preemptive} scheduling \aliD{implementation}. 
Even though the technique with the best performance is selected \aliD{upon a new call to \sil{}}, the execution of already scheduled loop iterations can not be preempted to be resumed with the newly selected DLS.

\begin{figure}[]
	\centering
	\subfloat[PSIA\_TS native performance on 128 cores]{%
		\includegraphics[clip, trim=0cm 0cm 0cm 0cm, width = 0.9\textwidth]{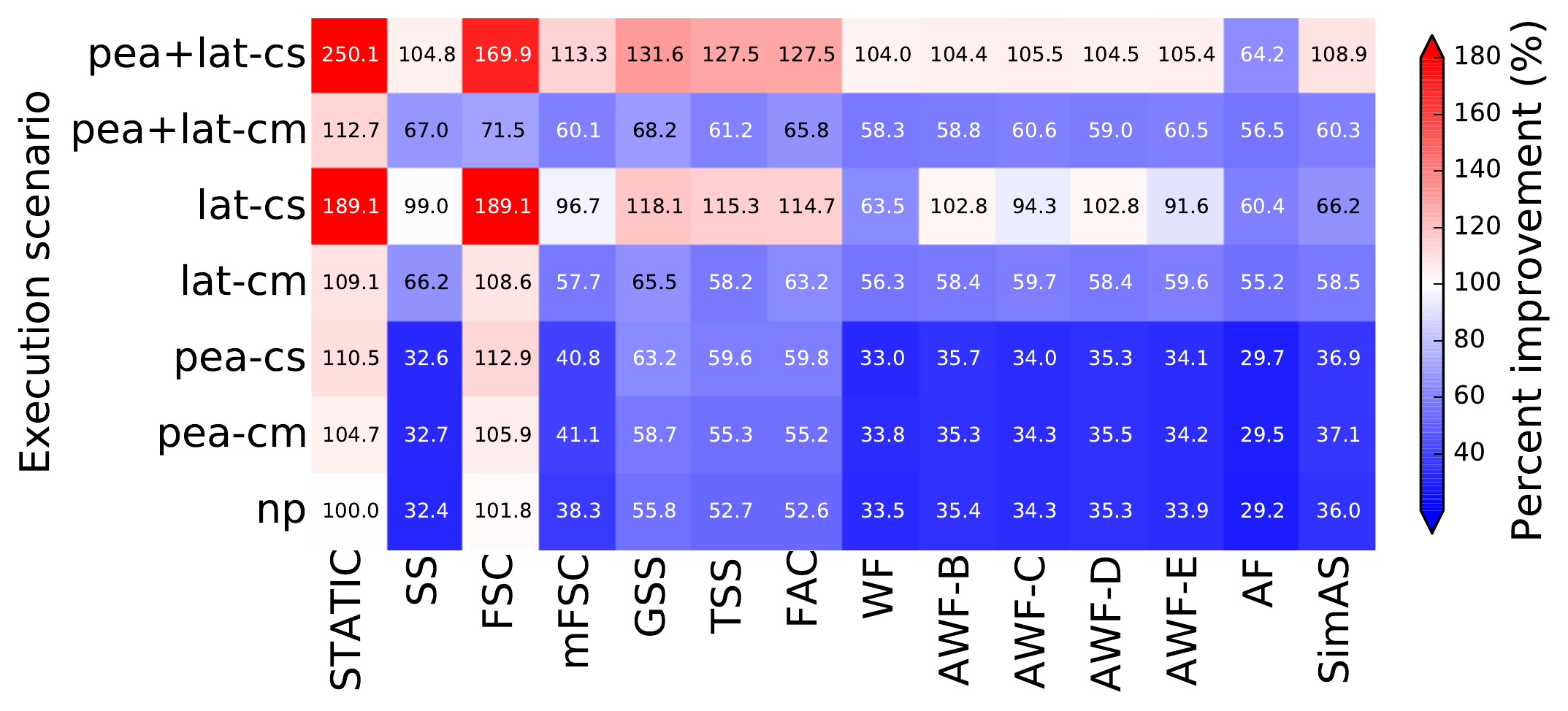}%
		\label{subfig:PSIA_TS_128_native_heatmap}%
	} \\
	\subfloat[Percentage of counts DLS techniques are selected by \sil{} ]{%
		\includegraphics[clip, trim=0cm 0cm 0cm 0cm, width = 0.9\textwidth]{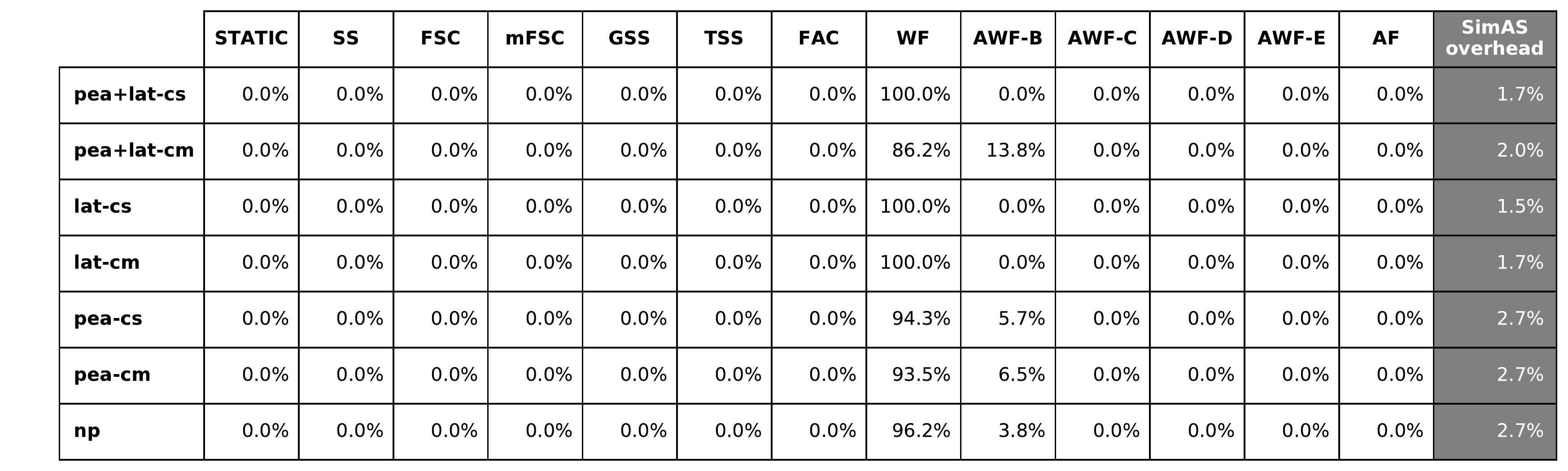}%
		\label{subfig:PSIA_TS_128_native_table}}
	\\
	\caption{ \textbf{Native} performance results of PSIA\_TS without (denoted with np) and with (the rest) perturbations using \sil{} and other thirteen loop scheduling techniques on miniHPC. Percent performance improvement normalized to STATIC in np scenario (baseline case without any perturbations and baseline load balancing method). White, red, and blue denote baseline ($=100\%$), degraded ($>100\%$), and improved performance ($<100\%$), respectively.
		Each table shows the DLS techniques dynamically selected by \sil{} and the percent of execution time spent in \sil{} calls.} 
	\label{fig:SimAS_PSIA_TS_native}
\end{figure}

\begin{figure}[]
	\centering
	\subfloat[Mandelbrot\_TS native performance on 128 cores]{%
		\includegraphics[clip, trim=0cm 0cm 0cm 0cm, width = 0.9\textwidth]{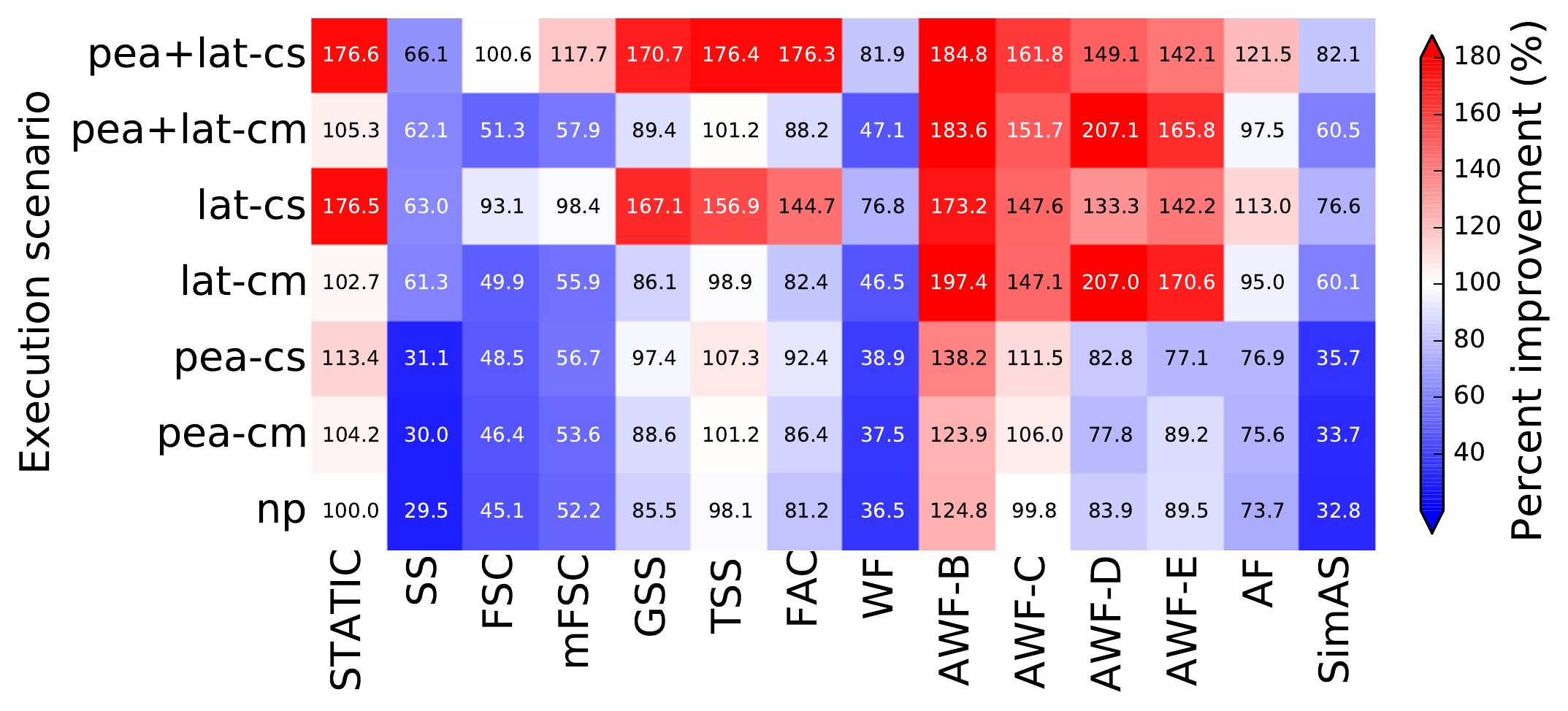}%
		\label{subfig:Mandelbrot_TS_128_native_heatmap}%
	} \\
	\subfloat[Percentage of counts DLS techniques are selected by \sil{} ]{%
		\includegraphics[clip, trim=0cm 0cm 0cm 0cm, width = 0.9\textwidth]{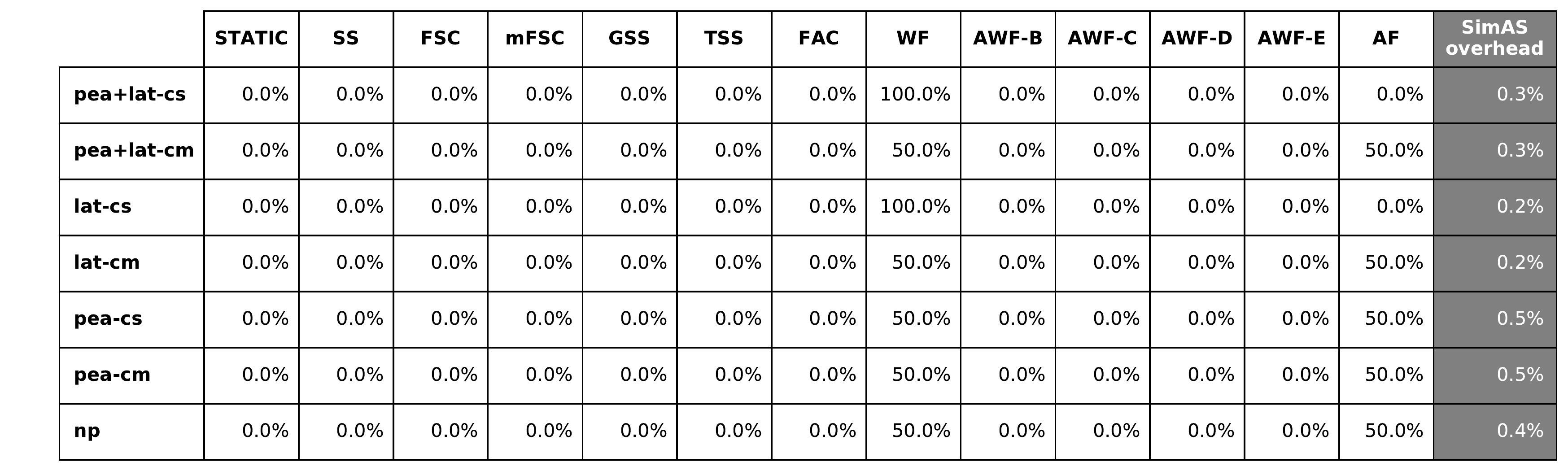}%
		\label{subfig:Mandelbrot_TS_128_native_table}}
	\\
	\caption{ \textbf{Native} performance results of Mandelbrot\_TS without (denoted with np) and with (the rest) perturbations using \sil{} and other thirteen loop scheduling techniques on miniHPC. Percent performance improvement normalized to STATIC in np scenario (baseline case without any perturbations and baseline load balancing method). White, red, and blue denote baseline ($=100\%$), degraded ($>100\%$), and improved performance ($<100\%$), respectively.
		The table shows the DLS techniques dynamically selected by \sil{} and the percent of execution time spent in \sil{} calls.} 
	\label{fig:SimAS_Mandelbrot_TS_native}
\end{figure}

To show the applicability of \sil{} approach to scientific applications, \mbox{time-stepping} versions of PSIA and Mandelbrot are also executed under perturbations with and without \sil{}.
In \mbox{time-stepping} applications, i.e., PSIA\_TS and Mandelbrot\_TS, \sil{} starts a new simulation at the beginning of each time step. 
WF is used as the default DLS technique in these experiments or the same DLS from the previous \mbox{time-steps} until the simulations are finished.
\sil{} selects the best performing DLS techniques based on the prediction from simulations for the current \mbox{time-step}.
This represents another \mbox{use-case} of \sil{} in \mbox{time-stepping} applications, which is frequently encountered in scientific applications.
The results of the \mbox{time-stepping} applications are shown in~\figurename{~\ref{fig:SimAS_PSIA_TS_native}} and \figurename{~\ref{fig:SimAS_Mandelbrot_TS_native}}.
Similar to the \mbox{non-time-stepping} versions, \sil{} improved the performance of applications in most of the cases. 
We note that no single DLS technique always achieves the best performance. Therefore, a dynamic selection of the DLS technique according to the current perturbations in the system is needed.
The \sil{} overhead is, in general, below $0.5\%$ of the execution time, except for PSIA\_TS, which has the overhead of $2.7\%$ at the most.
This is due to the short execution time of the \mbox{time-stepping} version of the PSIA compared to the \mbox{non-time-stepping} version.

\subsection{Discussion}
\label{subsec:discussion}
Even though the applications considered are \mbox{computationally-intensive} and only communicate loop indices with the master, \emph{perturbations in network latency had a significant impact on performance}. 
The implementation choice of the scheduling techniques, such as STATIC, implemented in a \mbox{self-scheduling} fashion, led to degrading its performance in scenarios with network perturbations. 
In most experiments, all the adaptive DLS techniques perform comparably. 
However, in certain instances, e.g., AWF-C and AF in \figurename{~\ref{fig:SimAS_Mandelbrot_native}} in~\texttt{lat-cm} and~\texttt{lat-cs}, their performance was significantly poorer compared to other adaptive DLS techniques.
This poor performance is due to the short execution time of the Mandelbrot application and the high variability of the loop iteration execution times, in addition to the added perturbations, which does not allow the core weights learned by these techniques to converge to the correct value.

Selecting the most performing DLS technique before execution might not deliver the best performance, as perturbations in the HPC system are unknown a priori.
For instance, the best DLS technique for Mandelbrot that could be identified before execution, i.e., in \textit{np} execution scenario, is SS, which is outperformed by \sil{} in \textit{lat-cs} and \textit{pea+lat-cs} in~\figurename{~\ref{fig:SimAS_Mandelbrot_native}}. 
A similar change in the best DLS technique is observed in the results of Mandelbrot\_TS in~\figurename{~\ref{fig:SimAS_Mandelbrot_TS_native}}.
Since there is no high load imbalance in the PSIA or PSIA\_TS, there is no high variation in the performance of different DLS techniques.
\emph{Since the best DLS technique can not be known before execution, \sil{} improved the performance by dynamically selecting the DLS with the best performance based on the simulation predictions.}

In general, DLS techniques are designed to be efficient.
However, efficiency prevents robustness due to the low tolerance of efficient techniques to uncertain events.
Uncertainty is ineradicable, and it manifests in HPC systems as perturbations.
This highlights the importance of the careful choice of DLS techniques for each application, system size, and execution scenario.
Dynamic selection of DLS techniques ensures that each DLS technique is employed where it is the most efficient.

The \sil{} approach can proactively select the best suited DLS before any perturbations manifest in the system, whenever perturbations can be predicted in advance.
The \sil{} leverages the use of already developed simulators, instead of needing the development of novel prediction techniques.
The DLS selection decisions taken by \sil{} can then be used to create a \mbox{rule-based} DLS selection mechanism for a combination of application, system, and execution scenarios, to improve application performance dynamically without the need of online simulation.

Running \sil{} simulations and the dynamic selection of DLS techniques incurs overhead. However, this overhead has a limited effect on applications' performance.
For example, the total time spent in \texttt{\sil{}\_setup} and \texttt{\sil{}\_update} functions is $3.49$ seconds out of $1147.55$ total application execution time for the PSIA on 128 cores in the \texttt{lat-cs} execution scenario.
However, due to the \mbox{non-preemptive} property of the DLS, the execution of already scheduled chunks of loop iterations is not preempted to be resumed with the newly selected DLS.
As shown in \figurename{~\ref{subfig:PSIA_128_native_table}}, even though the \sil{} selected DLS techniques with shorter execution times in the case of \texttt{lat-cs} with PSIA application on 128 cores, the execution time with \sil{} was even longer than that of SS, which was not selected by the \sil{}.

In \mbox{time-stepping} applications, the effect of frequent DLS technique switching and the \mbox{non-preemption} overhead is much less than the \mbox{single-sweep} applications. 
Therefore, the performance of \mbox{time-stepping} applications with \sil{} under perturbations is better than the \mbox{single-sweep} versions of the same applications as can be seen in~\figurename{~\ref{fig:SimAS_PSIA_TS_native}} and \figurename{~\ref{fig:SimAS_Mandelbrot_TS_native}}.
The preemption of scheduled (yet not executed) loop iterations may improve the performance while switching DLS techniques.

\section{Conclusion and Future Work}
\vspace{-0.35cm}
\label{sec:conc}
\aliC{
A new control-theoretic inspired approach, namely simulator-assisted scheduling (\sil{}) \aliD{approach}, was introduced to dynamically select a DLS that \aliD{is predicted to deliver} the best performance under unpredictable perturbations.
The performance of two \aliD{real} applications and five synthetic workloads \aliD{was} studied under perturbations and insights into the resilience of the DLS techniques to perturbations are provided. 
The performance results confirm the hypothesis that no single DLS technique can achieve the best performance in all the considered execution scenarios.
Furthermore, native DLS experiments under \aliD{system-induced} perturbations showed that even the computationally-intensive applications could be significantly affected by perturbations in the network \aliD{characteristics}.
The implementation choice of scheduling techniques, such as \aliD{STATIC implemented in a self-scheduling manner}, led to the degradation of its performance \aliD{under} network perturbations.
Using the \sil{} approach improved the performance of applications in most experiments.
\sil{} leverages state-of-the-art simulators to select the DLS predicted to result in the best performance of an application under perturbations.
\cut{The \sil{} can be asynchronously launched concurrently to the application execution.}{}
However, due to applications being non-preemptively scheduled, changing the DLS \aliD{technique} during execution may not \aliD{always} result in the best performance. 
It is planned in the future to experiment with preempting scheduled \aliD{yet not executed} loop iterations upon \aliD{a} change \aliD{in} the selected DLS technique by the \sil{} \aliD{approach}. 
Furthermore, experiments to investigate and enhance the performance of \sil{}, in terms of improving the DLS selection strategy and the period between \sil{} calls, are also planned as future work.
}
 {

\subsubsection*{Acknowledgments}
This work has been supported by the Swiss Platform for Advanced Scientific Computing (PASC) project SPH-EXA: Optimizing Smooth Particle Hydrodynamics for Exascale Computing  and by the Swiss National Science Foundation in 
the context of the Multi-level Scheduling in Large Scale High Performance Computers (MLS) grant number 169123. The authors gratefully acknowledge Ahmed Eleliemy for sharing an initial implementation of the PSIA application.

\bibliographystyle{ieeetr}
\bibliography{citedatabase}%

\end{document}